\documentclass[aps,prd,twocolumn,superscriptaddress,preprintnumbers,floatfix,showpacs]{revtex4-1}

\usepackage{graphicx}
\usepackage[caption=false]{subfig}
\usepackage{amsmath}
\usepackage{amssymb}
\usepackage{tikz}
\usepackage{siunitx}
\usepackage{color}
\usepackage[colorlinks=True]{hyperref}
\usepackage{standalone}
\usepackage{microtype}

\usetikzlibrary{patterns}

\newcommand{\ts}{\textsuperscript}

\newcommand{\beq}{\begin{equation}}
\newcommand{\eeq}{\end{equation}}

\newcommand*{\eq}[1]{Eq.\ \eqref{eq:#1}}
\newcommand*{\fig}[1]{Fig.\ \ref{fig:#1}}

\newcommand{\tp}{\tau}
\newcommand{\frot}{f_{\rm rot}}
\newcommand{\fgw}{f}

\newcommand{\hul}[1]{h_{\rm #1}^{95\%}}
\newcommand{\hT}{{h_{\rm t}}}
\newcommand{\hV}{{h_{\rm v}}}
\newcommand{\hS}{{h_{\rm s}}}

\newcommand{\data}{{\bf B}}
\newcommand{\datum}{{B_k}}
\newcommand{\datumifo}{{B_{i,k}}}

\newcommand*{\hyp}[1]{{{\cal H}_{\rm #1}}}
\newcommand*{\hypsub}[2]{{{\cal H}_{{\rm #1}_{#2}}}}
\newcommand{\hypn}{\hyp{N}}
\newcommand{\hyps}{\hyp{S}}

\newcommand*{\bayes}[2]{{\cal B}^{\rm #1}_{\rm #2}}
\newcommand{\bayessn}{\bayes{S}{N}}

\newcommand*{\odds}[2]{{\cal O}^{\rm #1}_{\rm #2}}
\newcommand{\oddssn}{\odds{S}{N}}

\newcommand{\p}{P}
\newcommand{\pdf}{p}

\newcommand{\nseg}{{N_{\rm S}}}
\newcommand{\ndet}{{N_{\rm D}}}
\newcommand{\npsr}{{N_{\rm P}}}
\newcommand{\nrun}{{N_{\rm R}}}

\newcommand{\xmax}{x_{\rm max}}
\newcommand{\xmin}{x_{\rm min}}
\newcommand{\xul}{x^{95\%}}

\newcommand{\posterior}{\pdf(\vec{\theta} \mid \data, \hyp{})}
\newcommand{\prior}{\pdf(\vec{\theta} \mid \hyp{})}
\newcommand{\evidence}{\p\left(\data\mid\hyp{}\right)}

\newcommand*{\red}[1]{#1}

\graphicspath{{./fig/}}

\begin{document}

\preprint{LIGO-P1600305}

\title{Probing Dynamical Gravity with the Polarization of Continuous Gravitational Waves}


\author{Maximiliano Isi}
\email[]{misi@ligo.caltech.edu}
\affiliation{LIGO Laboratory, California Institute of Technology,
Pasadena, California 91125, USA}

\author{Matthew Pitkin}
\affiliation{University of Glasgow, Glasgow G12 8QQ, Scotland, UK}

\author{Alan J.\ Weinstein}
\affiliation{LIGO Laboratory, California Institute of Technology,
Pasadena, California 91125, USA}


\date{\today}

\begin{abstract}
The direct detection of gravitational waves provides the opportunity to measure
fundamental aspects of gravity which have never been directly probed before,
including the polarization of gravitational waves. In the context of searches
for continuous waves from known pulsars, we present novel methods to detect
signals of any polarization content, measure the modes present and place
upper limits on the amplitude of nontensorial components. This will allow us
to obtain new model-independent, dynamical constraints on deviations from
general relativity. We test this framework on multiple potential sources
using simulated data from three advanced-era detectors at design sensitivity.
We find that signals of any polarization will become detectable and
distinguishable for characteristic strains $h\gtrsim 3\times10^{-27}
\sqrt{1~{\rm yr}/T}$, for an observation time $T$. We also find that our
ability to detect nontensorial components depends only on the power present in
those modes, irrespective of the strength of the tensorial strain.
\end{abstract}


\maketitle



\section{Introduction}

The recent detection of gravitational waves (GWs) by the advanced Laser
Interferometer Gravitational-Wave Observatory (aLIGO) heralds the beginning of
the long-awaited era of GW astronomy \cite{gw150914, gw151226}. One of the main
goals of this field is to use GWs as a probe of fundamental physics in the
highly dynamical and strong-field regimes of gravity, as predicted by the
general theory of relativity (GR). The first few GW detections have already
been used to place some of the most stringent constraints on deviations from GR
in this domain, which is inaccessible to laboratory, Solar System or
cosmological tests of gravity.

However, it has not been possible to use LIGO signals to learn about the
polarization content of GWs \cite{gw150914_tgr}, a measurement highly relevant
when comparing GR to many of its alternatives \cite{tegp, Will2006}. The reason
for this is that the relative orientation of the two LIGO detectors makes it
nearly impossible to unequivocally characterize the polarizations of transient
GW signals like the compact-binary coalescences (CBCs) observed so far.
In fact, at least five noncoaligned quadrupolar detectors would be needed
to break the degeneracies of all five nondegenerate polarizations allowed by
generic metric theories of gravity \cite{Eardley1973a, Eardley1973b}.

Existing observations that are usually taken to constrain the amount of allowed
non-GR polarizations can do so only in an indirect and strongly model-dependent
manner. For example, measurements of the orbital decay of binary systems are
sensitive to the total radiated GW power, but do not probe the waves directly
(see e.g.\ \cite{Weisberg2010, Freire2012}, or \cite{Stairs2003, Wex2014} for
reviews). In the context of specific alternative theories (e.g.\ scalar-tensor)
such observations can indeed constrain the power in extra polarizations;
however, they provide no direct, model-independent information on the actual
polarization content of the gravitational radiation. Thus, there may be
multiple theories, with different polarization content, that still predict the
correct observed GW emitted power. Because other traditional tests of GR (like
Solar System tests) have no bearing on GWs, there currently exist no direct
measurements of GW polarizations.

Unlike CBC transients, continuous gravitational waves (CWs) are, by definition,
long-lasting narrow-band signals. Although they have not yet been observed
\cite{einsthome2016, Aasi2015, cwallsky2016, cwallskybin2014, rome2016, o1cw},
CWs are expected to be emitted by stable systems, like spinning neutron stars
with an asymmetric moment of inertia \cite{Thorne1987}. If detected, such
signals would allow for tests of gravity complementary to those achievable with
transients, including the study of GW polarizations \cite{Isi2015}.

In \cite{Isi2015} we showed that it is possible to search for CWs in a
polarization-agnostic way and to disentangle the polarization content if a
signal is present. However, the data analysis methods proposed were based on a
frequentist approach to statistics and suffered from the associated
limitations. In this paper, we reframe the ideas of \cite{Isi2015} in a more
sophisticated Bayesian framework that allows us to achieve the following novel
goals:
\begin{enumerate}
\item {\em Model-independent detection:} determine whether a set of
GW detector data, prepared for any given known pulsar and from one or
multiple detectors, provides evidence for the presence of an astrophysical
signal of any polarization content.

\item {\em Model selection:} in the presence of a signal, determine whether the
data favor GR or a generic non-GR model, as well as comparing specific
alternative theories among themselves and to GR; combine data for multiple
sources into a single statement about the validity of GR.

\item {\em Inference:} if the data favor the presence of a GR signal, place
constraints on specific alternative theories using the tools of Bayesian
parameter estimation.
\end{enumerate}
Furthermore, while \cite{Isi2015} treated only the case of a single detector,
we are now able to consider the generic case of a network of detectors.

We present Bayesian methods to achieve the three goals above in the context of
searches targeted to known pulsars and present sensitivity estimates for the
advanced detector era, including the first generic estimates of sensitivity to
nontensorial CW polarizations ever published. In Sec.\ \ref{sec:background},
we review the basics of beyond-Einstein polarizations and the targeted pulsar
CW search. In Sec.\ \ref{sec:method}, we phrase our problem in the language
of model selection and explain the construction of hypotheses that will allow
us to distinguish GR from non-GR signals. In Sec.\ \ref{sec:analysis} we
specify the details of our analysis, and we explain our results in Sec.\
\ref{sec:results}. Finally, we summarize our findings and explain caveats in
section \ref{sec:conclusion}.

\section{Background} \label{sec:background}

\subsection{Polarizations} \label{sec:polarizations}

\begin{figure}
\includegraphics[width=0.33\columnwidth]{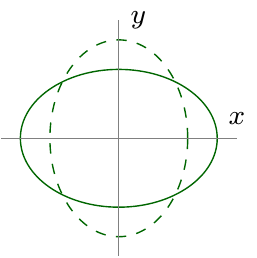}\hfill
\includegraphics[width=0.33\columnwidth]{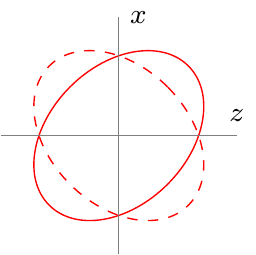}\hfill
\includegraphics[width=0.33\columnwidth]{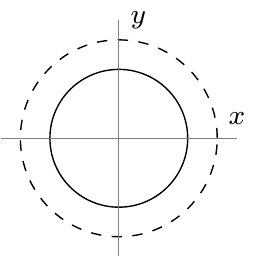}\\
\includegraphics[width=0.33\columnwidth]{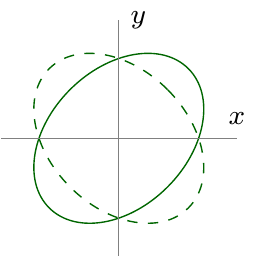}\hfill
\includegraphics[width=0.33\columnwidth]{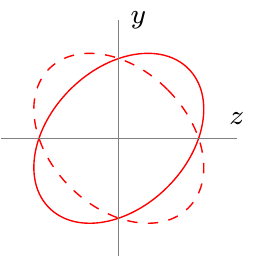}\hfill
\includegraphics[width=0.33\columnwidth]{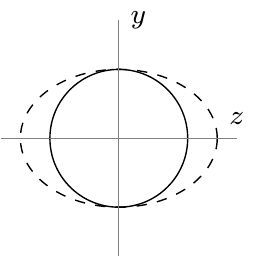}
\caption{{\em Effect of different GW polarizations on a ring of free-falling
test particles}. Plus (+) and cross ($\times$) tensor modes (green); vector-x
(x) and vector-y (y) modes (red); breathing (b) and longitudinal (l) scalar
modes (black). In all of these diagrams the wave propagates in the
\emph{z} direction. This decomposition into polarizations was first proposed
for generic metric theories in \cite{Eardley1973b}.}
\label{fig:circles}
\end{figure}

GWs can be decomposed into different polarizations, which arise from the
linearly independent components of the three-dimensional, rank-two tensor
representing the spatial metric perturbation \cite{Eardley1973b}. A generic
metric theory of gravity may thus allow any combination of up to six
independent modes: plus ($+$), cross ($\times$), vector x (x), vector y (y),
breathing (b) and longitudinal (l). The effect of each of these modes is
represented in \fig{circles}. The rotational properties of the fields
underlying any given theory determines which polarizations the theory supports:
$+$ and $\times$ correspond to tensor fields (helicity $\pm 2$), x and y to
vector fields (helicity $\pm 1$), and b and l to scalar fields (helicity $0$).

The components of the tensor and vector pairs are not separable, in the sense
that a signal model that includes one element of the group must also include
the other (e.g.\ it is not possible to have a model that allows plus $+$ but
not $\times$), because the distinction between $+$ and $\times$, or x and y, is
contingent on the frame of reference (e.g.\ relative orientation of source and
detector).

Einstein's theory only allows the existence of the $+$ and $\times$
polarizations. On the other hand, scalar-tensor and massive-graviton theories
may also predict the presence of some b and/or l component associated to the
theory's extra scalar field \cite{Will2006}. On top of tensor and scalar modes,
bimetric theories, like Rosen or Lightman-Lee theories, also predict vector
modes \cite{Chatziioannou2012}. Furthermore, less conventional theories might,
in principle, predict the existence of vector or scalar modes \emph{only},
while still possibly being in agreement with all other non-GW tests of GR (see
\cite{Mead2015} for an example). Although all these different theoretical
frameworks serve as motivation for our study, our approach to the measurement
of GW polarizations is phenomenological and, thus, theory-agnostic (Sec.\
\ref{sec:method}). It is important to underscore that the detection of a GW
signal with a non-GR polarization, no matter how small, is sufficient to
falsify GR (note the converse is not true, however).

Because different polarizations have geometrically distinct effects, GW
detectors will react differently to each mode. This is manifested in the
detector response function $F_p$ for each polarization $p$, which encodes the
effect of a linearly $p$-polarized GW with unit amplitude, $h_p=1$.
Ground-based GW detectors, like LIGO and Virgo are quadrupolar
antennas that perform low-noise measurements of the strain associated with the
differential motion of two orthogonal arms. Their detector response function
can thus be written as \cite{Nishizawa2009, Blaut2012, Isi2015, Poisson2014}:
\beq \label{eq:Fp}
F_{+} = \frac{1}{2} \left[ ({\bf w}_x \cdot {\bf d}_x)^2-({\bf w}_x \cdot
{\bf d}_y)^2 - ({\bf w}_y \cdot {\bf d}_x)^2+({\bf w}_y \cdot {\bf d}_y)^2
\right],
\eeq
\beq \label{eq:Fc}
F_{\times}=({\bf w}_x \cdot {\bf d}_x) ({\bf w}_y \cdot {\bf d}_x)-({\bf w}_x
\cdot {\bf d}_y) ({\bf w}_y \cdot {\bf d}_y),
\eeq
\beq \label{eq:Fx}
F_{\rm x}= ({\bf w}_x \cdot {\bf d}_x) ({\bf w}_z \cdot {\bf d}_x)- ({\bf w}_x
\cdot {\bf d}_y) ({\bf w}_z \cdot {\bf d}_y),
\eeq
\beq \label{eq:Fy}
F_{\rm y}= ({\bf w}_y \cdot {\bf d}_x) ({\bf w}_z \cdot {\bf d}_x)- ({\bf w}_y
\cdot {\bf d}_y) ({\bf w}_z \cdot {\bf d}_y),
\eeq
\beq \label{eq:Fb}
F_{\rm b}= \frac{1}{2} \left[ ({\bf w}_x \cdot {\bf d}_x)^2-({\bf w}_x \cdot
{\bf d}_y)^2+({\bf w}_y \cdot {\bf d}_x)^2-({\bf w}_y \cdot {\bf
d}_y)^2\right],
\eeq
\beq \label{eq:Fl}
F_{\rm l}=\frac{1}{2}\left[ ({\bf w}_z \cdot {\bf d}_x)^2- ({\bf w}_z
\cdot {\bf d}_y)^2 \right].
\eeq
Here, the spatial vectors ${\bf d}_x$, ${\bf d}_y$ have unit norm and point
along the detector arms such that ${\bf d}_z={\bf d}_x\times{\bf d}_y$ is the
local zenith; the direction of propagation of the wave from a source at known
sky location (specified by right ascension $\alpha$, and declination $\delta$)
is given by ${\bf w}_z$, and ${\bf w}_x$, ${\bf w}_y$ are such that ${\bf
w}_z={\bf w}_x\times{\bf w}_y$. We choose ${\bf w}_x$ to lie along the
intersection of the equatorial plane of the source with the plane of the sky,
and let the angle between ${\bf w}_y$ and the celestial north be $\psi$, the
{\em polarization angle}.

Because of their symmetries, the breathing and longitudinal modes are fully
degenerate to networks of quadrupolar antennas (see e.g.\ Sec.\ VI of
\cite{Chatziioannou2012}). This means that no model-independent measurement
with such a network can possibly distinguish between the two, so it is enough
for us to consider just one of them explicitly. We will refer to the scalar
modes jointly by the subscript ``s''. 

\begin{figure*}[p]
\subfloat[][Plus (+)]{\includegraphics[width=0.3\textwidth]{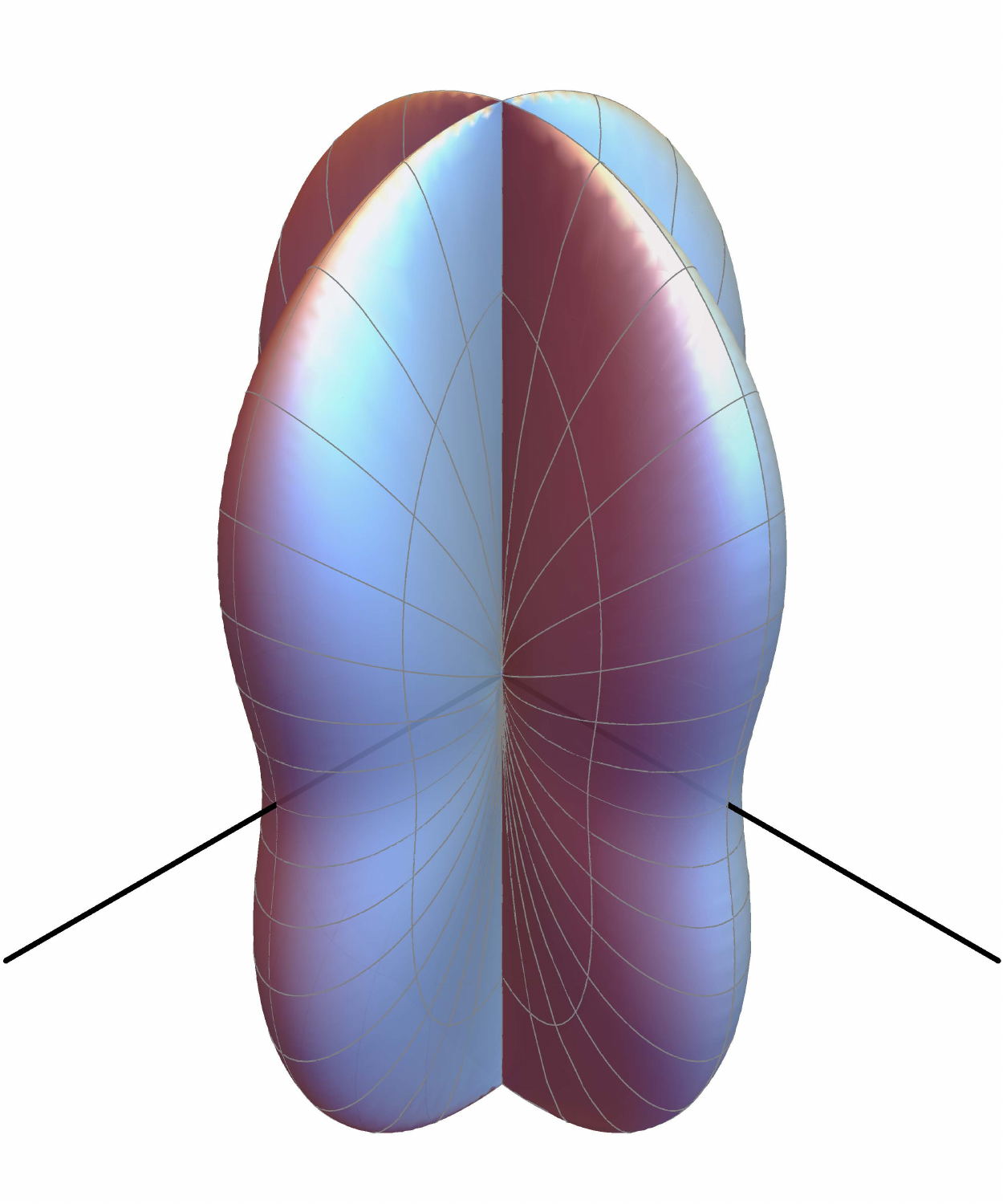}}\hspace{2cm}
\subfloat[][Cross ($\times$)]{\includegraphics[width=0.3\textwidth]{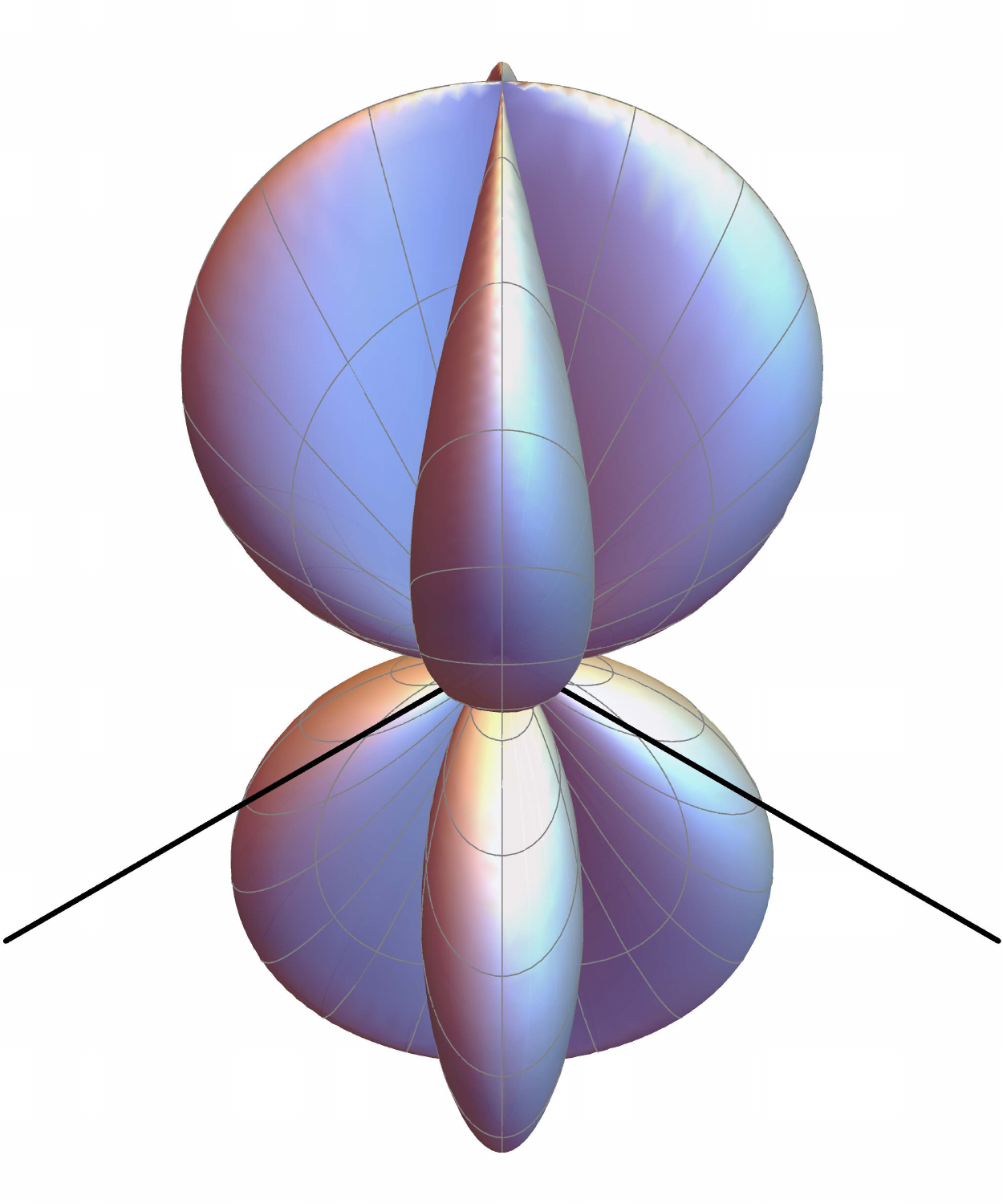}}\\
\subfloat[][Vector-x (x)]{\includegraphics[width=0.3\textwidth]{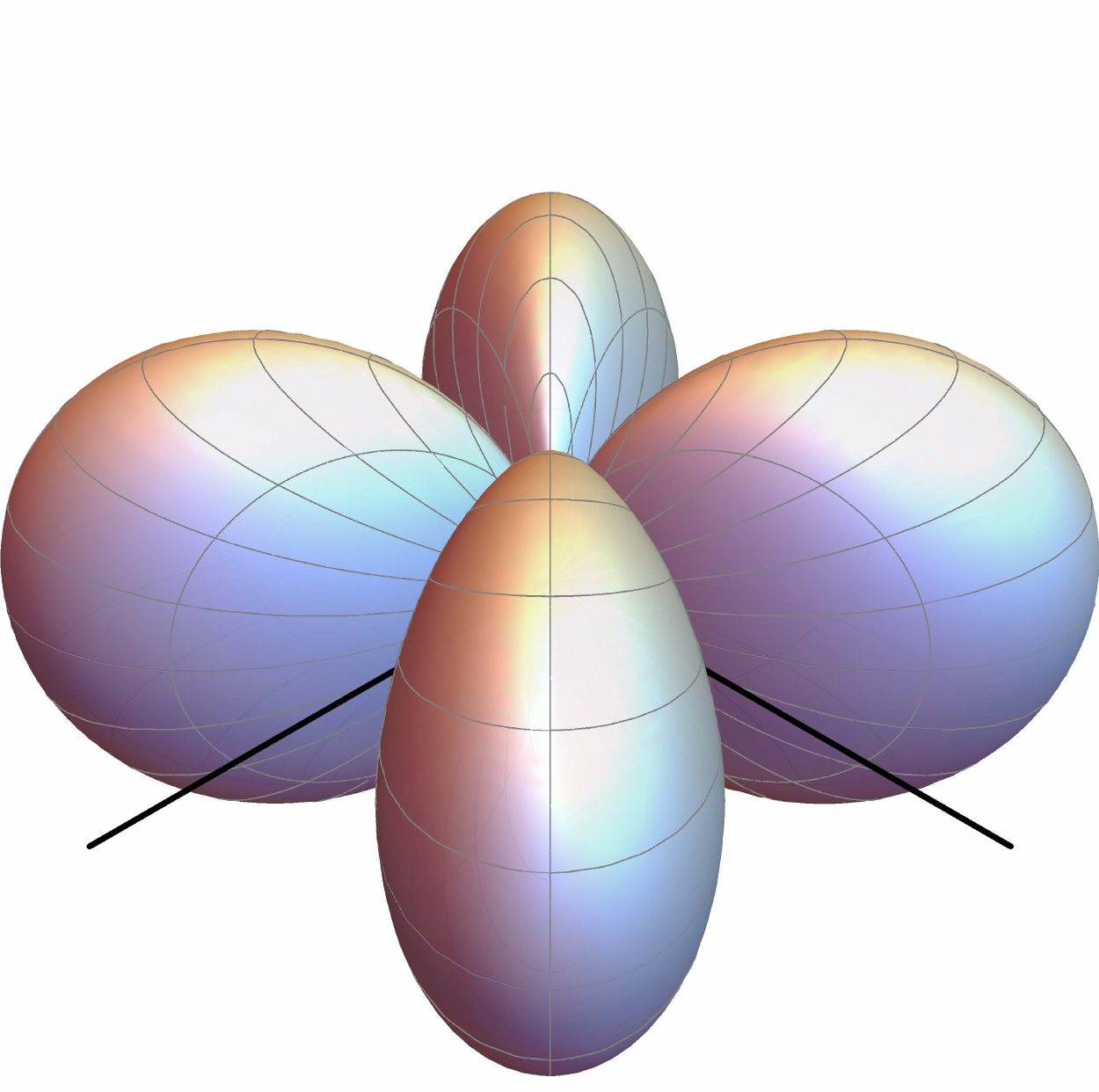}}\hspace{2cm}
\subfloat[][Vector-y (y)]{\includegraphics[width=0.3\textwidth]{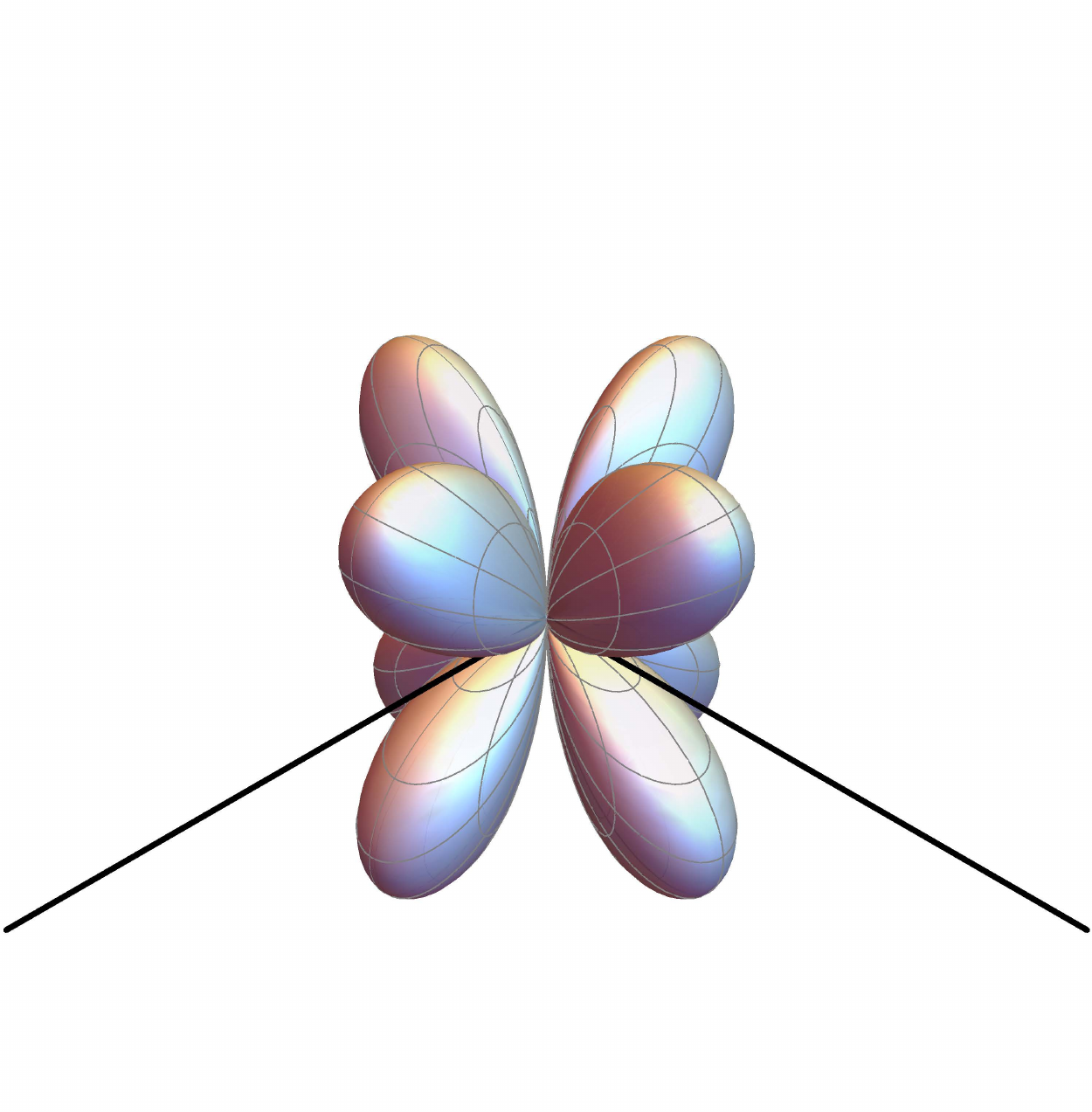}}\\
\subfloat[][Scalar (s)]{\includegraphics[width=0.3\textwidth]{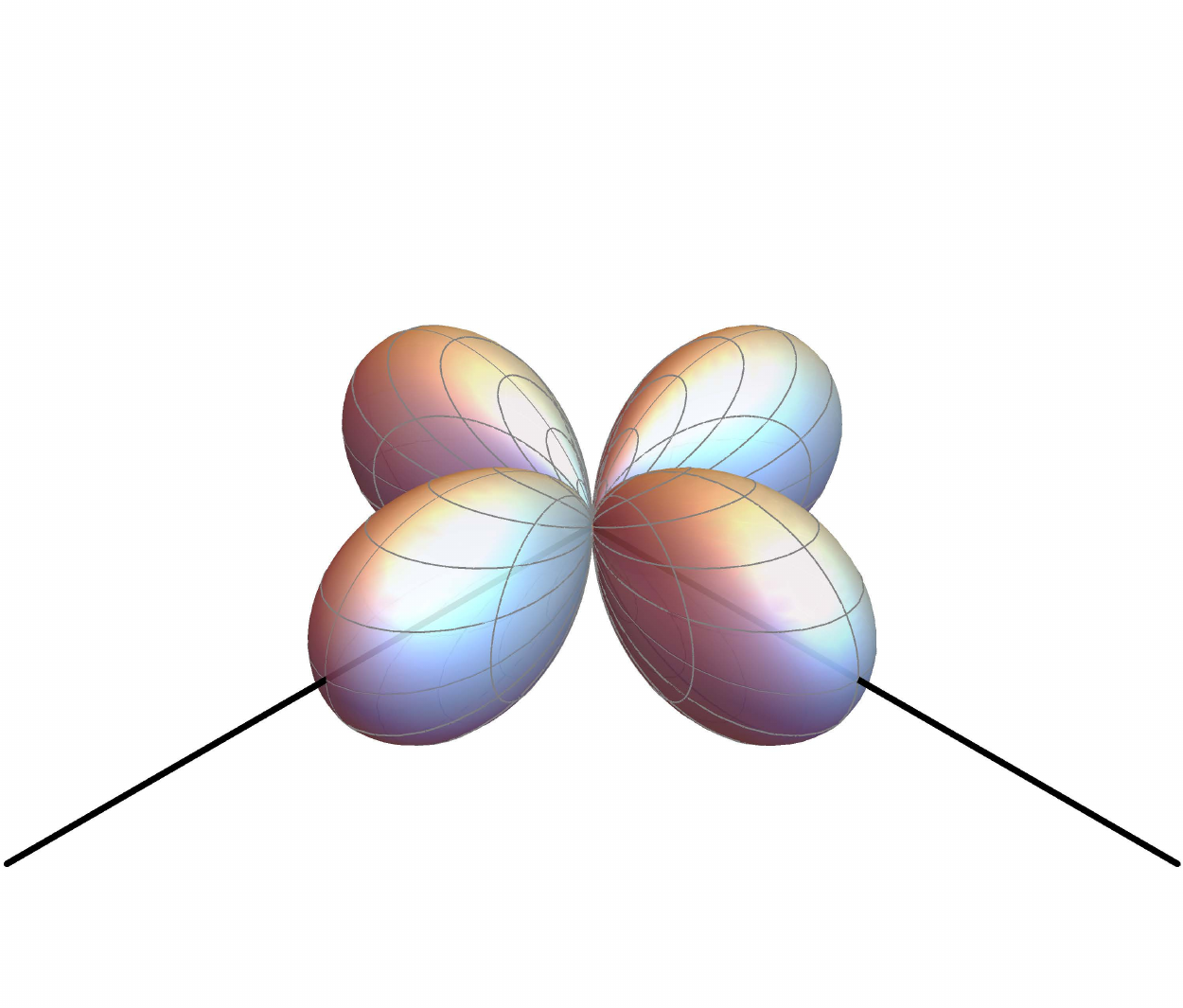}}
\vspace{0.5in}
\caption{{\em Angular response of a quadrupolar detector to each GW
polarization}. The radial distance represents the response of a single
quadrupolar antenna to a unit-amplitude gravitational signal of a tensor (top),
vector (middle), or scalar (bottom) polarization, i.e.\ $|F_p|$ for each
polarization $p$ as given by Eqs.\ (\ref{eq:Fp_ifo}--\ref{eq:Fs_ifo}). The
polar and azimuthal coordinates correspond to the source location with respect
to the detector, which is to be imagined as placed with its vertex at the
center of each plot and arms along the $x$ and $y$-axes. The response is
plotted to scale, such that the black lines representing the detector arms have
unit length in all plots. The response to breathing and longitudinal modes is
identical, so we only display it once and label it ``scalar''.} 
\label{fig:aps}
\end{figure*}

The response of gravitational detectors to signals of a given polarization and
direction of propagation can be represented, as in \fig{aps}, by a spherical
polar plot in which the radial coordinate corresponds to the sensitivity given
by Eqs.\ (\ref{eq:Fp}--\ref{eq:Fl}). In the frame of a given detector, this can
be written as [see e.g.\ Eqs.\ (13.98) in \cite{Poisson2014} with
$\psi=-\pi/2$, to account for the different wave-frame definition]:
\beq \label{eq:Fp_ifo}
F_+(\vartheta, \varphi\, ; \psi=0) = -\frac{1}{2}\left(1+\cos^2\vartheta \right)
\cos 2\varphi~,
\eeq
\beq
F_\times(\vartheta, \varphi\, ; \psi=0) = -\cos \vartheta \sin 2\varphi~,
\eeq
\beq
F_{\rm x}(\vartheta, \varphi\, ; \psi=0) = -\sin\vartheta \sin 2\varphi~,
\eeq
\beq
F_{\rm y}(\vartheta, \varphi\, ; \psi=0) = \sin\vartheta \cos\vartheta
\cos 2\varphi~,
\eeq
\beq \label{eq:Fs_ifo}
F_{\rm b/l}(\vartheta, \varphi\, ; \psi=0) = \mp \frac{1}{2} \sin^2\vartheta
\cos 2\varphi~,
\eeq
where $\vartheta$ and $\varphi$ are the polar an azimuthal coordinates of the
source with respect to the antenna at any given time (with detector arms along
the $x$ and $y$-axes), and we have fixed the wave frame so that $\psi=0$. The
representation of \fig{aps} makes it clear that quadrupolar detectors will
generally be more sensitive to some polarizations than others, although this
will vary with the sky location of the source. For example, for all but a few
sky locations, quadrupolar antennas will respond significantly less to a
breathing signal than a plus or cross signal.

For a given detector, polarization angle and sky location, the antenna patterns
of Eqs.\ (\ref{eq:Fp}--\ref{eq:Fl}) become simple, distinct functions of time
determined by the rotation of the Earth. This can be pictured by noting that,
as the Earth spins on its axis, the angular location of the source with respect
to detector will change, tracing an arc on the surfaces of \fig{aps} with
varying radial distance. As we explain in Sec.\ \ref{sec:modsel_single}, the
$F_p$'s of polarizations with different rotational properties can be
distinguished even in the absence of information on the source orientation; for
the minority of cases in which such information exists, it can be taken into
account to better distinguish among specific signal models (see Sec.\
\ref{sec:analysis}).

Because their characteristic period (a sidereal day) is much longer than the
CBC timescale (order of minutes or less), the $F_p$'s are treated as constants
in transient searches; however, this simplification is not valid for CW
searches, since their coherent-integration time can be of the order of months
to years. As we have pointed out before, this can be used to distinguish the
polarization content of a signal \cite{Isi2015}. Assuming wave frequency and
speed are the same for all modes, the only differences between CWs of different
polarizations arise from the sidereal-period amplitude modulations caused by
each antenna pattern.

\subsection{Continuous waves}

\subsubsection{Signal} \label{sec:signal}
A CW is an almost-monochromatic gravitational perturbation with constant
intrinsic amplitude and phase evolution $\phi(t)$. For arbitrary polarization
content, such a GW will induce a strain in a quadrupolar detector which can be
written as:
\beq \label{eq:cw}
h(t) = \sum_p F_p(t)h_p(t),
\eeq
where the sum is over the five independent polarizations, $p \in \{+,~\times,~
{\rm x},~{\rm y},~{\rm s}\}$; the $F_p$'s are those of Eqs.\
(\ref{eq:Fp}--\ref{eq:Fb}), and thus implicitly depend on the relative location
and orientation of source and detector by means of $\psi$, $\alpha$ and
$\delta$; the $h_p$ term encodes the amplitude and phase of the wave before
being projected onto the frame of the detector:
\beq \label{eq:cw_raw}
h_p(t) = a_p \cos \left( \phi (t) + \phi_p \right),
\eeq
where $a_p$ is a time-independent amplitude with a functional dependence on
source parameters determined by each particular theory of gravity; $\phi(t)$
the phase evolution, a consequence of the dynamics of the source in that
theory; and $\phi_p$ a phase offset for each polarization. The polarization
amplitudes $a_p$ and phases $\phi_p$ may take arbitrary values depending on
the specific theory of gravity and emission mechanism.

In GR, there are several ways in which a neutron star could emit CWs, but the
most likely is the presence of a nonaxisymmetry in the star's moment of
inertia \cite{Jones2002}. For this type of {\em triaxial}, nonprecessing
source, GR predicts:
\beq \label{eq:a_plus}
h_+(t) = h_0\frac{1}{2}(1+\cos^2 \iota) \cos \phi(t),
\eeq
\beq \label{eq:a_cross}
h_\times (t) = h_0 \cos \iota \sin \phi(t),
\eeq
\beq \label{eq:a_others}
h_{\rm x} = h_{\rm y} = h_{\rm s} = h_{\rm l} = 0,
\eeq
where $\iota$ is the inclination angle between the spin axis of the source and
the observing line-of-sight, and $h_0$ is an overall amplitude given by:
\beq \label{eq:h0}
h_0 = \frac{16 \pi^2 G}{c^4} \frac{\epsilon I_{zz}\frot^2}{r},
\eeq
where $r$ is the source distance, $\frot$ its rotation frequency
around the principal axis $z$, $\mathbf{I}$ the moment-of-inertia tensor and
$\epsilon\equiv (I_{xx}-I_{yy})/I_{zz}$ the equatorial ellipticity. For the
triaxial case, the GW frequency $\fgw$ is twice the rotational value $\frot$,
so that we can write:
\beq
\phi(t)=2\phi_{\rm rot} (t)+\phi_{({\rm GW}-{\rm EM})},
\eeq
where $\phi_{\rm rot}$ is the rotational phase as measured via electromagnetic
(EM) observations and $\phi_{({\rm GW}-{\rm EM})}$ is a potential, constant
phase offset between the GW and EM signals that can be absorbed into the
definition of the $\phi_p$'s in \eq{cw_raw}.

Note that other emission mechanisms may result in GW radiation at $\fgw=\frot$
\cite{Zimmermann1979}, or even noninteger powers $\frot$ \cite{Owen1998,
Bondarescu2009, Jones2015}. Furthermore, alternative theories of gravity may
(and, in general, will) support signals at any harmonic. Although in this paper
we only consider the case in which only the second rotational harmonic appears
in the GW phase, the analysis can be easily generalized to also include
contributions from the fundamental and other multiples of $\frot$ (see Sec.\
\ref{sec:conclusion}).

\subsubsection{Targeted search}

We would like to search a given set of data (from one or more detectors) for CW
signals coming from a specific candidate pulsar which has already been observed
and timed electromagnetically. Timing solutions are obtained through the pulsar
timing package TEMPO2 \cite{Edwards2006}. We want to achieve this regardless of
polarization content, and to reliably distinguish between the different modes
present.

If we assume all polarizations share the same phase evolution, then detector
response is the only factor distinguishing CW polarizations and, thus, all the
relevant information is encoded in the sidereal-day-period amplitude modulation
of the signal. This allows us to focus on a narrow frequency band around the
expected GW frequency by processing the data following the complex-heterodyne
method presented in \cite{Niebauer1993} and \cite{Dupuis2005}. This procedure
is summarized below.

A signal like Eqs.\ (\ref{eq:cw}, \ref{eq:cw_raw}) can be rewritten in the
form:
\beq 
h(t) = \Lambda(t) e^{i\phi(t)} + \Lambda^*(t) e^{-i\phi(t)}, 
\eeq
\beq \label{eq:het-signal}
\Lambda(t) \equiv \frac{1}{2} \sum \limits_{p=1}^5 a_p e^{i\phi_p}
F_p(t_k;\psi, \alpha, \delta),
\eeq
with $*$ indicating complex conjugation and $\phi(t)$ given by a Taylor
expansion around $\fgw=2\frot$:
\beq
\phi(t) = 2 \pi \left(2 f_{\rm rot} \tp + \dot{f}_{\rm rot}\tp^2 + ... \right),
\eeq
where $\tp$ is itself a function of time given by:
\beq \label{eq:tssb}
\tp (t) = t + \Delta_{\rm R} + \Delta_{\rm E} + \Delta_{\rm S} +
\Delta_{\rm binary}~.
\eeq
In the above, $\tp$ is the time measured by a clock inertial with respect to
the pulsar; $t$ is the time as measured at a given detector; $\Delta_{\rm R}$
is the Roemer delay; $\Delta_{\rm E}$ is the Solar-System Einstein delay;
$\Delta_{\rm S}$ is the Solar-System Shapiro delay; $\Delta_{\rm binary}$ is
the delay originating from the motion of the pulsar in its binary (a term that
vanishes for isolated sources) \cite{Dupuis2005}.

It is important to remember that, the $F_p$'s are functions of the source
orientation and sky location relative to the detector, so we have made this
dependence explicit in \eq{het-signal} by writing $F_p(t_k)$ as
$F_p(t_k;\psi,\alpha, \delta)$. Also, recall that these functions have a
characteristic period of a sidereal day (${\sim}10^{-5}$ Hz).

Because the phase evolution $\phi(t)$, including all corrections from
\eq{tssb}, is known (with known uncertainties) from electromagnetic
observations, we can digitally heterodyne the data by multiplying by
$\exp{\left[-i\phi(t)\right]}$ so that the signal therein becomes:
\beq \label{eq:het-data} 
h'(t)\equiv h(t) e^{-i\phi(t)} =\Lambda(t) + \Lambda^*(t) e^{-i2\phi(t)} 
\eeq
and the frequency modulation of the first term is removed, while that of the
second term is doubled. A series of low-pass filters can then be used to remove
the quickly varying term, which enables the down-sampling of the data by
averaging over minute-long time bins. As a result, $\Lambda(t)$ is the only
contribution from the original signal left in our data, and hence we can use
\eq{het-signal} as the template for our search. Note that, although we started
with real-valued data, after this process the data are now complex.

From \eq{het-data} we see that, in the presence of a signal, the heterodyned
and down-sampled noisy detector strain data $\datum$ for the $k$\ts{th} time
bin (which can also be labeled by the Earth-frame GPS time-of-arrival at the
detector, $t_k$) are expected to be of the form:
\beq \label{eq:data}
B_{\rm expected}(t_{k})= \Lambda(t_k) + n(t_{k}),
\eeq
where $n(t_{k})$ is the heterodyned, filtered and downsampled noise in bin $k$,
which carries no information about the GW signal. Note then that $\datum(t_k) -
\Lambda (t_k)$ should be expected to have the statistical properties of noise,
a fact that will be used below in defining likelihoods.

\section{Method} \label{sec:method}

\subsection{Model selection}

We use the tools of Bayesian model selection (also known as {\em second-level
inference}) to determine whether the data contain a signal and, if so, whether
that signal agrees with the GR prediction or not. Our procedure is
hierarchical and consists of the following stages:
\begin{enumerate}
\item {\em detection}: select between signal and noise models;
\item {\em test of GR}: if a signal is present, select between GR and non-GR
models;
\item {\em upper limits}: if GR is favored, place upper limits on nontensorial
strain amplitudes, in the context of specific alternative polarization models.
\end{enumerate}
This subsection covers only the first two items in this list, since the
placement of upper limits belongs in the section on parameter estimation. We
treat the case of a single data set in \ref{sec:modsel_single} and
\ref{sec:modsel_odds}, and we show how to combine results from multiple
analyses in \ref{sec:modsel_multiple}; we offer some considerations about how
to approach the problem of non-Gaussian noise in \ref{sec:nongaussian}.

\subsubsection{Hypotheses} \label{sec:modsel_single}

For any given pulsar, we would first like to use {\em reduced} (i.e.\
heterodyned, filtered and downsampled) GW data to decide between the following
two logically disjoint hypotheses:
\begin{enumerate}
\item {\em noise} ($\hypn$): no signal, the data are drawn from a Gaussian
distribution of zero mean and some (possibly slowly varying) standard
deviation;

\item {\em signal} ($\hyps$): the data contain noise drawn from a Gaussian
distribution and a signal with the assumed phase evolution and {\em any}
polarization content.
\end{enumerate}
In order to perform model selection, we need to translate these hypotheses into
the corresponding Bayesian models; this means setting a likelihood function
derived from the expected noise properties and picking a multidimensional prior
distribution over all parameters. It is important to underscore that a Bayesian
model is {\em defined} by the choice of these two probability distributions.

For $\hypn$, the construction of the likelihood is straightforward. First, let
$\sigma$ be the standard deviation of the detector noise at or near the
expected GW frequency; then, for each complex-valued data point $\datum$,
Gaussianity implies:
\beq \label{eq:likelihood_basic}
\pdf (\datum \mid \sigma, \hypn) = \frac{1}{2\pi\sigma^2} \exp
\left(-\frac{|\datum|^2}{2\sigma^2}\right).
\eeq
Here, and throughout this document, a lower-case $\pdf$ is used for probability
densities, while an uppercase $\p$ is used for discrete probabilities.

If the data are split into $\nseg$ segments of lengths $s_j$
($j=1,\dots,\nseg$) over which the standard deviation $\sigma_j$ is assumed to
remain constant, we can analytically marginalize over this parameter to obtain
a likelihood for the entire data set $\data$ in the form of a Student's
$t$-distribution \cite{Dupuis2005, Pitkin2017}:
\beq \label{eq:noise_likelihood_single_ifo}
\p(\data\mid \hypn) = \prod_{j=1}^\nseg A_j \left(
\sum_{k=\kappa_j}^{K_j} \left| \datum \right|^2 \right)^{-s_j},
\eeq
with $A_j=(s_j-1)!/2\pi^{s_j}$, $\kappa_j=1+\sum_{n=1}^j s_{n-1}$, $K_j =
\kappa_j+s_j-1$ and $s_0=0$. Data streams from $\ndet$ detectors can be analyzed
coherently by generalizing this to: 
\beq \label{eq:noise_likelihood}
\p(\data\mid \hypn) = \prod_{i=1}^\ndet \prod_{j=1}^{\nseg_i} A_{i,j} \left(
\sum_{k=\kappa_{i,j}}^{K_{i,j}} \left| \datumifo \right|^2 \right)^{-s_{i,j}},
\eeq
where $i$ indexes detectors, $\datumifo \equiv B_i(t_k)$ is the datum
corresponding to the $i$\ts{th} detector at time $t_k$, and $A_{i,j}$,
$\kappa_{i,j}$ and $K_{i,j}$ are defined analogously to $s_{j}$, $\kappa_j$
above. The splitting of the data into segments of constant standard deviation
may be achieved with a strategy similar to the Bayesian-blocks algorithm of
\cite{Scargle1998}, and explained in detail in \cite{Pitkin2017}.

Note that the likelihood $\pdf(\data\mid\vec{\theta},\hyp{})$ of some
hypothesis $\hyp{}$, is the probability of observing the data $\data$ assuming
$\hyp{}$ is true and given a specific choice of free parameters $\vec{\theta}$
from the model's parameter space $\Theta$. However, in the case of the noise
(``null'') hypothesis, as defined by the Student's $t$ likelihood above, there
are no free parameters. Consequently, $\Theta=\varnothing$ and
$\pdf(\data\mid\vec{\theta},\hyp{N}) = \p(\data\mid\hyp{N})$.

The case of $\hyps$ requires more careful attention. One could be tempted to
use \eq{data} to define a likelihood like \eq{noise_likelihood} with the
substitution $|\datum|\rightarrow|\datum-\Lambda_k|$, for
$\Lambda_k\equiv\Lambda(t_k)$ including all polarizations like in
\eq{het-signal}; the priors would reflect uncertainties in measured source
parameters and extend over reasonable ranges for $a_p$ and $\phi_p$. However,
for most realistic prior choices, that would correspond to a hypothesis that
assigns most of the prior probability to regions of parameter space for which
$a_p \neq 0$ for all $p$, thus downweighting more conservative models
(including GR) that we would like to prioritize. This is simply because the
subspace in parameter space corresponding to any of these smaller subhypotheses
(which, for example, fix one of the $a_p$'s to be zero) has infinitely less
volume (i.e.\ it offers infinitesimally less support) than its complement;
hence any practical choice of prior probability density will also assign this
subspace infinitely less weight, and so the prior for the corresponding
subhypothesis will be vanishingly small.

Formally, the inadequacy of the naive construction of $\hyps$ as proposed in
the previous paragraph is related to the logical independence of nested
hypotheses. We refer to this important point multiple times in the following
sections; in particular, we discuss it in the context of odds computations in
the text surrounding \eq{hyps_post}. We refer readers not familiar with this
line of reasoning to a similar discussion in \cite{Li2012}, or, more generally,
to Ch.\ 4 in \cite{Sivia2006} or Ch.\ 28 in \cite{MacKay2005}.

Instead, we will construct $\hyps$ from two logically disjoint component
hypotheses:
\begin{enumerate}
\item {\em GR signal} ($\hyp{GR}$ or $\hyp{t}$): the data contain Gaussian
noise and a tensorial signal with the assumed $\phi(t)$;
\item {\em non-GR signal} ($\hyp{nGR}$): the data contain Gaussian noise and a
signal with non-GR polarization content, but with the assumed $\phi(t)$.
\end{enumerate}

The tensorial hypothesis is embodied most generally by a signal model such
that
\beq \label{eq:lambda_t}
\Lambda_{\rm t}(t) = \frac{1}{2}\left[a_+ e^{i \phi_+} F_+(t;\psi=0) + a_\times e^{i \phi_\times}F_\times(t; \psi=0)\right],
\eeq
where $a_+$, $a_\times$, $\phi_+$ and $\phi_\times$ are free parameters, and we
pick a specific polarization frame by setting $\psi=0$ (we are allowed to do
this because of a degeneracy between $\psi$ and $a_+$, $a_\times$ explained in
Appendix \ref{ap:tensor_modes}). An alternative parametrization can be derived
from the triaxial emission model of Eqs.\ 
(\ref{eq:a_plus}--\ref{eq:a_others}), namely
\begin{align} \label{eq:lambda_gr}
\Lambda_{\rm GR}(t) = \frac{1}{2} h_0 e^{i \phi_0}
&\left[\frac{1}{2}(1+\cos^2\iota)F_+(t; \psi) -i \vphantom{\frac{1}{2}}
\cos\iota F_\times(t; \psi)\right] ,
\end{align}
where the free parameters are now $h_0$, $\phi_0$, $\iota$ and $\psi$ [in the
notation of Eqs.\ (\ref{eq:cw_raw}, \ref{eq:het-signal}), $\phi_+ = \phi_0$ and
$\phi_\times = \phi_0-\pi/2$]. This is the parametrization used in most
traditional GR-only searches (see e.g.\ \cite{o1cw, Pitkin2017}).

The templates of \eq{lambda_t} and \eq{lambda_gr} span the same signal space;
therefore, if we pick parameter priors properly related by their Jacobian, the
respective hypotheses ($\hyp{t}$ and $\hyp{GR}$) will be logically equivalent
(i.e.\ $\hyp{t}\equiv \hyp{GR}$). However, we will sometimes want to restrict
$\psi$ or $\iota$ in \eq{lambda_gr} to incorporate measurements of the source
orientation (see Table 3 in \cite{tcw2013}), and compare those results to the
unconstrained model of \eq{lambda_t}. In such cases, $\hyp{t}$ and $\hyp{GR}$
are no longer equivalent: the former corresponds to a {\em free-tensor} signal,
while the latter now corresponds to a GR {\em triaxial} signal for some given
source orientation [i.e.\ a signal with the functional dependence on $\iota$
and $\psi$ of \eq{lambda_gr}]. Because of lack of any orientation information,
this is a distinction without a difference for most pulsars. (See Appendix
\ref{ap:tensor_modes} for more details.)

The non-GR hypothesis, $\hyp{nGR}$, can itself be seen as a composite
hypothesis encompassing all the signal models that depart from GR in some way,
i.e.\ models that include polarizations other than $+$ and $\times$. We denote
such subhypotheses with a subscript listing the polarizations included in the
signal. For example, ``st'' (meaning ``scalar plus tensor'') corresponds to a
model with unrestricted scalar and tensor contributions:
\begin{align}
\Lambda_{\rm st}(t) = \frac{1}{2}&\left[a_+ e^{i \phi_+} F_+(t; \psi=0) +
a_\times e^{i \phi_\times} F_\times(t; \psi=0) \right. \nonumber \\
&\left. + ~a_s e^{i \phi_s}F_s(t;
\psi=0)\right].
\end{align}
With this notation extended to the names of the relevant hypotheses, we may
then write $\hyp{nGR}$ as the logical union (``or'' junction, $\lor$)
\begin{align} \label{eq:hypngr}
\hyp{nGR} \equiv~&\hyp{s} \lor 
\hyp{v} \lor \hyp{st} \lor \hyp{sv} \lor \hyp{tv} \lor \hyp{stv} \nonumber\\
 =~& \bigvee_{m\in \tilde{M}}{\cal H}_m ,
\end{align}
where, for convenience, we have defined the non-GR subscript set $\tilde{M}$:
\begin{align} \label{eq:tildem}
\tilde{M}\equiv&\left\{{\rm s},~{\rm v},~{\rm st},~{\rm sv},~{\rm tv},~{\rm
stv} \right\}.
\end{align}
Just as before, we may equivalently use the triaxial parametrization,
\eq{lambda_gr}, for the tensor modes in the non-GR hypotheses by instead
defining $\tilde{M}$ as
\beq \label{eq:tildem_gr}
\tilde{M} = \{{\rm s,~ v,~ sv,~ GR+s,~ GR+v,~ GR+sv}\},
\eeq
where, for example, GR+s denotes a signal template like
\begin{align}
\Lambda_{\rm GR+s}(t) &= \frac{h_0}{2} e^{i \phi_0}
\left[\frac{1}{2}(1+\cos^2\iota)F_+(t;\psi) -i \cos\iota
F_\times(t;\psi)\right] \nonumber \\ &+ \frac{1}{2} a_b e^{i \phi_b}F_b(t;
\psi),
\end{align}
and similarly for GR+v and GR+sv, with the added vector modes. Again, the two
definitions of $\tilde{M}$, Eqs.\ (\ref{eq:tildem}, \ref{eq:tildem_gr}), are
equivalent unless orientation information is incorporated in the way explained
above.

By the same token, the signal hypothesis can be built from the logical union of
$\hyp{GR}$ or $\hyp{t}$, and $\hyp{nGR}$:
\beq \label{eq:hyps}
\hyps \equiv \hyp{GR/t} \lor \hyp {nGR} = \bigvee_{m\in M} {\cal H}_m,
\eeq
with $M$ defined similarly to $\tilde{M}$, but also including the tensor-only
hypothesis, $\hyp{GR}$ or $\hyp{t}$:
\begin{align} \label{eq:model_set}
M \equiv \tilde{M} \cup \{{\rm GR/t}\}.
\end{align}

The validity of Eqs.\ (\ref{eq:hypngr}, \ref{eq:hyps}) is contingent on the
mutual logical independence of all the ${\cal H}_m$'s. This requirement is
satisfied by construction, since each of the ${\cal H}_m$'s is defined to
exclude regions of parameter space that would correspond to other hypotheses
nested within it (e.g.\ $\hyp{GR+s}$ is defined over all values of the scalar
amplitude except $a_{\rm s}=0$, to avoid including $\hyp{GR}$). In practice,
however, it is not necessary to explicitly exclude these infinitesimal regions
of parameter space, as will be explained in the following section.

\subsubsection{Odds} \label{sec:modsel_odds}

\begin{figure}
\centering
\includegraphics[width=\columnwidth]{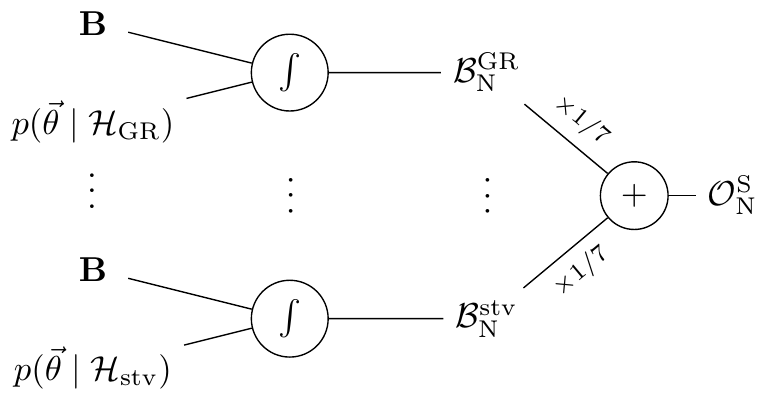}
\caption{{\em Computation of $\oddssn$}. First, the Bayes factor ${\cal
B}^m_{\rm N}$ is obtained from the data $\data$ and corresponding priors
$p(\theta|{\cal H}_m)$ for each model $m\in M$, by evaluating the integral of
\eq{evidence} using a nested sampling algorithm that samples over
$\vec{\theta}$ (step indicated by integral sign); these values are then added
and multiplied by $\p({\cal H}_m)/\p(\hypn)$ to obtain $\oddssn$, as in
\eq{O_S_N}. (Note that here we have set $\p({\cal H}_m)/\p(\hypn)=1/7$, as
explained Sec.~\ref{sec:analysis}.) The computation of $\odds{nGR}{GR}$ is
analogous.}
\label{fig:hyp_tree}
\end{figure}

We can construct a Bayesian model for $\hyps$ starting from its components:
for each subhypothesis ${\cal H}_m$ for $m\in M$, we use a likelihood function
like \eq{noise_likelihood} with the substitution
$|\datumifo|\rightarrow|\datumifo-\Lambda_{m,i,k}|$, i.e.\ 
\beq \label{eq:signal_likelihood}
\pdf(\data\mid \vec{\theta}, {\hyp{}}_m) = \prod_{i=1}^\ndet
\prod_{j=1}^{\nseg_i} A_{i,j} \left( \sum_{k=\kappa_{i,j}}^{K_{i,j}} \left|
\datumifo-\Lambda_{m,i,k} \right|^2 \right)^{-s_{i,j}}
\eeq
(where $\Lambda_{m,i,k}$ is the template corresponding to model $m$, for
detector $i$ and time-bin $k$), and suitable priors on the model parameters
$\vec{\theta}_m \in \Theta_m$; then, we combine the posteriors with priors on
the models themselves to obtain the posterior for $\hyps$. This last step
allows us to incorporate our {\em a priori} beliefs about the validity of each
of the components. This procedure is represented schematically in
\fig{hyp_tree} and fleshed out below.

The choice of model priors can be made clearer by considering the posterior
probability for the signal model. Given some set of detector data $\data$ and
underlying assumptions $I$ (suppressed from the following expressions), the
posterior probability for $\hyps$ is:
\beq \label{eq:hyps_post}
\p(\hyps\mid\data) = \sum_{m\in M} \p({\cal H}_m\mid\data)
\eeq
by \eq{hyps} and because the components are all logically independent [i.e.\
${\cal H}_{m_1}\land{\cal H}_{m_2}={\rm False}$, hence $P({\cal H}_{m_1} \land
{\cal H}_{m_2}\mid\data)=0$ for any $m_1,~m_2 \in M$ such that $m_1\neq m_2$].
Note that this is true even for hypotheses that may contain each other as
special cases. For instance, even though the GR template can be obtained from
GR+s by setting the scalar amplitude to $a_{\rm s}=0$, the points in the GR+s
parameter space satisfying this condition define an infinitesimally-thin slice
in parameter space that offers no support to the prior distribution and is thus
assigned no weight (see similar discussion in \cite{Li2012}).

We can expand each term on the RHS of \eq{hyps_post} using Bayes' theorem:
\beq
\p(\hyps\mid\data) = \sum_{m\in M}\p({\cal H}_m)\p(\data\mid{\cal
H}_m)/\p(\data).
\eeq
Each of the terms on the RHS is made up of three factors: a marginalized
likelihood $\p(\data\mid{\cal H}_m)$, a prior on the model $\p({\cal H}_m)$,
and a normalization constant $\p(\data)$. 

The marginalized likelihood (also known as {\em evidence}) is computed from the
data:
\beq \label{eq:evidence}
\p(\data\mid{\cal H}_m) = \int_{\Theta_m} \pdf(\data\mid\vec{\theta}_m, {\cal
H}_m)~ \pdf(\vec{\theta}_m\mid{\cal H}_m)~ {\rm d}\vec{\theta}_m,
\eeq
where $\pdf(\data\mid\vec{\theta}_m, {\cal H}_m)$ is itself the likelihood. The
evaluation of the multidimensional integral of \eq{evidence} is the most
computationally intensive part of our analysis (see Sec.\ \ref{sec:analysis}
for details).

We are free to choose the model priors (discussed in Sec.\ \ref{sec:analysis}),
as long as we satisfy the constraint:
\beq \label{eq:sig_prior}
\p(\hypn) + \sum_{m\in M} \p({\cal H}_m) = 1.
\eeq
This is a statement about the exhaustiveness and disjointedness of the
hypotheses we are considering: we assume that reality will agree with {\em one
and only one} of the hypotheses at hand. (As we will see in Sec.\
\ref{sec:conclusion}, this assumption might not hold; for example, the noise
may not be Gaussian.) The particular choice of prior for each model will encode
our expectations about the corresponding theory (before seeing the data), and
thus allow for some degree of subjectivity. 

Note that we cannot directly compute $\p(\data)$ in a straightforward manner
and without assuming that our hypothesis set is indeed exhaustive (which is not
the case for non-Gaussian detector noise, see Sec.\ \ref{sec:nongaussian}).
However, the need for this computation can be avoided by looking at relative
probabilities, i.e.\ {\em odds}. The odds for $\hyps$ versus $\hypn$ is defined
as:
\beq
{\cal O}^{\rm S}_{\rm N} \equiv \frac{\p(\hyps \mid \data)}{\p(\hypn
\mid\data)}.
\eeq
Using Bayes' theorem again and canceling the $\p(\data)$ factors, this 
simplifies to:
\beq \label{eq:O_S_N}
\oddssn = \frac{\sum \p({\cal H}_m) \p(\data\mid{\cal
H}_m)}{\p(\hypn)\p(\data\mid\hypn)} = \sum_{m\in M}{\frac{\p({\cal
H}_m)}{\p(\hypn)} {\cal B}^m_{\rm N}},
\eeq
where, in the second equality, we have used the definition of the {\em Bayes
factor}:
\beq
{\cal B}^{i}_{j} \equiv \frac{\p(\data\mid{\cal H}_{i})}{\p(\data\mid{\cal H}_{j})},
\eeq
for any two hypotheses ${\cal H}_i$, ${\cal H}_j$.

The odds in \eq{O_S_N} can be used as a detection statistic to determine
whether it is likely that the data contain a signal (of any polarization) or
not. Once the presence of a signal has been established, a similar ratio
can be constructed to assess agreement with GR:
\beq \label{eq:O_nGR_GR}
\odds{nGR}{GR} = \frac{\p(\hyp{nGR}\mid\data)}{\p(\hyp{GR}\mid\data)} 
= \sum_{m\neq GR} \frac{\p({\cal H}_m)}{\p(\hyp{GR})} {\cal B}^m_{\rm GR}.
\eeq
This ratio encodes the relative probability that there is a GR violation.
Because it is now assumed that there is a signal in the data, $\p(\hypn)=0$ and
the model priors must instead satisfy:
\beq
\sum_{m\in M} \p({\cal H}_m) = 1.
\eeq

We can reduce the number of computations needed to obtain $\oddssn$ and
$\odds{nGR}{GR}$ by using the fact that:
\beq
{\cal B}^i_j = \frac{\p(\data\mid{\cal H}_{i})}{\p(\data\mid{\cal H}_{j})} =
\frac{\p(\data\mid{\cal H}_{i})}{\p(\data\mid\hypn)}
\frac{\p(\data\mid\hypn)}{\p(\data\mid{\cal H}_{j})} =
\frac{{\cal B}^i_{\rm N}}{{\cal B}^j_{\rm N}}.
\eeq
This means that we need to evaluate an integral like \eq{evidence} seven times
per set of data, to compute ${\cal B}^m_{\rm N}$ for each $m$ in $M$. Those
seven numbers, together with the evidence for $\hypn$, are enough to compute
all the quantities of interest.

Instead of asking about a generic deviation from GR, we may also compare GR to
a particular alternative theory. For such purpose, we will usually assign equal
prior weight to GR and its alternative to compute:
\beq
{\cal O}^j_{\rm GR} = \frac{\p({\cal H}_j)}{\p(\hyp{GR})} {\cal B}^j_{\rm GR} = 
 {\cal B}^j_{\rm GR},
\eeq
where ${\cal H}_j$ may be any of the hypotheses in $\tilde{M}$ or an
even more specific hypothesis. (The latter case demands an extra execution of 
the inference code.)

\subsubsection{Multiple data sets} \label{sec:modsel_multiple}

So far we have assumed that the data $\data$, corresponding to one or more GW
detectors, can be analyzed coherently; however, there are cases in which we
would like to combine results from sets of data analyzed incoherently. Examples
are data sets corresponding to different sources or observation periods. Our
Bayesian framework makes it possible to combine the respective odds in order to
make an overall model selection statement (in our case, about the presence of
signal or the validity of GR).

For instance, we may analyze data for $\npsr$ pulsars and ask about the
probability that any of them contain a signal; treating each as an independent
observation, the combined probability can be constructed from the odds above.
Letting ${\cal H}_{{\rm S}_i},~{\cal H}_{{\rm N}_i}$ respectively denote signal
and noise hypotheses for the $i$\ts{th} source, while $\hyp{S_{any}}$
corresponds to a signal being present in {\em any} of the sources and
$\hyp{N_{all}}$ corresponds to Gaussian noise in data for {\em all} sources:
\begin{align} \label{eq:O_S_N_psr}
^{(\npsr)} \odds{S_{any}}{N_{all}} 
&=\frac{\p(\hyp{S_{any}}\mid\data)}{\p(\hyp{N_{all}}\mid\data)}
=\frac{1-\p(\hyp{N_{all}}\mid\data)}{\p(\hyp{N_{all}}\mid\data)} \nonumber\\
&=\frac{1}{\p(\bigwedge_i {\cal H}_{{\rm N}_i}\mid\data)} - 1
=\left[\prod^\npsr_{i=1}\frac{1}{\p({\cal H}_{{\rm N}_i}\mid\data_i)}\right]
- 1\nonumber\\
&=\left[ \prod^\npsr_{i=1}\frac{\p({\cal H}_{{\rm S}_i}\mid\data_i)+\p({\cal
H}_{{\rm N}_i}\mid\data_i)}{\p({\cal H}_{{\rm N}_i}\mid\data_i)}\right]
-1\nonumber\\
&=\left[\prod^\npsr_{i=1} \left({\cal O}^{{\rm S}_i}_{{\rm N}_i}+1\right)
\right] - 1,
\end{align}
where we have used the exclusivity and exhaustiveness of the signal and noise
hypotheses, i.e.
\beq
\p(\hyp{S_{any}}\mid\data)+\p(\hyp{N_{all}}\mid\data) = 1,
\eeq
\beq
\p({\cal H}_{{\rm S}_i}\mid\data_i)+\p({\cal H}_{{\rm N}_i}\mid\data_i) = 1,
\eeq
with $i$ indexing data sets. Note that the data sets for different sources
($\data_i$'s) are {\em not} conditionally independent under $\hyp{S_{any}}$ or
$\hyp{N_{all}}$. Also, \eq{O_S_N_psr} does not enforce the requirement that, if
signals are present in multiple sources, they all correspond to the same model
from \eq{model_set}; such a constraint could be implemented at this stage, but
is more easily enforced by examining individual values of ${\cal O}^{m}_{\rm
N}$ when necessary.

The construction of \eq{O_S_N_psr} implicitly assigns model priors to each of
the meta-hypotheses $\hyp{S_{any}}$ and $\hyp{N_{all}}$ such that:
\beq
\frac{\p(\hyp{S_{any}})}{\p(\hyp{N_{all}})} = \left[\frac{\p(\hyps)}{\p(\hypn)}
+ 1 \right]^\npsr - 1,
\eeq
where we have assumed the priors for signal vs noise are equal for all sources,
i.e.\ $\p(\hypsub{S}{i})=\p(\hyps)$ and $\p(\hypsub{N}{i})=\p(\hypn)$ for all
$i$. When making combined statements for multiple sources, we may wish to
choose $\p(\hyps)/\p(\hypn)$ such as to produce any desired value of
$\p(\hyp{S_{any}})/\p(\hyp{N_{all}})$, say
$\p(\hyp{S_{any}})=\p(\hyp{N_{all}})$. Furthermore, one may wish to weight each
pulsar differently within $\hyp{S_{any}}$ by incorporating information about
the source distance (or other parameters) into the priors via a parametrization
like \eq{h0}; this may improve the sensitivity of the ensemble odds to weak
signals in the set, as suggested in \cite{Fan2016}. However, using such a
parametrization generally implies committing to a specific gravitational theory
(or family of theories). We choose not to take such approach in this study.

Besides combining data for multiple pulsars, for a given source, we could also
(incoherently) combine the results of analyses using data from different
observation periods. Since the astrophysical CWs we are considering should
either be present in all $\nrun$ observation runs or in none of them, the
relevant odds, generalizing \eq{O_S_N}, are:
\begin{align}
^{[\nrun]}\oddssn &= \frac{\p(\hyps \mid \data)}{\p(\hypn \mid \data)} =
\sum_{m\in M} \frac{\p(\hyp{}_m \mid \data)}{\p(\hypn \mid \data)} \nonumber \\
&= \sum_{m\in M} \frac{\p(\data \mid \hyp{}_m) \p(\hyp{}_m)}{\p(\data \mid
\hypn) \p(\hypn)} \nonumber \\
&= \sum_{m \in M} \frac{\p(\hyp{}_m)}{\p(\hypn)} \prod_{j=1}^\nrun \left({\cal
B}^{m}_{{\rm N}}\right)_j,
\end{align}
where we have again used $\data=\{\data_j\}_{j=1}^\nrun$ to refer to the
totality of data, with $j$ indexing observation runs. The independence of the
$\data_j$'s, conditional on ${\cal H}_m$ and $\hypn$, is applied on the last
line to write the result in terms of the individual Bayes factors for each run,
$\left({\cal B}^{m}_{{\rm N}}\right)_j$.

Similarly, we can use multiple data sets to make a single statement about
deviations from GR. Once we have made $\npsr$ detections from different
sources, the odds for a GR violation is:
\beq \label{eq:ensemble_odds_ngr_gr}
^{(\npsr)}\odds{nGR}{GR} = \sum_{m \in \tilde{M}}
\frac{\p(\hyp{}_m)}{\p(\hyp{GR})} \prod_{i=1}^\npsr \left({\cal B}^{m}_{{\rm
GR}}\right)_i,
\eeq
where, again, $i$ indexes sources; this is a generalization of \eq{O_nGR_GR}.
(See Sec.\ IIID of \cite{Li2012} for an analogous derivation.)

\subsubsection{Non-Gaussian noise} \label{sec:nongaussian}

Up to this point, like most other CW studies, we have assumed that the detector
noise is Gaussian. However, although previous work has indicated that this is
generally a very good approximation \cite{Isi2015, o1cw}, it is not exactly
true for actual detector noise (for some frequencies more so than others).
Happily, most of the model selection statements expounded so far are valid also
in the presence of non-Gaussian instrumental noise, after some light
reinterpretation.

If the assumption of Gaussianity does not hold, the hypotheses constructed in
Sec.\ \ref{sec:modsel_single} are no longer exhaustive: the data may not only
be explained by Gaussian noise or a signal (GR or otherwise), but also by
non-Gaussian artifacts that are impossible to satisfactorily model.
Nevertheless, the computation and interpretation of evidences and odds remain
unchanged for all the hypotheses under consideration.

Because ``noise" no longer just means ``Gaussian noise", $\oddssn$ (which
compares the signal model vs {\em Gaussian} noise) has to be treated more
carefully for detection purposes. Indeed, instrumental features that are
clearly non-Gaussian (e.g.\ a loud, narrow-band artifact wandering across the
frequency of interest) will generally result in a relatively large value of
$\oddssn$, even if there is no detectable astrophysical signal in the data.
This issue affects the standard GR searches as well \cite{o1cw}, although
perhaps to a lesser degree due to the reduced signal parameter space.

It is possible to mitigate this problem by constructing a hypothesis that
captures some key characteristic of instrumental features and helps
discriminate those from real astrophysical signals. Perhaps the best way to do
this is to take advantage of the fact that an astrophysical CW must manifest
itself coherently across detectors, while the same is not true for detector
artifacts \cite{Keitel2014}. We can thus define an {\em instrumental feature
hypothesis} ($\hyp{I}$) to encompass the cases in which the data are composed
of Gaussian noise, or features that look like astrophysical signals but are not
coherent across detectors (viz.\ they do not have a consistent phase evolution
and they are best described by different waveform parameters). 

Formally, we define $\hyp{I}$ by: \beq \label{eq:hyp_i} \hyp{I} \equiv
\bigwedge_{d=1}^\ndet \left( \hypsub{S}{d} \lor \hypsub{N}{d} \right), \eeq
where the subscript $d$ identifies detectors, and $\land$ is the logical
``and'' junction. This definition does not explicitly encompass instrumental
features that are coherent across some subset of the detectors. Also, note that
\eq{hyp_i} implicitly contains a term equivalent to the usual noise hypothesis
$\hypn=\bigwedge_d \hypsub{N}{d}$. Similarly, it also contains a term
corresponding to the presence of signals in all detectors ($\bigwedge_d
\hyp{S}_d$). Importantly, such an {\em incoherent} term is not equivalent to the
{\em coherent} signal hypothesis $\hyps$, as given by the multidetector
likelihood of \eq{signal_likelihood}:
\beq \label{eq:hyp_coh_vs_inc}
\hyps \neq \bigwedge_{d=1}^\ndet \hyp{S}_d.
\eeq
While the evidence integral of \eq{evidence} factorizes into single-detector
terms for $\hypn$ (due to the null parameter space), the same is not true for
$\hyps$. Furthermore, because it does not demand detector coherence, the RHS of
\eq{hyp_coh_vs_inc} is associated with a considerably larger parameter space
than the LHS. Thus, in the presence of an astrophysical signal, model selection
will favor $\hyps$ due to its smaller Occam's penalty. The same is true, of
course, when comparing $\hyps$ to $\hyp{I}$ as a whole.

From \eq{hyp_i}, it is straightforward to write the evidence for $\hyp{I}$ as
\begin{align} \label{eq:incoherent_evidence}
\p(\data\mid\hyp{I}) &=\prod_{d=1}^\ndet \left[ \p(\data_d\mid\hypsub{S}{d})
\p(\hypsub{S}{d}\mid\hyp{I}) \right. \nonumber \\
&\left. \hphantom{\prod_{d=1}^\ndet [} +\p(\data_d\mid\hypsub{N}{d})
\p(\hypsub{N}{d}\mid\hyp{I}) \right]
\end{align}
and use this to construct the odds comparing against $\hyps$:
\begin{align}
\odds{S}{I} 
&=\frac{\p(\hyps)}{\p(\hyp{I})} \frac{\bayessn}{\prod_{d=1}^\ndet \left[
\p(\hypsub{S}{d}\mid\hyp{I}) ({\cal B}^{{\rm S}_d}_{{\rm N}_d} - 1)+1 \right]}.
\end{align}
Here we have used \eq{incoherent_evidence}, together with the fact that
$\p(\hypsub{S}{d}\mid\hyp{I})+\p(\hypsub{N}{d}\mid\hyp{I}) = 1$ and $\p(\data
\mid \hypn) = \prod_d \p(\data \mid \hyp{N}_d)$, to write $\odds{S}{I}$ as a
function of the detector-coherent signal vs noise Bayes factor $\bayessn$, the
single-detector signal vs noise Bayes factors ${\cal B}^{{\rm S}_d}_{\rm N}$,
and model priors $\p(\hyps)$, $\p(\hyp{I})$ and $\p(\hypsub{S}{d}\mid\hyp{I})$. 

As usual, we are free to choose the model priors to give more or less weight to
different hypotheses. For example, we recover the choice of \cite{o1cw}
(Appendix A3) by setting $\p(\hypsub{S}{d}\mid\hyp{I})=0.5$ for all $d$ and
$\p(\hyps)=\p(\hyp{I}) \times 0.5^\ndet$ such that:
\beq \label{eq:logodds_s_i}
\ln\odds{S}{I} = \ln\bayessn - \sum^\ndet_{d=1} \ln\left({\cal B}^{{\rm
S}_d}_{{\rm N}_d} +1 \right).
\eeq
(When comparing to Appendix A3 of \cite{o1cw}, however, note that in that work
``I" is used to denote both the background information and the
``incoherent-signal-or-noise" hypothesis, which can be identified with our
$\hyp{I}$.)

There is reason to believe that $\ln\odds{S}{I}$, with model priors as in
\eq{logodds_s_i}, is quite good at picking out instrumental features, even for
data from just two instruments \cite{o1cw}. (Note that we would expect the
discriminatory power of $\ln\odds{S}{I}$ to grow with the number of detectors
available.) However, at the end of the day, we can never be fully confident
that $\hyp{I}$ will indeed capture all nonastrophysical disturbances. To
address this, we may always treat $\ln\oddssn$ and $\ln\odds{S}{I}$ as any
generic detection statistic and use estimates of the background distribution to
establish significance.

\subsection{Parameter estimation} \label{sec:method_pe}

Besides choosing between different models, we can use Bayesian statistics to
obtain posterior probability density functions (PDFs) on the parameters of a
given template ({\em first--level inference}). In the absence of a loud signal,
this can be used to obtain credible intervals that yield upper--limits on the
amplitudes of GR deviations.

For a model $\hyp{}{}$ with $N$ parameters, an $N$-dimensional posterior PDF
covering the parameter space $\Theta$ can be obtained from Bayes' theorem:
\beq
\pdf (\vec{\theta}\mid\data,{\cal H}) = \frac{\pdf(\data\mid\vec{\theta},
\hyp{})~ \pdf (\vec{\theta}\mid\hyp{})} {\p (\data\mid\hyp{})},
\eeq
for $\vec{\theta}$ in $\Theta$, and with $\pdf (\vec{\theta}\mid\hyp{})$ the
prior over $\Theta$. To obtain a one-dimensional PDF for a single parameter
(call it $\theta_i$), the $N$-dimensional distribution must be marginalized
over all nuisance parameters (viz.\ all parameters except $\theta_i$):
\begin{align} \label{eq:post}
\pdf(\theta_i\mid\data,{\cal H}) &= \int_{\Theta'} \pdf(\vec{\theta} \mid
\data, {\cal H})~{\rm d}^{N-1}\theta_j \nonumber\\
&\propto \int_{\Theta '} \pdf(\data\mid\vec{\theta}, {\cal H})~
\pdf (\vec{\theta}\mid{\cal H})~ {\rm d}^{N-1}\theta_j,
\end{align}
where $0<j\leq N$, such that $j\neq i$, and $\Theta'$ denotes the parameter
space $\Theta$ with the $i$\ts{th} dimension removed. Note that the equality
has been replaced by a relation of proportionality because we have excluded the
evidence $\p(\data\mid\hyp{})$ from the expression. (Although of great
importance for model selection, this quantity is uninteresting for the purposes
of parameter estimation and can be treated as a simple normalization constant.)
As discussed in Sec.\ \ref{sec:analysis}, we evaluate \eq{post} with the same
algorithm used to compute the evidence.

\eq{post} can be used to place upper limits on model parameters; in particular,
we will use it to place limits on the amplitude of GR deviations. Consider, for
instance, the case of a scalar-tensor theory that can be encapsulated by our
GR+s model as described in the previous section; the $95\%$-credible upper
limit on the strength of the breathing mode is $\hul{s}$, defined by:
\beq \label{eq:hbul}
0.95=\int_{\min(h_{\rm s})}^{\hul{s}} \pdf (h_{\rm s} \mid \data, 
\hyp{GR+s})~{\rm d}h_{\rm s},
\eeq
where $\min(h_{\rm s})$ is the minimum value of $h_{\rm s}$ allowed by the
prior.

Note that there may be reasons to compute posteriors under different priors
than when computing Bayes factors. In particular, it is conventional to present
upper limits obtained using a uniform prior over some broad range of the
amplitude parameters. With a uniform prior, the posterior is trivially related
to the likelihood. This approach produces a more conservative upper limit than
other choices, e.g.\ a Jeffreys prior (see Appendix \ref{app:priors}).

\begin{figure}
\centering
\includegraphics[width=0.4\columnwidth]{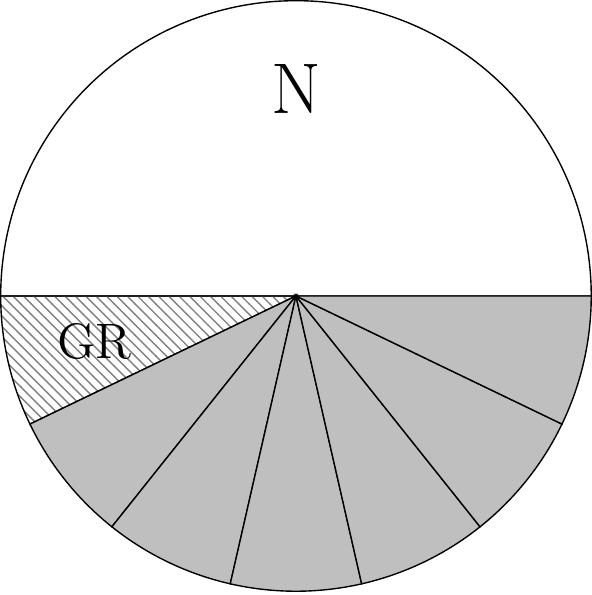}\qquad
\includegraphics[width=0.4\columnwidth]{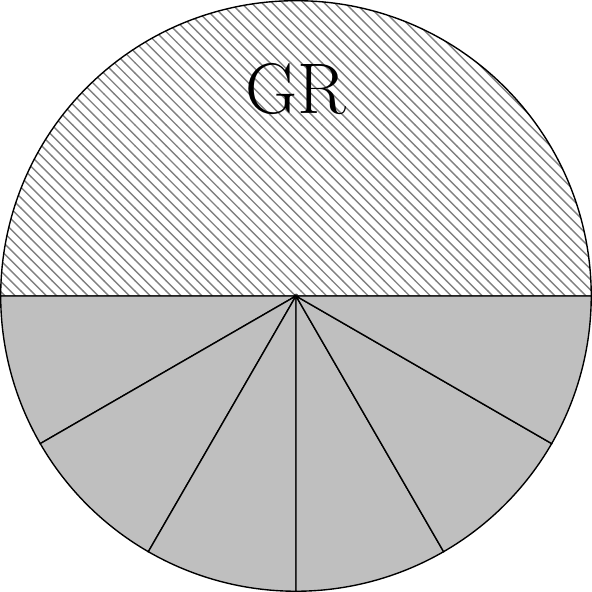}
\caption{{\em Model priors}. Distribution of prior probability over
subhypotheses for the construction of $\oddssn$ (left) and $\odds{nGR}{GR}$
(right), according to Eqs.\ (\ref{eq:O_S_N}) and (\ref{eq:O_nGR_GR})
respectively. For $\oddssn$, we assign equal weight to the $\hypn$ (white) and
$\hyps$ (gray); as in \eq{model_priors_s}, we make no {\em a priori}
distinction between non-GR models (solid) and GR (hatched). For
$\odds{nGR}{GR}$, we set equal prior probability for $\hyp{GR}$ and
$\hyp{nGR}$, distributing the prior equally among non-GR models, as in
\eq{model_priors_ngr}.}
\label{fig:prior_pies}
\end{figure}

\section{Analysis} \label{sec:analysis}

We quantify our ability to use Bayesian model selection to detect CW signals
and determine their polarization content as described above. To do this, we use
one year of simulated data from three advanced interferometric detectors at
design sensitivity: LIGO Hanford (H1), LIGO Livingston (L1) and Virgo (V1).
Detector noise is simulated by drawing from a Gaussian distribution with zero
mean and variance corresponding to the power spectral density (PSD) of each
detector at the GW frequency of the pulsar. (Previous work has shown that these
are good assumptions for actual reduced detector data \cite{Isi2015, o1cw}.)

As mentioned in the previous section, the key step in our analysis is the
computation of the evidence integral of \eq{evidence} for the hypotheses under
consideration (one noise model, plus seven signal submodels; see Sec.\
\ref{sec:modsel_single}). We carry this out using a version of the Bayesian
inference code used for the targeted pulsar search in
\cite{Pitkin2012,Pitkin2017}, which we modified to handle signals from theories
other than GR. This inference code is itself built on the implementation of
Skilling's nested-sampling algorithm \cite{Skilling2006} in the
\texttt{LALInference} package \cite{Veitch2015}, part of the LIGO Algorithm
Library Suite \cite{lalsuite}. This is the same inference software used for CBC
analyses, including GW150914 \cite{gw150914_pe}.

In computing likelihoods, we take source location, frequency and frequency
derivatives as known quantities (relevant uncertainties are negligible for this
analysis). Unless otherwise stated, priors uniform in the logarithm are used
for amplitude parameters ($h_0$ or $h_p$'s), since these are the least
informative priors for scaling coefficients (also known as ``Jeffreys priors")
\cite{Jaynes1968}; we make the somewhat arbitrary choice of restricting the
strain amplitudes to the $10^{-28}$--$10^{-24}$ range (this is of little
consequence for model selection, as explained in Appendix \ref{app:priors}).
Flat priors are placed over all phase offsets ($\phi_0$ and all the
$\phi_p$'s). 

All plots for the Crab pulsar (PSR J0534+2200) in Sec.\ \ref{sec:results} are
produced using known values of its orientation parameters, $\cos\iota$ and
$\psi$, and with the triaxial parametrization of tensor modes; for other
pulsars, however, the free-tensor parametrization is used instead. (See Sec.\
\ref{sec:modsel_single} and Appendix \ref{ap:tensor_modes}.)

We follow common practice by adopting the principle of indifference (see e.g.\
Ch.\ 5 of \cite{Sivia2006}) in assigning equal prior probability to the signal
and noise models, i.e.\ we let 
\beq
\p(\hyps)=\p(\hypn)=1/2~.
\eeq
We must also decide how to split the prior among the different ${\cal H}_m$'s 
when computing $\oddssn$ and $\odds{nGR}{GR}$. In the former case we choose
to distribute the prior weight uniformly among all signal models, so that:
\beq \label{eq:model_priors_s}
\p({\cal H}_m)=|M|^{-1}/2=1/14,
\eeq
with $|M|=7$ the cardinality of $M$ [i.e.\ the number of signal models that go
into the construction of $\hyps$, see \eq{model_set}]. In the latter, however,
we prioritize GR by setting:
\beq
\p(\hyp{GR} \mid \hyps) = 1/2,
\eeq
\beq \label{eq:model_priors_ngr}
\p({\cal H}_m \mid \hyps) = |\tilde{M}|^{-1}/2 = 1/12.
\eeq
This distribution is illustrated schematically in \fig{prior_pies}. Note that
these are not the only justifiable options; for example, we might want to
prioritize $\hyp{GR}$ when constructing $\hyps$ in order to better handle a
noise background that does not conform to our assumption of Gaussianity. (Other
strategies to tackle non-Gaussian noise are discussed in Sec.\
\ref{sec:nongaussian}.) In any case, the code is sufficiently flexible to make
different choices for the model priors if desired.

To study our method in the presence of signal, we perform several injections of
scalar, vector and tensor polarizations (and combinations thereof) for all the
\red{200} pulsars analyzed in \cite{o1cw}. The simulated signals have a range
of signal-to-noise ratios (SNRs), which we proxy below by their {\em effective
strain amplitudes}. We define these in terms of the $a_p$'s from \eq{cw_raw}
by:
\beq \label{eq:ht}
\hT \equiv \sqrt{a_+^2 + a_\times^2},
\eeq
\beq \label{eq:hv}
\hV \equiv \sqrt{a_{\rm x}^2 + a_{\rm y}^2},
\eeq
\beq \label{eq:hs}
\hS \equiv a_{\rm s},
\eeq
for tensor, vector and scalar signals respectively. Each simulated signal is
generated with a random value of the nuisance phase parameters ($\phi_0$ or
$\phi_p$'s). GR injections are always carried out using the triaxial template
of \eq{lambda_gr}, with random orientation parameters ($\psi$ and $\iota$) when
those are not known. Location is always taken to be fixed at the known value
for each pulsar.

\begin{figure}
\includegraphics[width=\columnwidth]{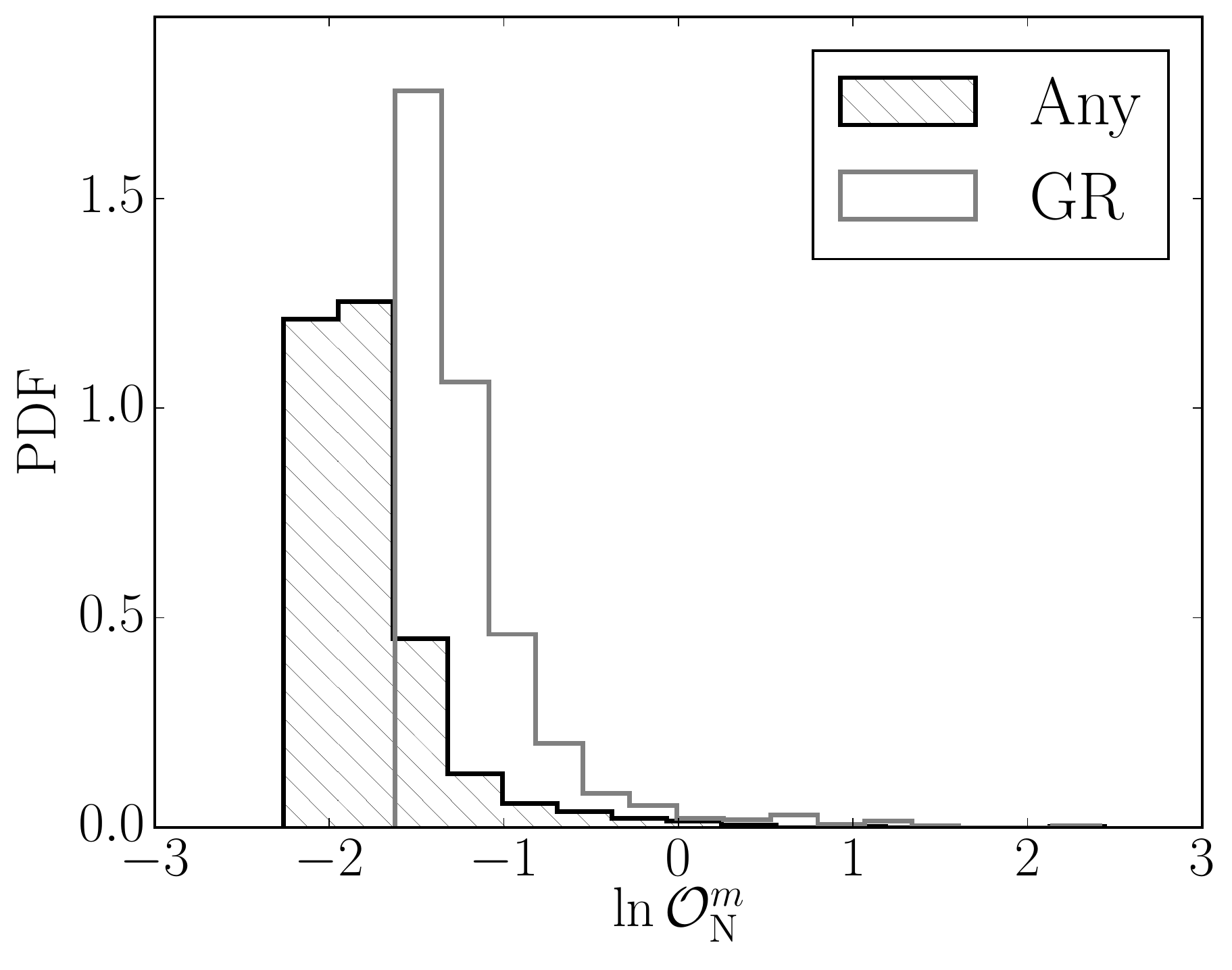}
\caption{{\em Signal vs noise log-odds background distributions for any-signal
and GR hypotheses}. Histograms of $\ln\oddssn$ (black line, hatched) and
$\ln\odds{GR}{N}$ (gray line) over an ensemble of \red{1000} simulated
noise instantiations corresponding to the Crab pulsar. For each instantiation,
three time series of Gaussian noise were produced using the design noise
spectra of H1, L1 and V1, as outlined in Sec.\ \ref{sec:analysis}; the data
are analyzed coherently across detectors. (Note that here $\ln\odds{GR}{N} =
\ln\bayes{GR}{N}$, since we assign equal weight to both models.)} 
\label{fig:crab_hist_i-none}
\end{figure}

\begin{figure*}
\centering
\includegraphics[width=\textwidth]{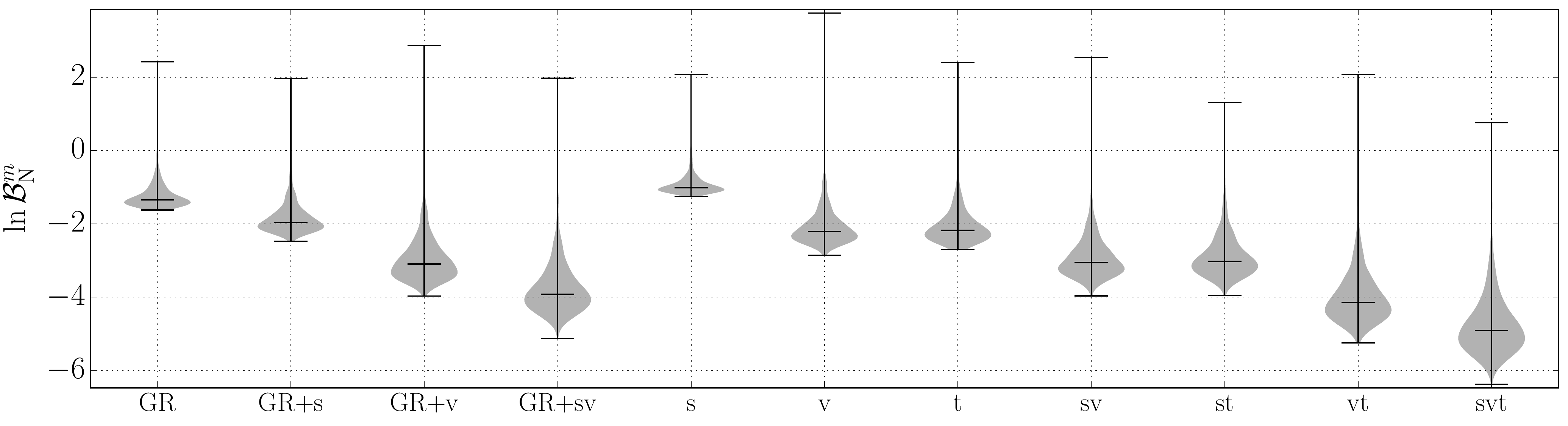}
\caption{{\em Signal vs noise log-Bayes background distributions for all
subhypotheses}. Violin plots representing histograms of the log-Bayes of
several models versus noise, computed over an ensemble of \red{1000} simulated
noise instantiations each corresponding to H1, L1 and V1 design data prepared
for the Crab pulsar; the data are analyzed coherently across detectors. The
labels on the $x$-axis indicate which hypothesis is being compared against
noise; the ``GR'' label indicates tensor modes parametrized by \eq{lambda_gr}
with fixed $\psi$ and $\iota$. Black lines mark the range and median of each
distribution. (The gray histogram in \fig{crab_hist_i-none} corresponds to the
leftmost distribution here.)}
\label{fig:crab_violin}
\end{figure*}

\section{Results} \label{sec:results}

\subsection{Model selection}

\subsubsection{Signal versus noise}

We first show that $\oddssn$, as defined in \eq{O_S_N}, can be used to
discriminate signals of any polarization from Gaussian noise, without
significant loss of sensitivity to GR signals. The black histogram in
\fig{crab_hist_i-none} shows the distribution of the natural logarithm of this
quantity (henceforth, ``log-odds''), obtained from the analysis of an ensemble
of noise instantiations corresponding to a single source---in this case, the
Crab pulsar. For comparison, the gray (unhatched) histogram in
\fig{crab_hist_i-none} is the analogous distribution for $\ln\bayes{GR}{N}$
[note that $\bayes{GR}{N} = \odds{GR}{N}$ if we assign equal priors to the GR
and Gaussian noise models, cf.\ \eq{O_S_N} with $m={\rm GR}$]; this is the
value computed in regular, GR-only targeted pulsar searches, although with
different signal amplitude priors \cite{o1cw}. Note that odds carry an
intrinsic probabilistic meaning in terms of gambling probabilities, and a
background histogram like this is not required to interpret their value (see
e.g.\ \cite{Sivia2006}).

For both quantities shown in \fig{crab_hist_i-none}, a negative value marks a
preference for the noise model ($\hyp{N}$, as defined at the beginning of
section \ref{sec:modsel_single}). However, note that a conservative (as
determined by the priors) analysis should not be expected to strongly favor
$\hyp{N}$, since the presence of a weak signal below the noise threshold cannot
be discarded; this explains why the ranges in the plots of
\fig{crab_hist_i-none} do not extend to more negative values. Generally
speaking, the magnitude of the signal prior volume (viz.\ the volume of
parameter space allowed by the signal model, weighted by the prior function)
will determine the mean of background distributions like
\fig{crab_hist_i-none}, which will be more negative the greater the signal
volume. This is a manifestation of an implicit Occam's penalty.

The relationship between the Bayes factors for different signal hypotheses vs
noise is illustrated in \fig{crab_violin}, which shows violin plots
representing the noise-ensemble distributions of $\ln{\cal B}^m_{\rm N}$ for
all models discussed in \ref{sec:modsel_single}. The values for $m\in\{\rm s,
v, sv, GR, GR+s, GR+t, GR+sv\}$ are combined to produce $\ln\oddssn$ in
\fig{crab_hist_i-none}. As explained above, the ``GR" label indicates that the
tensor modes have been parametrized using the triaxial model of \eq{lambda_gr},
with orientation parameters fixed at the known values for the Crab pulsar; on
the other hand, the ``t" label corresponds to the free-tensor template of
\eq{lambda_t}. We include both parametrizations to demonstrate the effect of
assuming a triaxial emission mechanism and restricting the orientation of the
source (see also Appendix \ref{ap:tensor_modes}).

Interestingly, \fig{crab_violin} reveals the relationship between $\ln{\cal
B}^m_{\rm N}$ and the number of degrees of freedom (a proxy for the prior
volume) of model $m$: models with more degrees of freedom have a greater prior
volume and are correspondingly downweighted, resulting in more negative values
of $\ln{\cal B}^m_{\rm N}$; this is a manifestation of the Occam's penalty
automatically applied by the Bayesian analysis (see e.g.\ Ch.\ 28 in
\cite{MacKay2005}). We underscore that this feature arises naturally from the
computation of the evidence integral, and not from manually downweighting
either model {\em a priori}.

\begin{figure*}
\centering
\includegraphics[width=0.5\textwidth]{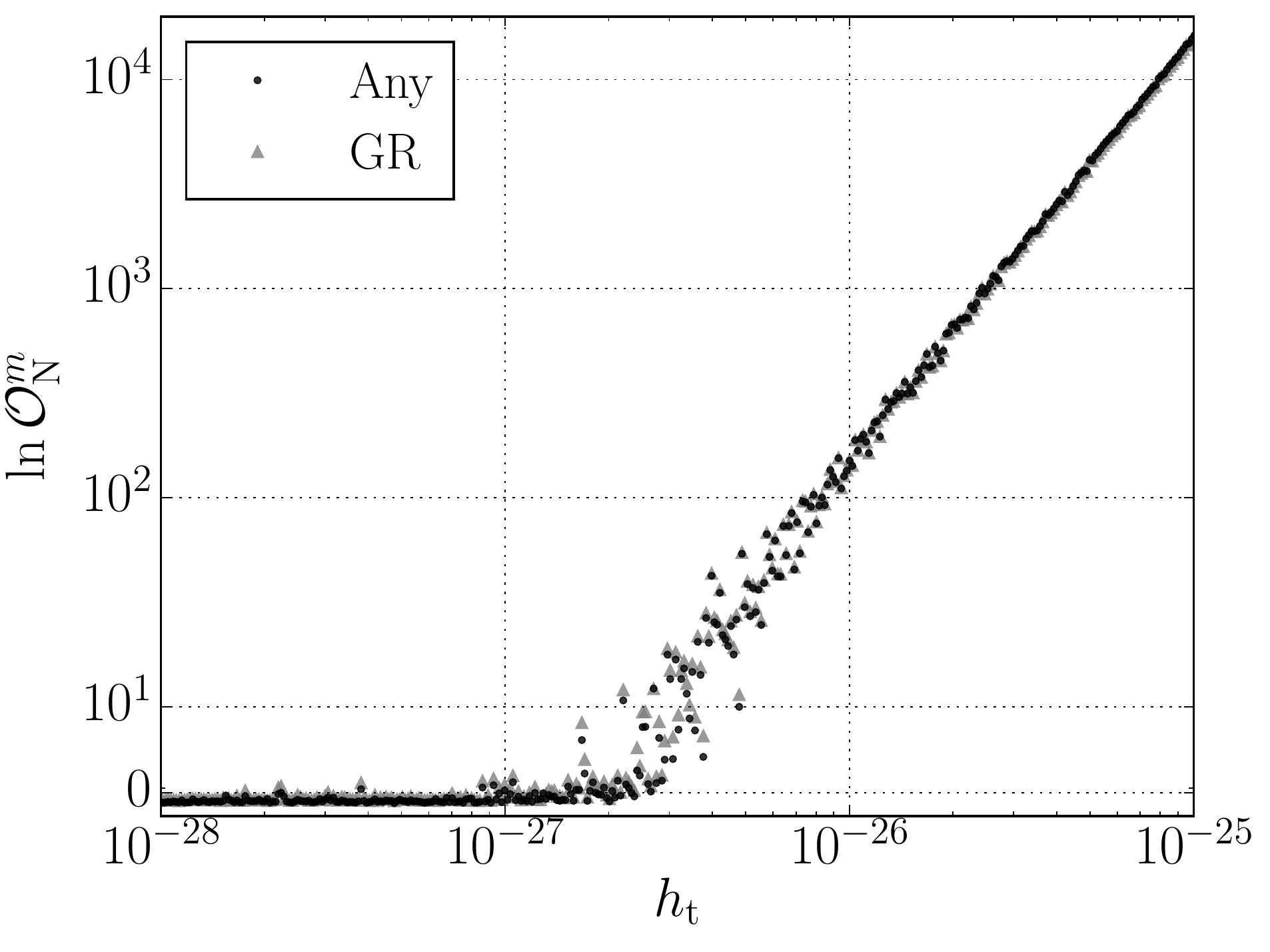}\hfill
\includegraphics[width=0.5\textwidth]{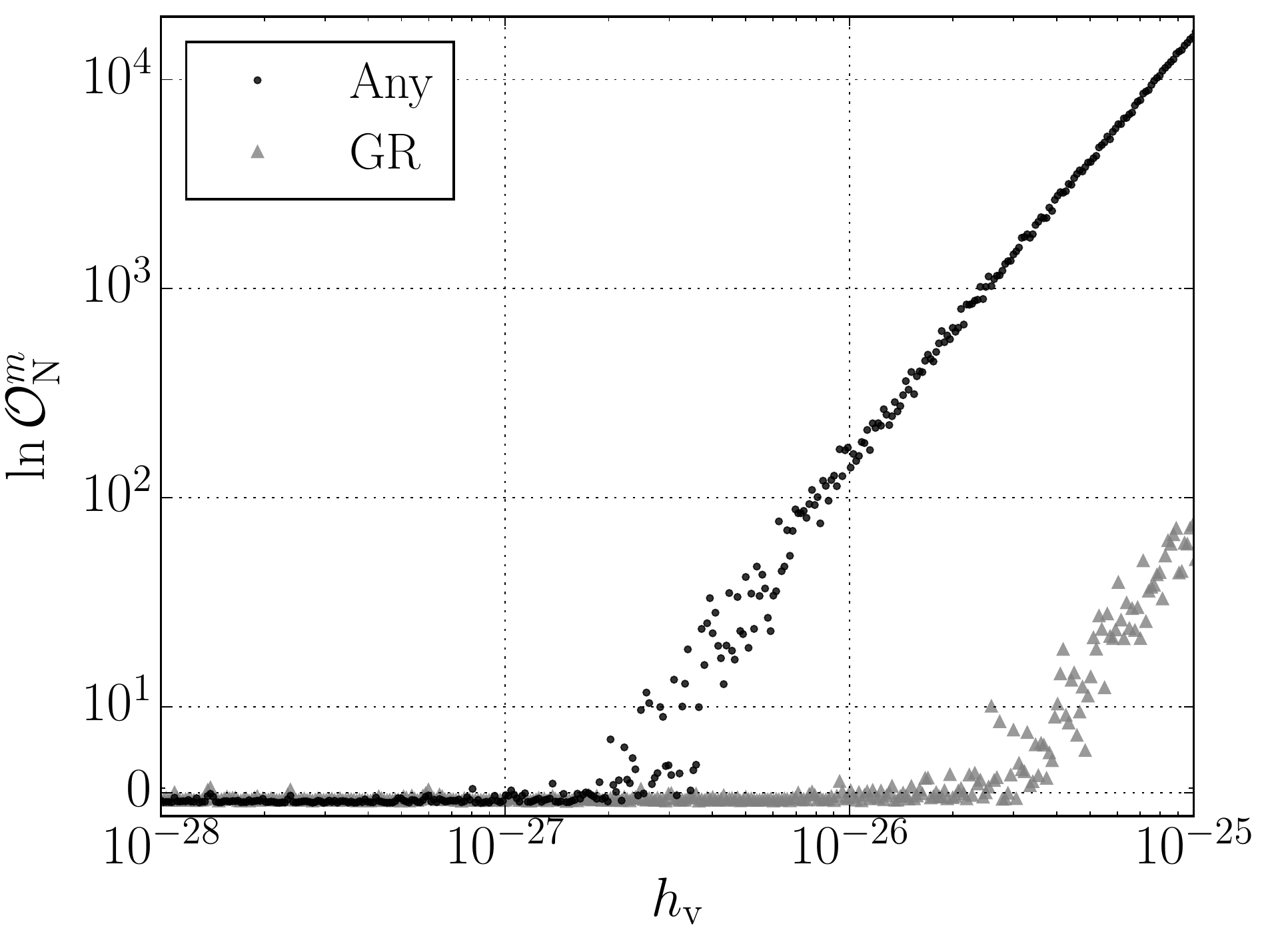}
\caption{{\em Expected sensitivity to GR and vector injections}. Log-odds of
any-signal ($\hyp{S}$, black circles) and GR ($\hyp{GR}$, gray triangles)
versus noise ($\hyp{N}$) hypotheses, as a function of injection amplitude, for
signals corresponding to both GR (left) and the vector-only model from
\cite{Mead2015} (right). The any-signal odds is defined in \eq{O_S_N}. Each of
the \red{500} points corresponds to a data instantiation (one time series for
each detector: H1, L1 and V1) made up of Gaussian noise plus a simulated
Crab-pulsar signal of the indicated strength. The injections were performed
with random values of the nuisance phase parameters, and the data were analyzed
coherently across detectors. A logarithmic scale is used for the $y$-axis,
except for a linear stretch corresponding to the \red{first decade}.}
\label{fig:crab_scat}
\end{figure*}

\begin{figure*}
\centering
\subfloat[][Any]{\includegraphics[width=0.333\textwidth]{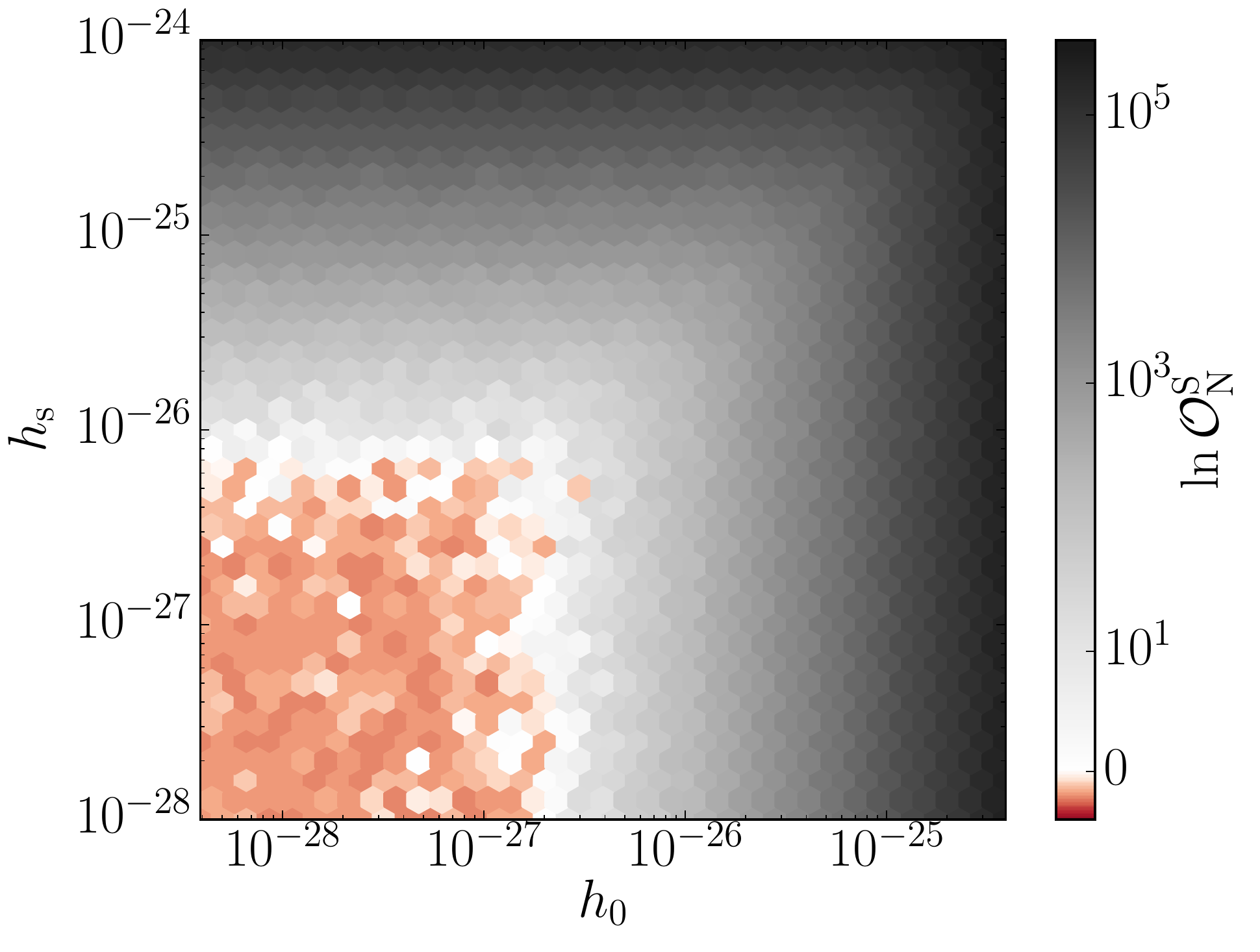}}%
\hfill
\subfloat[][GR]{\includegraphics[width=0.333\textwidth]{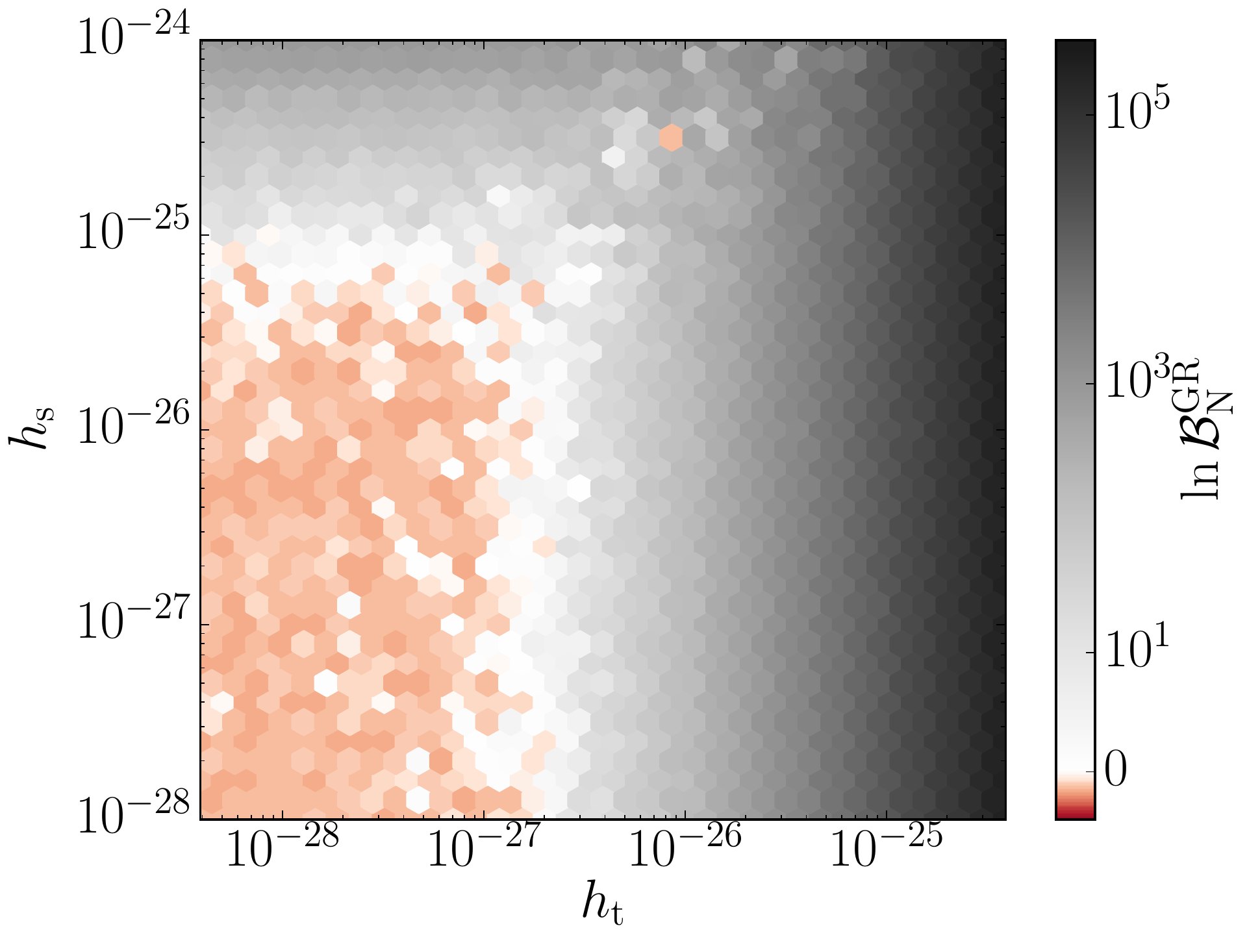}}%
\hfill
\subfloat[][GR+s]{\includegraphics[width=0.333\textwidth]{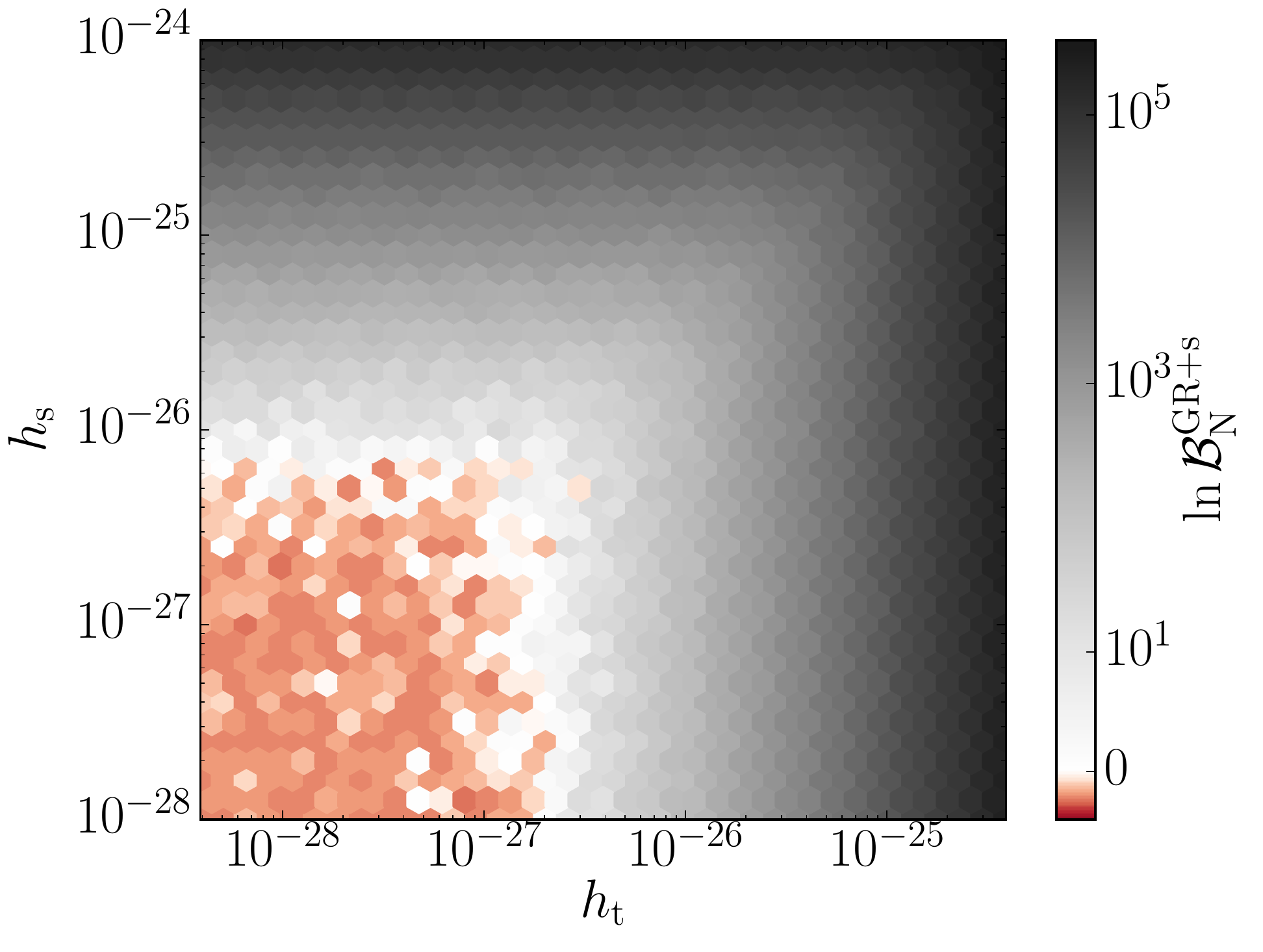}}
\caption{{\em Expected sensitivity to scalar-tensor injections}. Log-odds of
any-signal ($\hyp{S}$, left), GR ($\hyp{GR}$, center) and GR+s ($\hyp{GR+s}$,
right) hypotheses versus noise. The any-signal odds is defined in \eq{O_S_N}.
Each plot was produced by analyzing \red{2500} instantiations of data (one time
series for each detector: H1, L1 and V1) made up of Gaussian noise plus a
simulated Crab-pulsar GR+s signal of the indicated tensor ($x$-axis) and scalar
($y$-axis) amplitudes. The color of each hexagon represents the average value
of the log-odds in that region of parameter space; color is normalized
logarithmically, except for a linear stretch in the \red{$(-1,1)$} range.}
\label{fig:st2d_signal_noise}
\end{figure*}

If the data contain a sufficiently loud signal of any polarization, the
evidence for $\hyps$ will surpass that for $\hypn$, and this can be used to
establish a detection. \fig{crab_scat} shows the response of $\ln\oddssn$ and
$\ln\bayes{GR}{N}$ to the presence of GR and non-GR signals. In particular, the
second panel in \fig{crab_scat} shows results for injected signals of the
vector-only model of \cite{Mead2015}, but the behavior would be the same for
scalar-only signals. The general features of these plots confirm our
expectations that for weak, subthreshold signals, the analysis should not be
able to distinguish between the signal and noise models, yielding a Bayes
factor close to unity (more precisely, a value of $\ln\oddssn$ consistent with
the background distributions of \fig{crab_hist_i-none}). Note that, in
agreement with \fig{crab_hist_i-none}, the noise baseline for $\ln\oddssn$ lies
below that of $\ln\bayes{GR}{N}$, due to its greater prior volume. 

\begin{figure*}
\centering
\includegraphics[width=\textwidth]{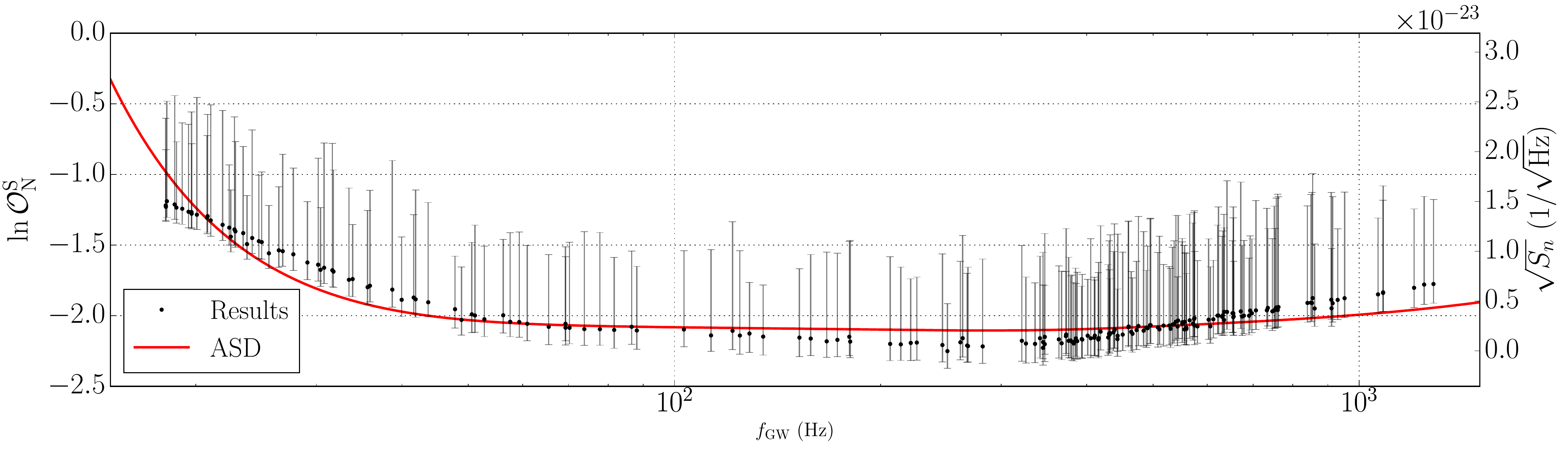}
\caption{{\em Signal log-odds vs GW frequency for noise-only data}. Circles
mark the mean of the distribution of $\ln\oddssn$, as a function of the
expected GW frequency for each pulsar in our set; vertical lines indicate
one-sided standard deviations for each source. Each data point and corresponding
bars summarize the shape of a distribution like \fig{crab_hist_i-none} for each
of the pulsars, but produced from only \red{100} runs per source. The effective
noise amplitude spectral density $\sqrt{S_n}$ (ASD, red curve), corresponding
to the harmonic mean of each detector PSD, is overlaid for comparison (scaling
obtained from a linear regression).}
\label{fig:lnbvsf}
\end{figure*}

For stronger (detectable) signals, the basic form of our likelihood functions,
\eq{likelihood_basic}, leads us to expect $\ln\oddssn$ to scale linearly with
the square of the signal-to-noise ratio (SNR):
\beq
\ln {\cal B}^m_{\rm N} \sim \left(\data \cdot {\boldsymbol \Lambda}_{\rm MP} -
|{\boldsymbol \Lambda}_{\rm MP}|^2/2\right)/\sigma^2 \propto \left(h_{\rm
inj}/\sigma\right)^2,
\eeq
where the variance $\sigma^2$ proxies the PSD and we let ${\boldsymbol
\Lambda}_{\rm MP}$ be the time-series vector corresponding to the maximum
probability template; for a stationary PSD, this implies $\ln {\cal B}^m_{\rm
N} \propto h_{\rm inj}^2$, as observed in \fig{crab_scat}. The spread around
the trendline is due to the individual features of each noise instantiation and
(much less so) to numerical errors in the computation of the evidence,
\eq{evidence}. For details on numerical uncertainty, see Appendix
\ref{app:errors}. 

From the left panel of \fig{crab_scat}, we conclude that $\ln \odds{S}{N}$ can
be as good an indicator of the presence of GR signals as $\ln \bayes{GR}{N}$
itself; this implies that we may include non-GR polarizations in our search
without significantly sacrificing sensitivity to GR signals. However, the
power of $\ln\oddssn$ lies in responding also to non-GR signals. As an example
of this, the right panel in \fig{crab_scat} shows $\ln\oddssn$ and
$\ln\bayes{GR}{N}$ as a function of the amplitude of a fully non-GR injection.
Here, we have chosen to inject a particular model of vector signal developed in
\cite{Mead2015}, but the results are generic. 

Note that, for sufficiently loud signals, $\hyp{GR}$ becomes preferable over
$\hyp{N}$ (hence $\ln\bayes{GR}{N}>0$), even when the injection model does not
match the search; this is because the noise evidence drops faster than GR's and
becomes very small (i.e.\ the data do not look at all like Gaussian noise,
although they do not match the expected GR signal well either). The particular
SNR at which this occurs will depend on the overlap between the antenna
patterns of the injection and those of GR, and will consequently vary among
sources.

For the interesting case of scalar-tensor theories (here, templates composed of
GR plus an extra breathing component, and denoted ``GR+s"), the behavior is
slightly different. This is both because GR+s has an extra amplitude degree of
freedom ($a_{\rm s}$) and, as discussed in Sec.\ \ref{sec:modsel_single},
because $\hyp{GR}$ can be recovered as a special case of $\hyp{GR+s}$ (namely,
when $a_{\rm s} \rightarrow 0$). In \fig{st2d_signal_noise}, we present the
log-odds of signal versus noise hypotheses as a function of injected GR
($x$-axis) and scalar ($y$-axis) strengths. These plots divide the $\hS$--$\hT$
plane in roughly two regions where the associated signal model ($\hyps$,
$\hyp{GR}$ or $\hyp{GR+s}$) is preferred (black) and where it is not (red). The
latter corresponds to the area of parameter space associated with subthreshold
signals that cannot be detected. 

As expected, the best coverage is obtained when analyzing the data using the
model matching the injection, GR+s, (rightmost plot) or the all-signal model
(leftmost plot). In both these cases, the results improve with either scalar or
tensor SNR. In contrast, the GR analysis (center plot) is sensitive to
tensor strain, but, as evidenced by the extended red region in the central
plot, it misidentifies strong scalar signals as noise. Nevertheless, if the
scalar component is larger than ${\sim}5\times 10^{-26}$, the GR analysis will
disfavor the noise hypothesis, even for a small tensor component, as in the
right panel of \fig{crab_scat}; this is the same behavior observed in
\fig{crab_scat}. In contrast, the any-signal analysis is sensitive to the total
power of the injected signal, regardless of polarization.

We have produced distributions of background $\ln\oddssn$, like those of
\fig{crab_hist_i-none}, for all \red{200} known pulsars in the sensitive band
of the three detectors under consideration (same set analyzed in \cite{o1cw}.
In \fig{lnbvsf}, these are represented by their respective means and one-sided
standard deviations as a function of the pulsar's GW frequency. The frequency
dependence is explained by variations in the instrumental noise spectra. This
is explained by the fact that, for a particular prior choice, more information
is gained from the data if the noise floor is lower: with less noise it is
possible to discard the presence of weaker signals, so the value of
$\ln\oddssn$ decreases.

\subsubsection{GR vs non-GR}

\begin{figure*}[p]
\centering
\includegraphics[width=\columnwidth]{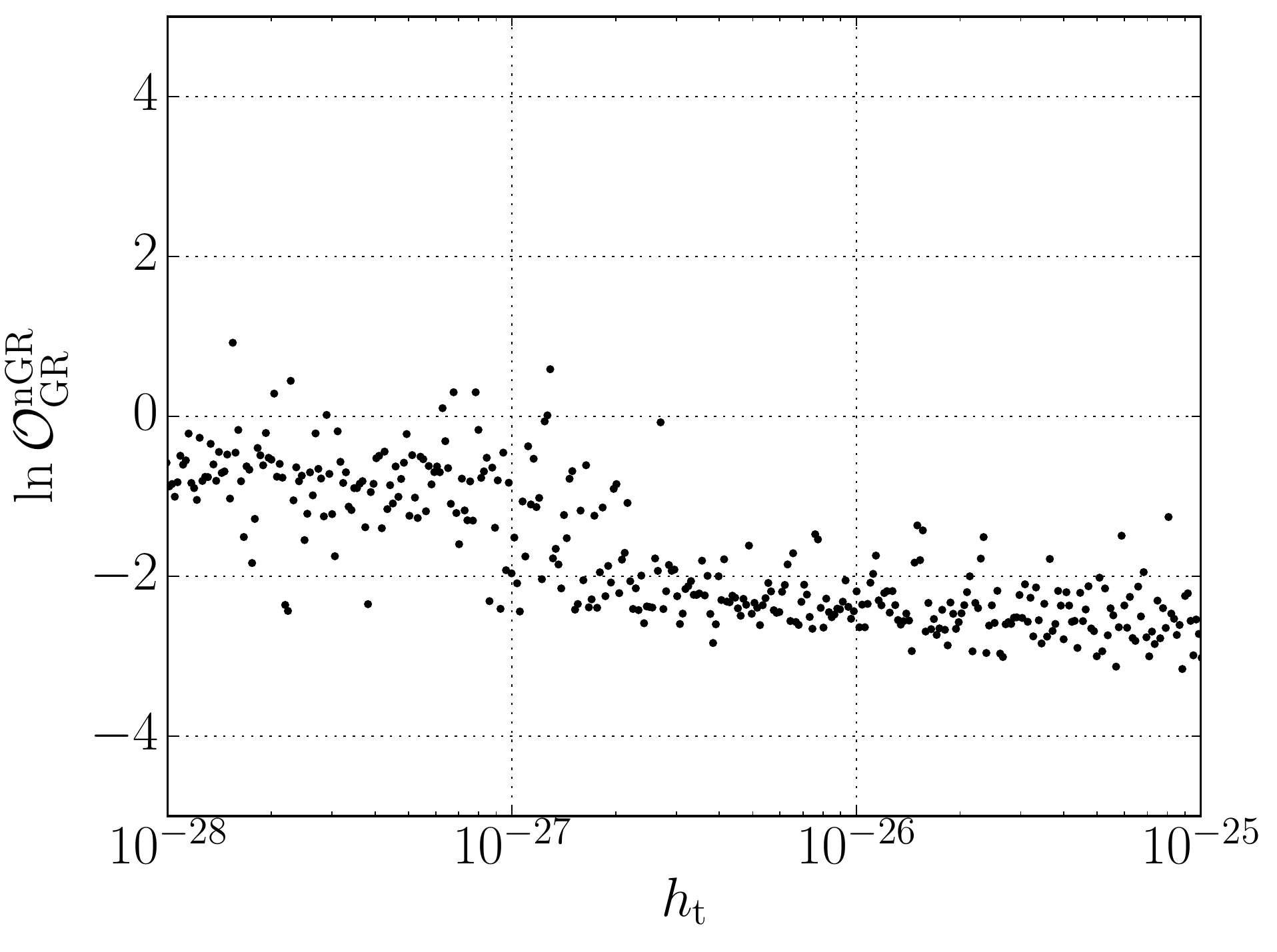}%
\hfill
\includegraphics[width=\columnwidth]{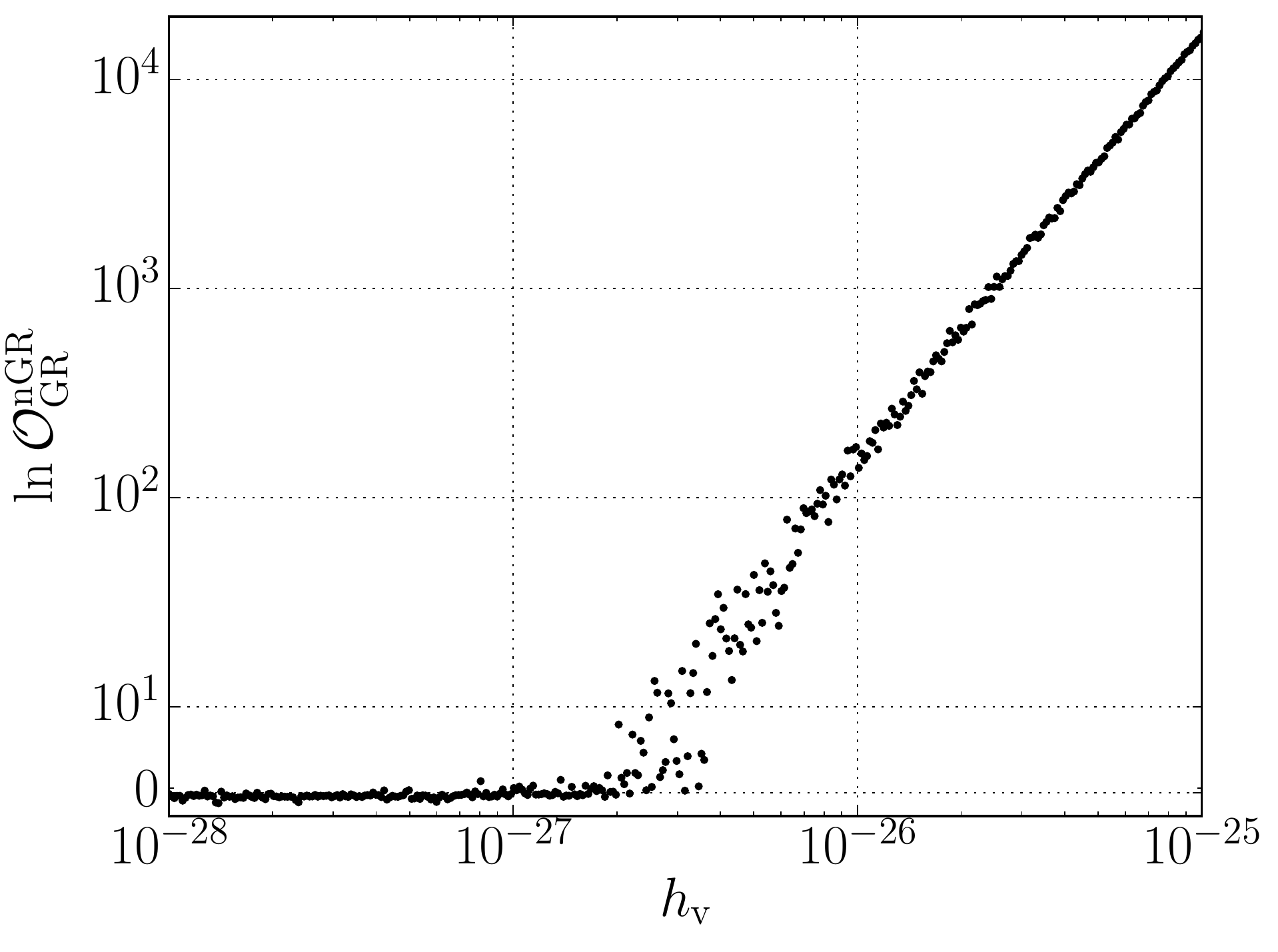}
\caption{{\em Categorizing tensor and vector injections} ($\hyp{nGR}$ vs
$\hyp{GR}$). Non-GR vs GR log-odds, as a function of effective injection
amplitude, for both GR (left) and the vector-only model from \cite{Mead2015}
(right). Each of the \red{500} points corresponds to a data instantiation (one
time series for each detector: H1, L1 and V1) made up of Gaussian noise plus a
simulated Crab-pulsar signal of the indicated strength. The injections were
performed with random values of the nuisance phase parameters, and the data
were analyzed coherently across detectors. Note that, on the right, a
logarithmic scale is used for the $y$-axis, except for a linear stretch
corresponding to the \red{first decade}. }
\label{fig:crab_nGR_GR}
\end{figure*}

\begin{figure*}[p]
\centering
\includegraphics[width=0.5\textwidth]{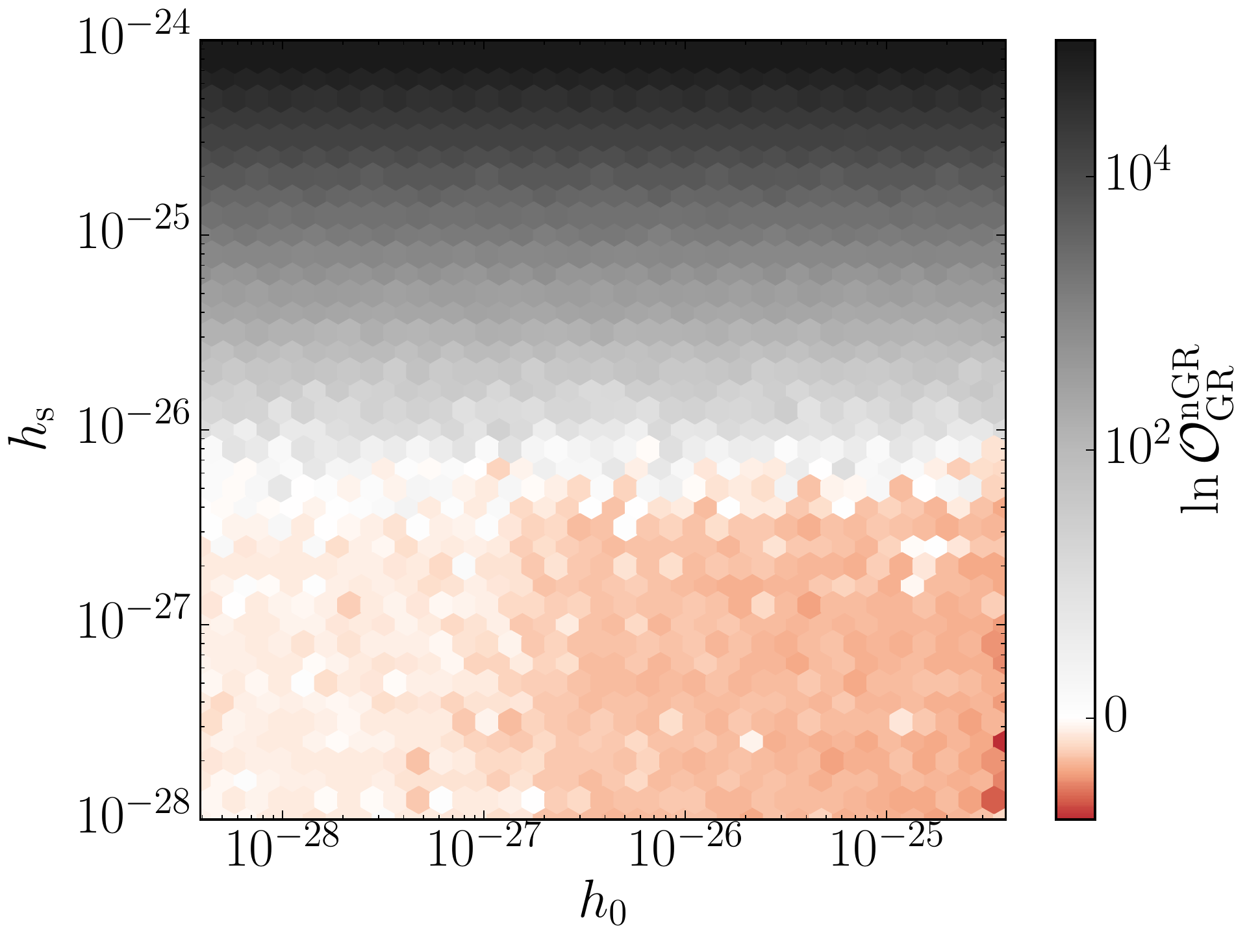}\hfill
\includegraphics[width=0.5\textwidth]{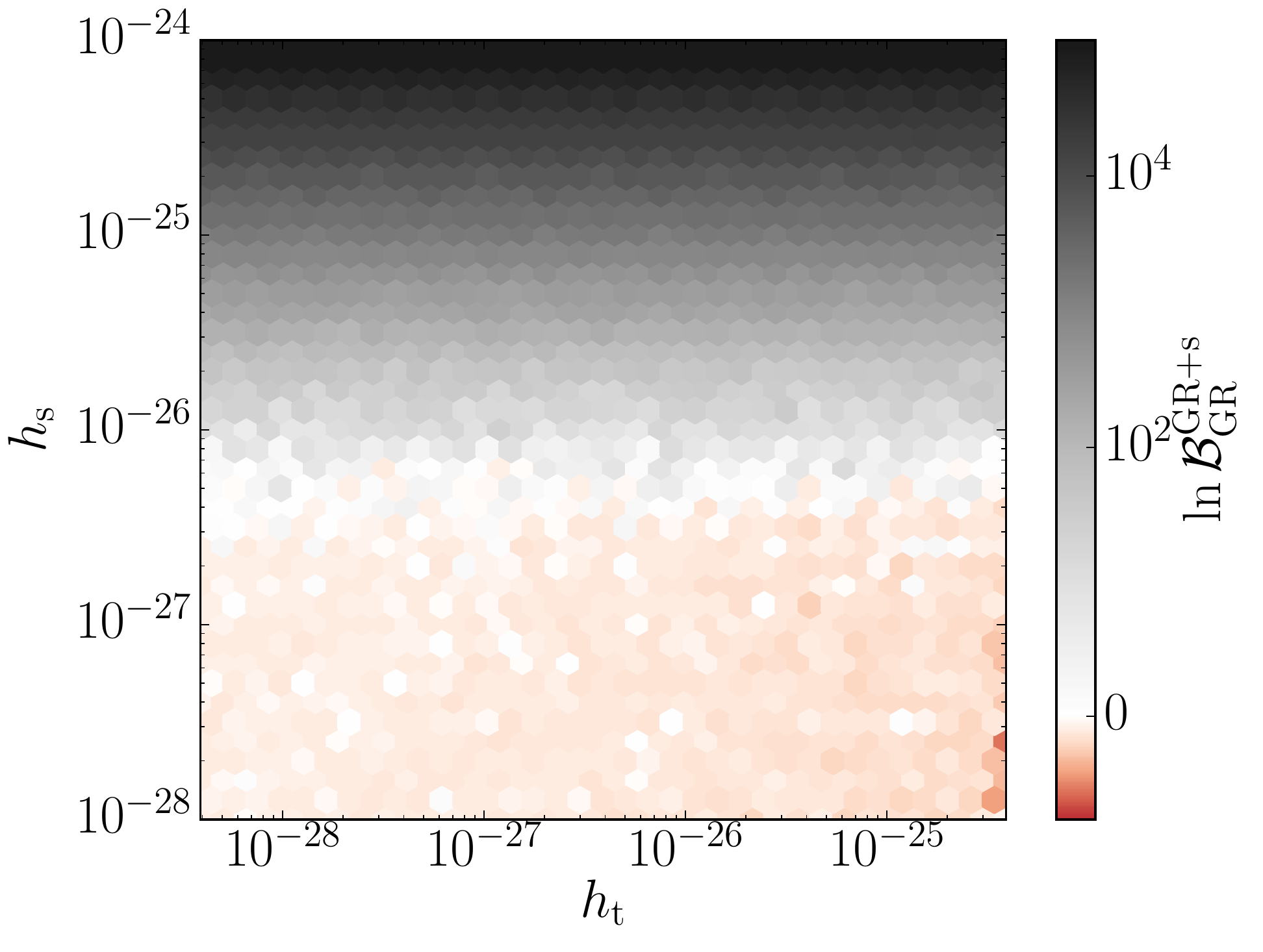}
\caption{{\em Categorizing scalar-tensor injections} ($\hyp{nGR}$ \&
$\hyp{GR+s}$ vs $\hyp{GR}$). Log-odds comparing the non-GR and GR+s hypotheses
to GR. The non-GR odds is defined in \eq{O_nGR_GR}. Each plot was produced by
analyzing \red{2500} instantiations of data (one time series for each detector:
H1, L1 and V1) made up of Gaussian noise plus a simulated Crab-pulsar GR+s
signal of the indicated tensor ($x$-axis) and scalar ($y$-axis) amplitudes.
The color of each hexagon represents the average value of the log-odds in that
region of parameter space; color is normalized logarithmically, except for a
linear stretch in the \red{$(-1,1)$} range.}
\label{fig:crab_nGR_GR_i-ST}
\end{figure*}

In the presence of a signal, $\odds{nGR}{GR}$, as defined by \eq{O_nGR_GR},
indicates whether there is reason to believe there is a GR violation or not.
Because there could always be an unresolvably small departure from GR, we do
not expect our analysis (with priors as chosen) to ever strongly favor the GR
hypothesis; rather, in the presence of a GR signal we will find that
$\ln\odds{nGR}{GR}$ remains relatively close to zero, simply meaning that there
is no strong evidence for or against non-GR features. This is indeed the
behavior observed in the left panel of \fig{crab_nGR_GR}, where
$\ln\odds{nGR}{GR}$ is shown to be roughly insensitive to tensor injection
amplitude. For values of $\hT$ below certain threshold (which, in this case, is
around $3\times 10^{-27}$), the search does not detect a signal and,
consequently, no information is gained for or against $\hyp{GR}$, i.e.\
$\ln\odds{nGR}{GR}\sim0$. The difference between the two populations (below and
above threshold) is determined mainly by the choice of amplitude priors.

The behavior of $\odds{nGR}{GR}$ is less ambiguous in the presence of a non-GR
signal. For instance, if the data contain a detectable signal that completely
lacks tensor components, then $\odds{nGR}{GR}$ will unequivocally reflect this.
This is evidenced by the growth of $\ln\odds{nGR}{GR}$ with injected
nontensorial SNR in the right panel of \fig{crab_nGR_GR}. In other words,
while the analysis is inconclusive for GR injections because it cannot discard
the presence of subthreshold non-GR components hidden by the noise, vector
signals are are clearly identified as not conforming to GR. This is a
reflection of the fact that, as mentioned in the introduction, any evidence of
a nontensorial component is fatal for GR, but absence of non-GR components
does not mean Einstein's theory is necessarily correct.

\begin{figure*}
\centering
\includegraphics[width=\columnwidth]{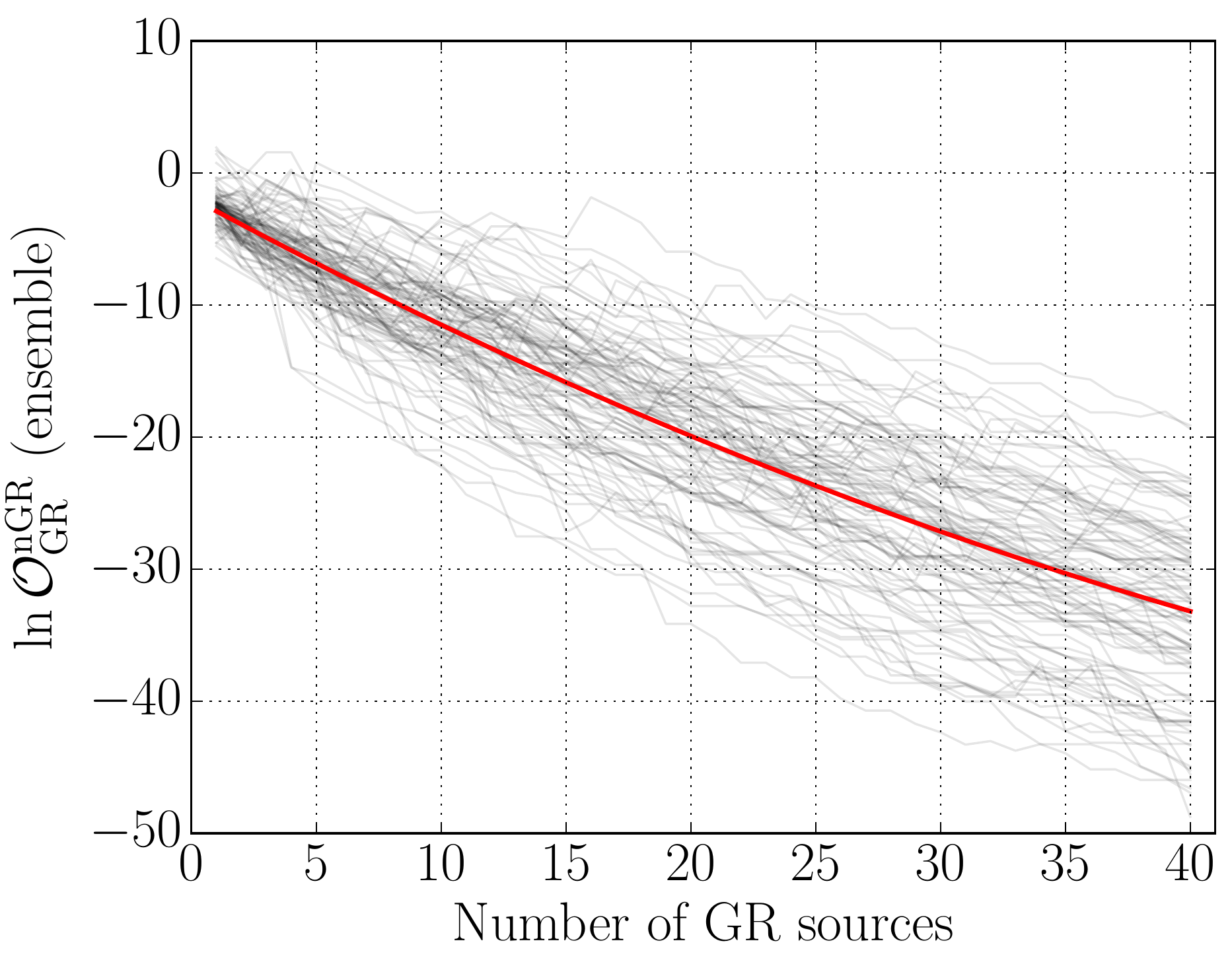}%
\hfill
\includegraphics[width=\columnwidth]{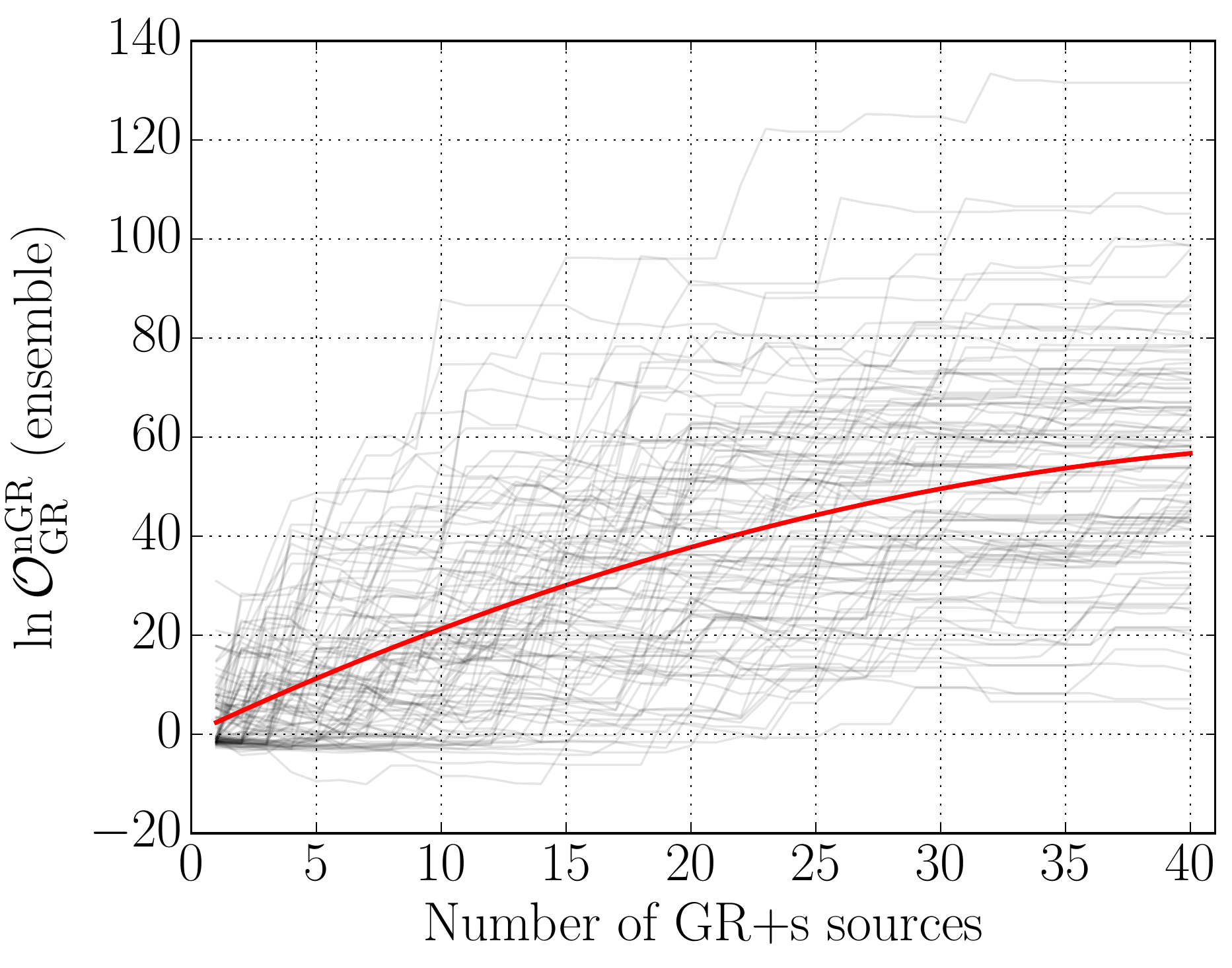}
\caption{{\em Ensemble non-GR vs GR log-odds}. Non-GR vs GR log-odds computed
from data for multiple sources, vs the number of sources in the set. Each
light-gray trace marks a possible progression of the ensemble log-odds as new
sources are added; the red line corresponds to the best quadratic fit. For each
pulsar, we chose an arbitrary data instantiation containing a GR (left) or GR+s
(right); GR signals are restricted to \red{$10^{-27} < \hT < 10^{-26}$}, while
GR+s signals also satisfy \red{$0.3 < \hS/\hT< 1$}. We compute the value of
$\ln\odds{nGR}{GR}$ for each signal in the set and combine them according to
\eq{ensemble_odds_ngr_gr} to obtain the ensemble value plotted in the
$y$-axis.} 
\label{fig:ensemble_ngr_gr}
\end{figure*}

As might be expected, $\odds{nGR}{GR}$ responds to non-GR signals that include
a tensor component with a combination of features from both panels of
\fig{crab_nGR_GR}. As an example, the left plot of \fig{crab_nGR_GR_i-ST} shows
$\odds{nGR}{GR}$ in the presence of GR+s injections, as a function of injected
tensor and scalar amplitudes. This plot can be split into three clearly
demarcated regions: one in which the signal is not detected (light red, bottom
left), one in which the signal is detected and the non-GR model is preferred
(black, top), and one which the signal is detected but where the evidence for a
deviation from GR is not clear due to the predominance of the tensorial
component (darker red, bottom right). The first corresponds to the subthreshold
population on either side of \fig{crab_nGR_GR}, while the second and third
correspond to the above-threshold populations on the right and left sides of
\fig{crab_nGR_GR} respectively; indeed, note that a horizontal slice taken over
the red region of the left plot produces a series of points like those in the
left panel of \fig{crab_nGR_GR}. For reference, \fig{crab_nGR_GR_i-ST} also
includes the direct comparison of GR+s and GR on the right.

We can make a stronger statement about the agreement of the data with GR by
making use of signals from multiple sources, as discussed in Sec.\
\ref{sec:modsel_multiple}. The power of combining multiple signals is
illustrated in \fig{ensemble_ngr_gr}, where $\ln\odds{nGR}{GR}$, as defined in
\eq{ensemble_odds_ngr_gr}, is plotted vs number of GR (left) and GR+s (right)
signals detected. Note that this presumes that, for each source, the presence
of a signal has already been established from the value of $\ln\oddssn$.
Computing the ensemble $\ln\odds{nGR}{GR}$, as done here, is a good way of
summarizing the information contained in the data about the relative
likelihoods between the two models, but it provides no information not already
present in the set of individual single-source odds.

\subsection{Parameter estimation}

When no conclusive evidence for a CW is found in the data, we are still
interested in placing upper limits on the strength of possible signals (up to
some credibility), and this is done as explained in Sec.\ \ref{sec:method_pe}.
By the same token, if a signal consistent with GR is detected, we can always
place an upper limit on the amplitude of non-GR modes, even if the odds
indicate there is no clear sign of a GR violation.

\begin{figure*}[p]
\centering
\includegraphics[width=\columnwidth]{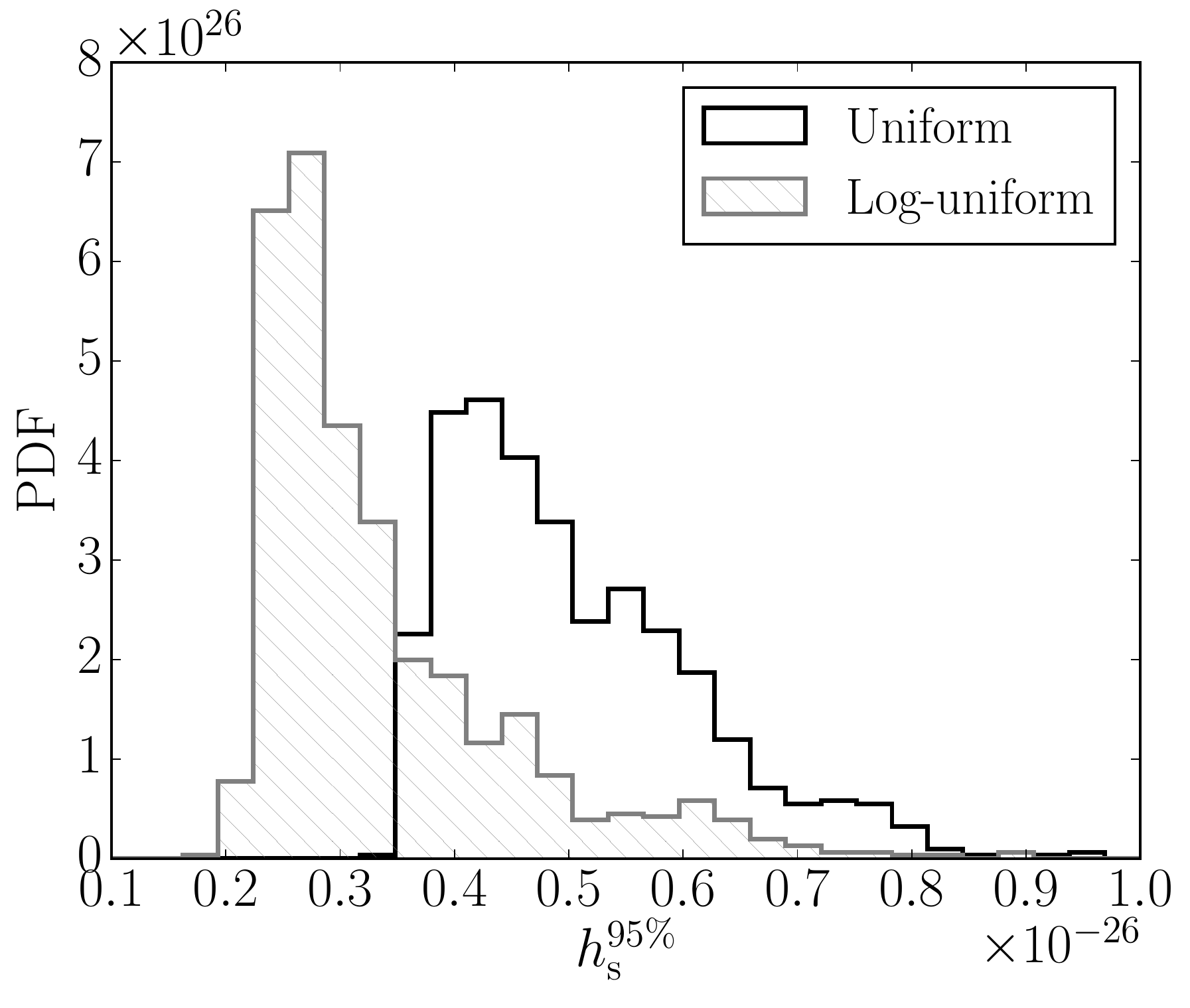}\hfill
\includegraphics[width=\columnwidth]{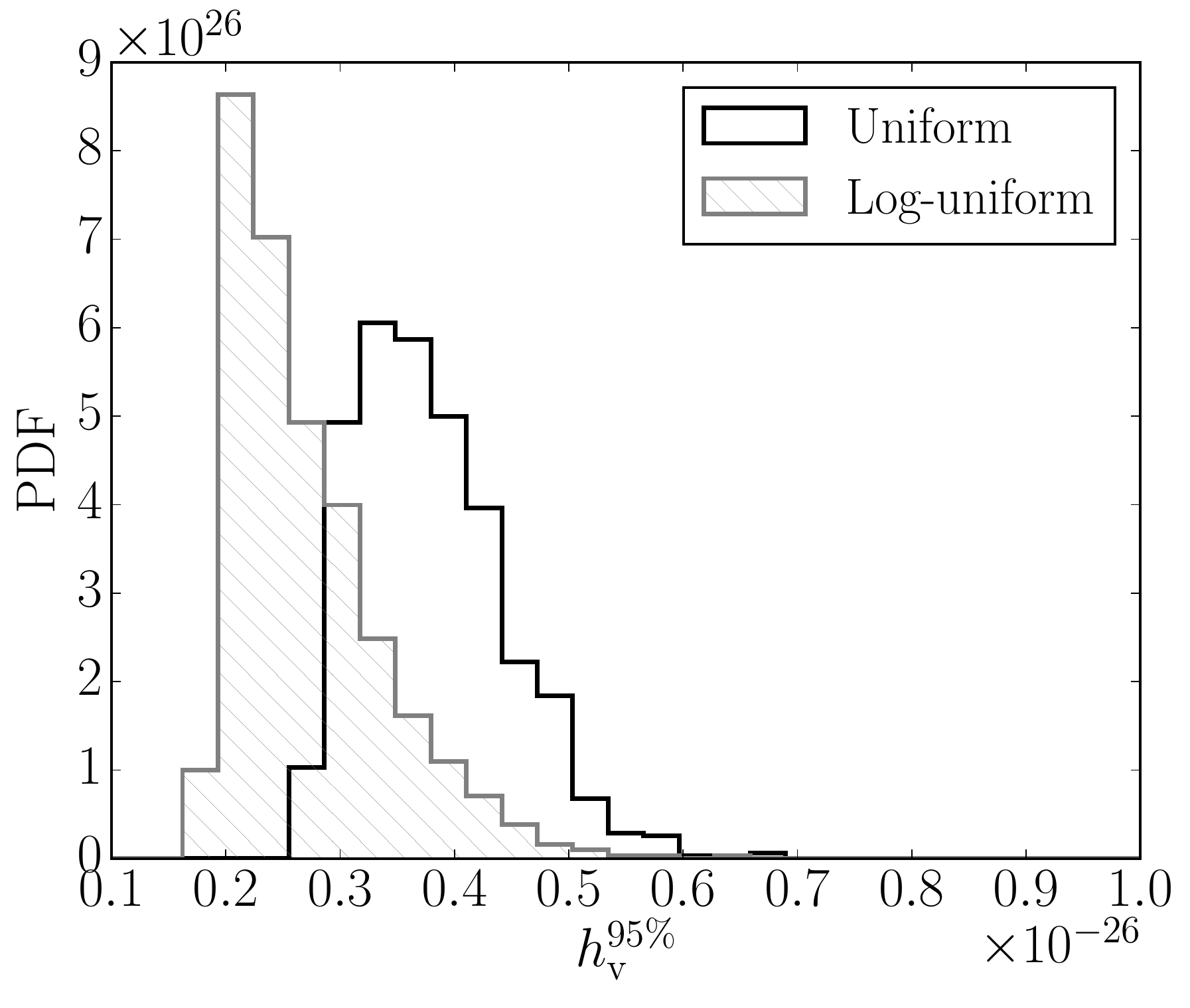}
\caption{{\em Expected Crab non-GR upper limits in absence of signal}.
Histogram of 95\%-credible upper limits for the scalar (left) and vector
(right) amplitudes, for a set of \red{1000} noise-only data sets, computed
using priors uniform in the amplitude (black) or uniform in the logarithm of
the amplitude (hatched gray); the differences between these two priors are
discussed in detail in Appendix \ref{app:priors}. Each instantiation (one time
series for each detector: H1, L1 and V1) is made up of simulated Gaussian noise
with standard deviation given by the advanced design PSDs. Scalar and vector
upper limits are produced using the GR+s and GR+v models respectively.}
\label{fig:crab_hul_i-none}
\end{figure*}

For instance, we can get a quantitative estimate of our sensitivity to scalar
modes from a given source by looking at the distribution of $\hul{s}$, defined
in \eq{hbul}, computed for a set of noise-only data instantiations. Such
distribution for the Crab pulsar is presented in the left panel of
\fig{crab_hul_i-none}. Similarly, the right panel presents estimates for the
sensitivity to vector modes coming from the Crab pulsar, assuming a
vector-tensor model. In this case, however, the quantity plotted is the upper
limit on total, effective vector strain amplitude $\hV$, \eq{hv}. These plots
include distributions produced using the same log-uniform prior used to obtain
Bayes factors, as well as more conservative ones obtained using uniform
amplitude priors (see Appendix \ref{app:priors}). In either case, the magnitude
of $\hul{v}$ is comparable to that of $\hul{s}$.

\begin{figure*}[p]
\centering
\includegraphics[width=0.5\textwidth]{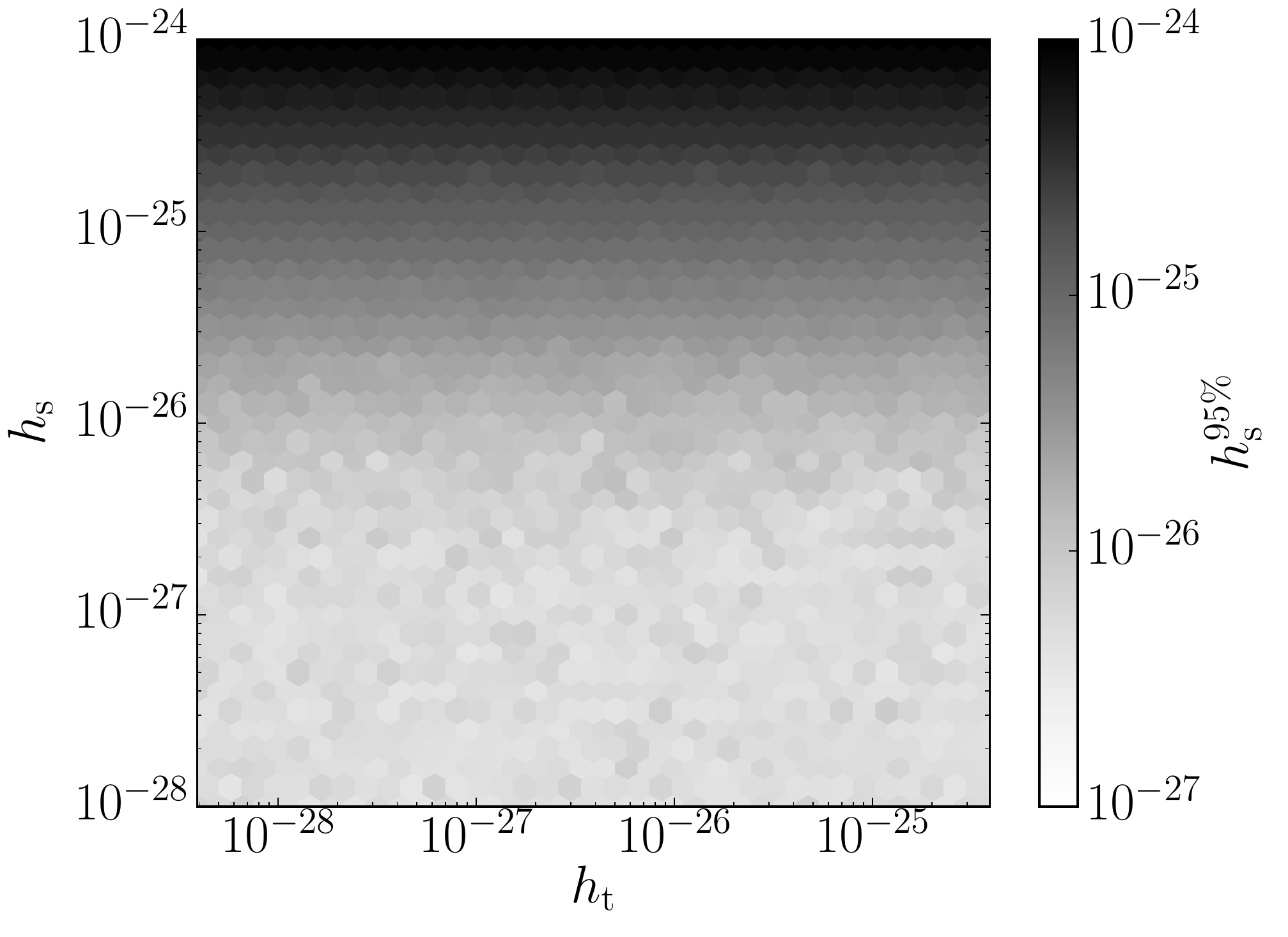}\hfill
\includegraphics[width=0.5\textwidth]{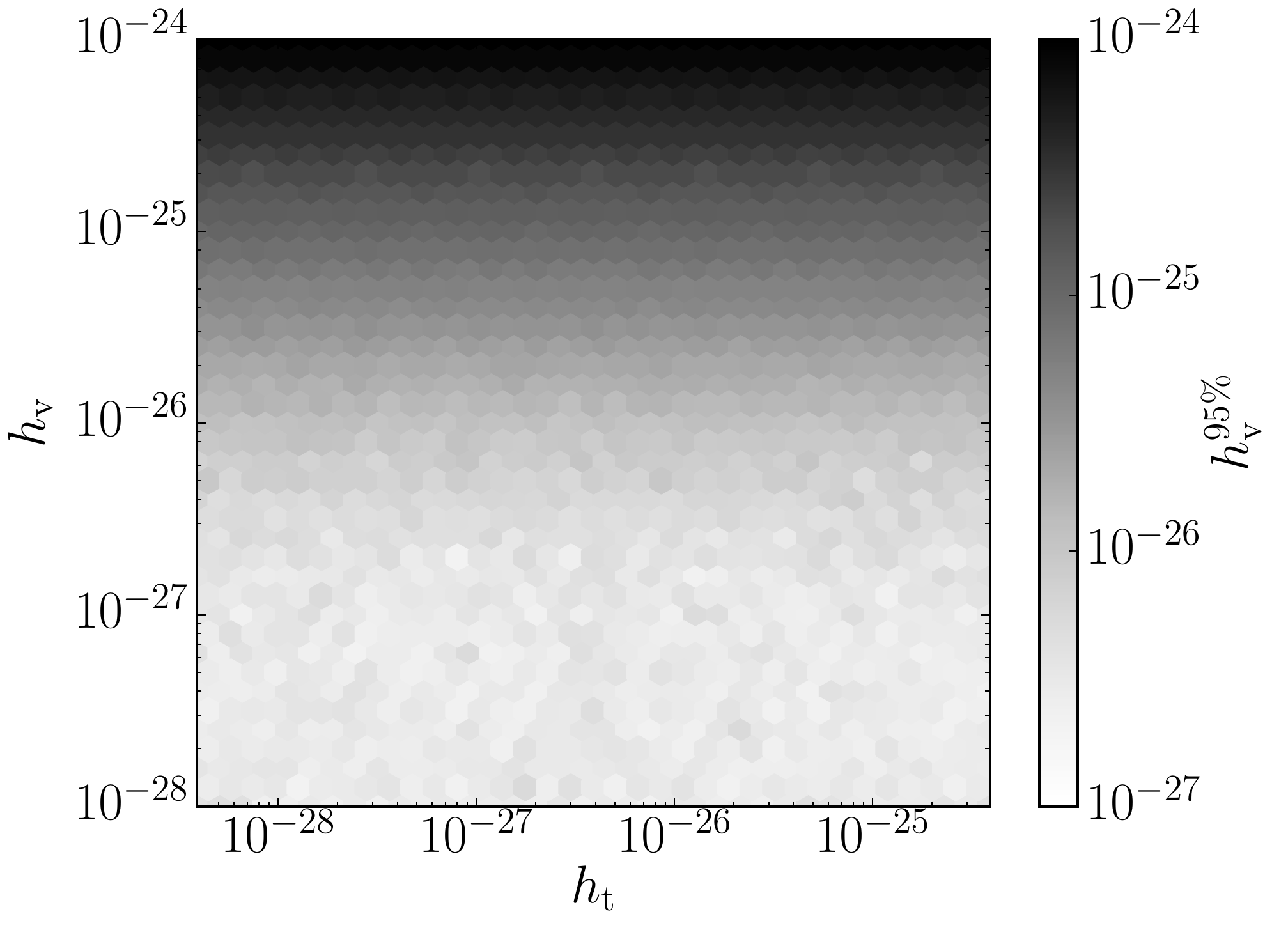}
\caption{{\em Expected Crab scalar and vector upper limits in presence of GR+s
and GR+v signals}. Shading represents the 95\%-credible upper limit for the
scalar ($\hul{s}$, left) and vector ($\hul{v}$, right) amplitudes, vs the
amplitude of injected GR ($x$-axis) and corresponding non-GR
($y$-axis) components.
Each plot was produced by analyzing \red{2500} instantiations of data (one time
series for each detector: H1, L1 and V1) made up of Gaussian noise plus a
simulated Crab-pulsar GR+s (left) or GR+v (right) signal with indicated
strains. The color of each hexagon represents the average value of
the upper limit in that region of parameter space.}
\label{fig:crab_hbul_i-ST-VT}
\end{figure*}

Interestingly, our ability to measure scalar and vector amplitudes is
unaffected by the presence of other modes. We illustrate this for the Crab
pulsar in \fig{crab_hbul_i-ST-VT}, which results from analyzing data with GR+s
(left) and GR+v (right) injections. There we plot $\hul{s}$ as a function of
scalar and tensor injection amplitudes on the left, and $\hul{v}$ as a function
of vector and tensor injection amplitudes on the right. From these plots, one
can conclude that $\hul{s}$ and $\hul{v}$ are sensitive only to the
corresponding scalar and vector components, and not by $\hT$. (It is worth
emphasized that the upper limits, $\hul{s}$ and $\hul{v}$, are well-defined
even when the non-GR component is strong enough to be detected, as is the case
for the darker-colored regions.)

\begin{figure*}[p]
\subfloat[Scalar-tensor]{\includegraphics[width=\textwidth]{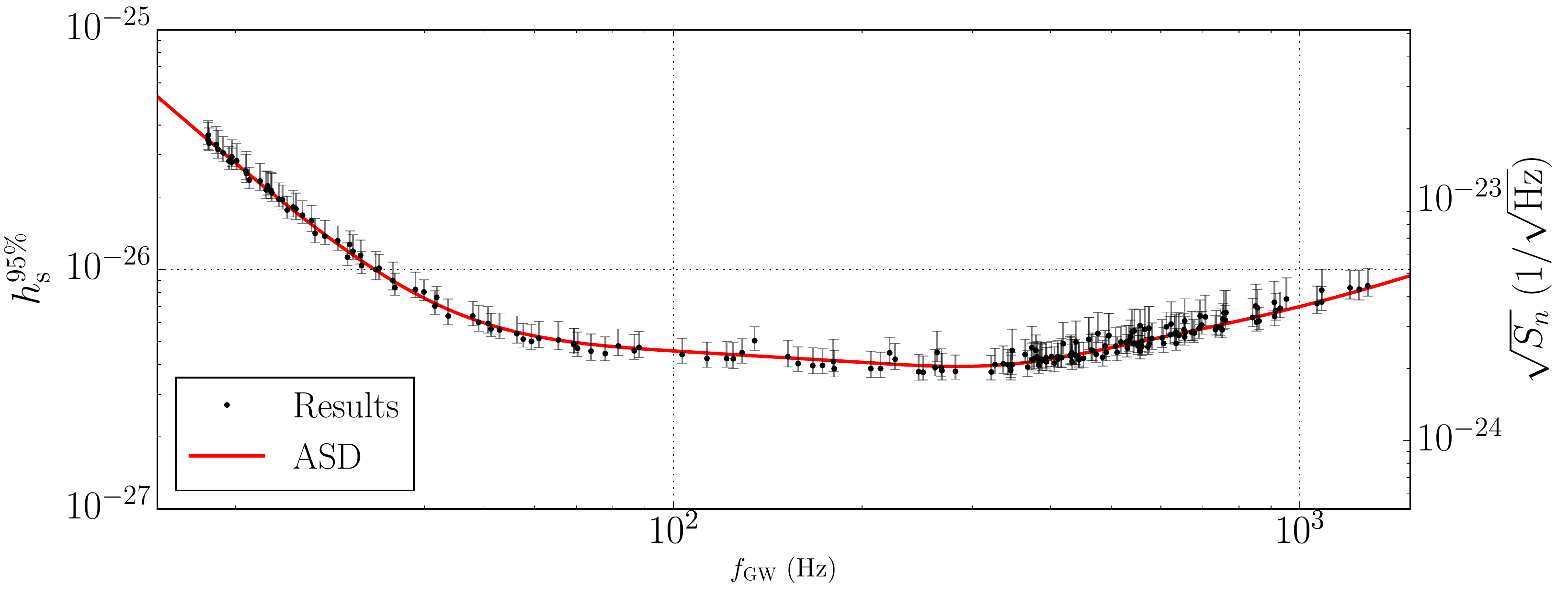}} \label{fig:hbulvsf}%
\\
\subfloat[Vector-tensor]{
\includegraphics[width=\textwidth]{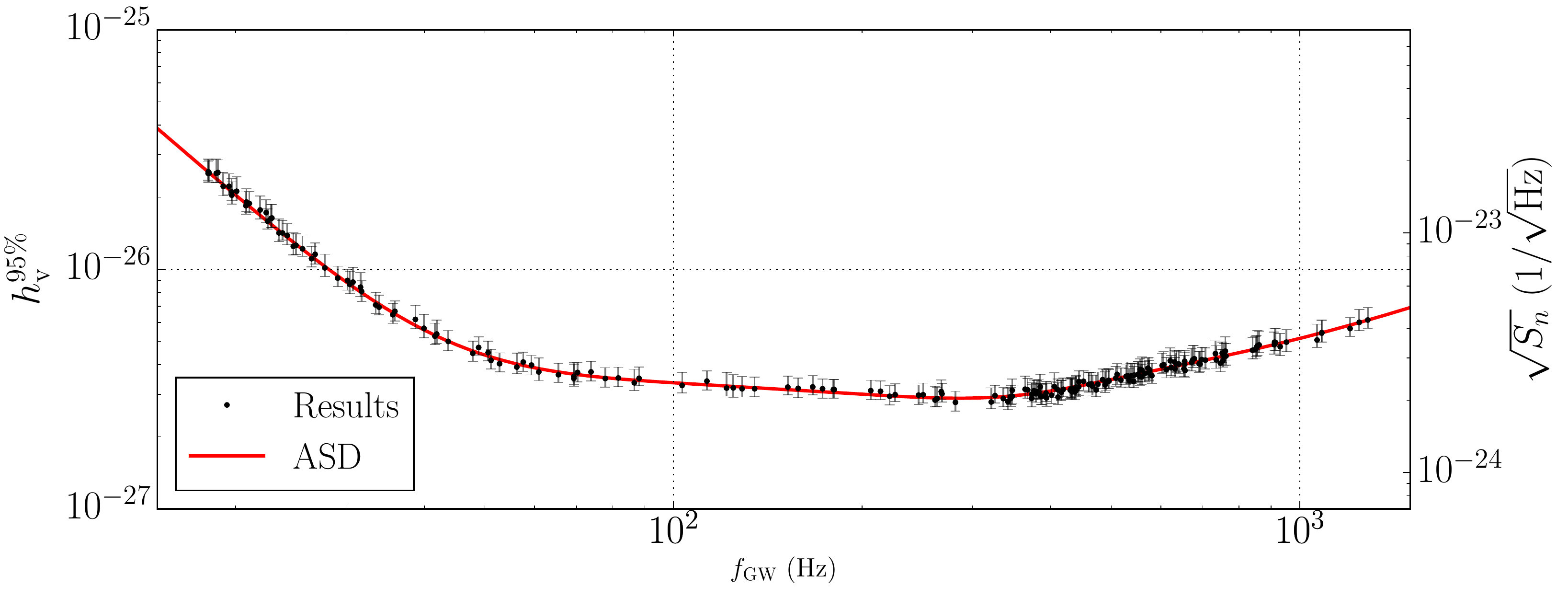}}%
\\
\subfloat[Tensor]{\includegraphics[width=\textwidth]{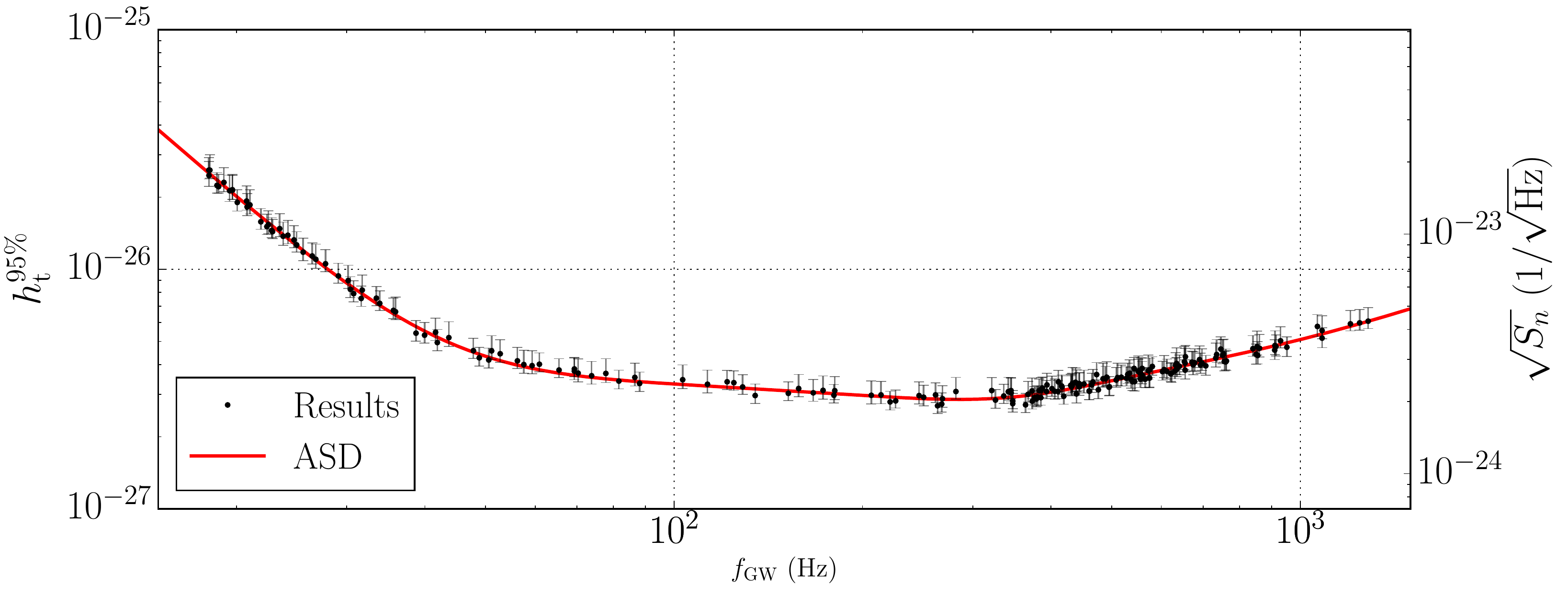}}
\label{fig:hvulvsf}
\caption{{\em Expected upper limits in absence of signal vs GW frequency}.
Circles mark the mean of the distribution of $\hul{s}$ (top), $\hul{v}$
(middle) and $\hul{t}$ (bottom), as a function of expected GW frequency for
each pulsar in our set; vertical lines mark one-sided standard deviations for
each source. Each data point and corresponding bars summarize the shape of a
distribution like those of \fig{crab_hul_i-none}, but produced from \red{100}
noise instantiations each. The scalar, vector and tensor upper limits were
produced assuming st, vt and t models respectively. We use \red{uniform} priors
in all amplitude parameters (see \fig{crab_hul_i-none} and Appendix
\ref{app:priors}).}
\label{fig:hulvsf}
\end{figure*}

As shown previously in the literature, the mean of distributions like those of
\fig{crab_hul_i-none} will scale with $\sqrt{S_{\rm n}(\fgw)/T}$, where $S_{\rm
n}(\fgw)$ is the effective PSD of the detector noise at the expected GW
frequency $\fgw$, and $T$ is the integration time (cf.\ Eq.\ (26) and Fig.\ 1
of \cite{Dupuis2005}). Because of this, the mean of this distribution will vary
with the source's expected GW frequency, as shown in \fig{hulvsf}.  Following
convention, these upper limits are computed using uniform amplitude priors,
which means they are a factor of a few less stringent than those obtained with
a log-uniform prior (see \fig{crab_hul_i-none} and Appendix \ref{app:priors}).
Also, for completeness, \fig{hulvsf} also includes the expected tensor
upper limits, $\hul{t}$. Note that those values are {\em not} the same as would
be obtained by the standard GR-only search, because that analysis looks at the
triaxial $h_0$ of \eq{lambda_gr}, rather than $\hT$.

\begin{figure}
\includegraphics[width=\columnwidth]{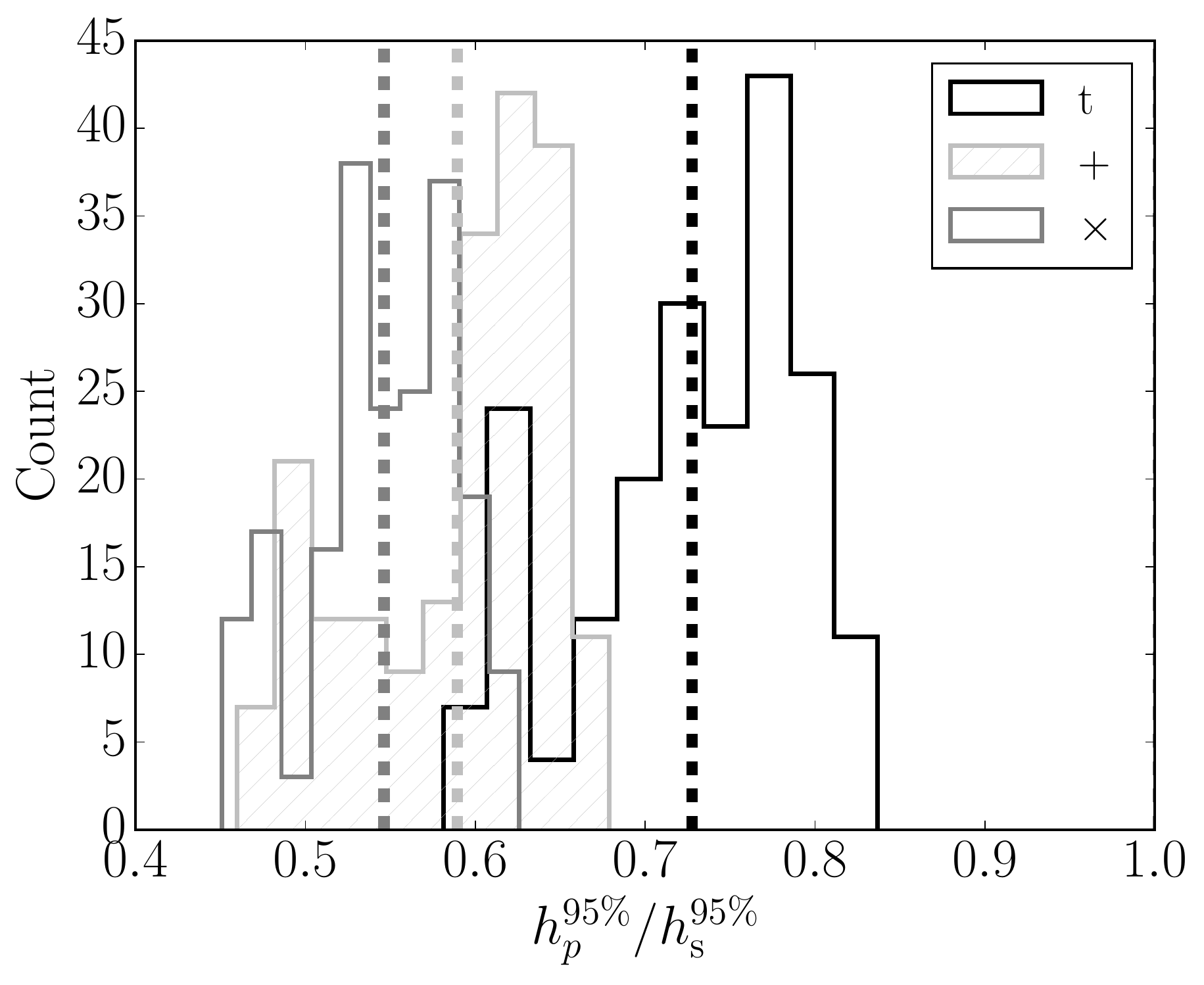}\\
\includegraphics[width=\columnwidth]{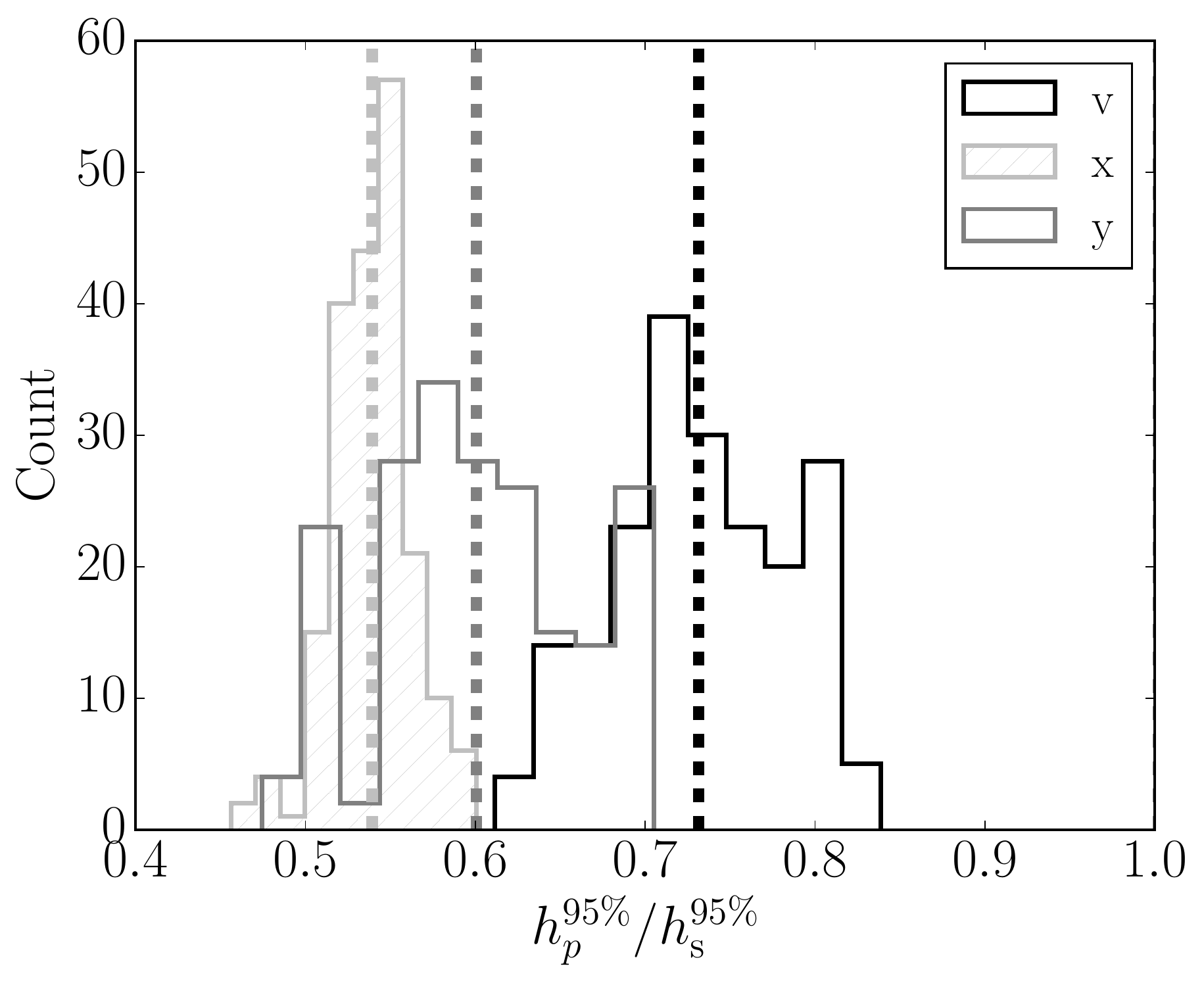}
\caption{{\em Tensor and vector upper limits as a ratio of scalar
upper limits}. Histogram of tensor (top) and vector (bottom) upper limits
divided by the scalar upper limit for each pulsar. The top plot shows ratios
for $\hul{t}$ (black), $\hul{+}$ (light gray, hatched), and $\hul{\times}$
(dark gray); the bottom plot shows ratios for $\hul{v}$ (black), $\hul{x}$
(light gray, hatched), and $\hul{x}$ (dark gray). Vertical dashed lines mark
the mean of each distribution.}
\label{fig:ulcomp}
\end{figure}

In order to compare our sensitivity to the different polarizations, in
\fig{ulcomp} we histogram the the tensor and vector upper limits as a ratio of
the scalar upper limits---this includes the $\hul{t}$ and $\hul{v}$ values
shown in \fig{hulvsf}, as well as the limits on the individual amplitudes from
which they are constructed ($\hul{+}$, $\hul{\times}$, $\hul{x}$ and
$\hul{y}$). The mean of these distributions (vertical dashed lines in
\fig{ulcomp}) indicate that, for most pulsars, the scalar upper limit is
slightly larger in magnitude than those for the $+$, $\times$, x or y modes;
this systematic effect is a manifestation of the decreased sensitivity of
quadrupolar detectors to scalar waves, which was discussed in Sec.\
\ref{sec:polarizations} (see, in particular, \fig{aps}). The fact that the
difference between $\hul{s}$ and $\hul{t}$, or $\hul{v}$, is less pronounced
can be easily be explained as a statistical factor arising from the definitions
of $\hT$ and $\hV$ as square-roots of sums of squares, Eqs.\ (\ref{eq:ht},
\ref{eq:hv}). Both these scalings are discussed in more detail in Appendix
\ref{app:uls}.

\section{Conclusion} \label{sec:conclusion}

We have developed a Bayesian framework to detect CW signals from known sources
regardless of polarization content, to disentangle the modes present in a given
signal, and to constrain the amplitudes of extra polarizations that may be
hiding under the noise. We have implemented this as an extension of LIGO's
Bayesian targeted CW search pipeline \cite{Pitkin2017}, and thus benefit from
the power of the nested sampling algorithm on which it is based.

We have tested our methods on one year of simulated noise for three
advanced-era detectors at design sensitivity (H1, L1, V1), and prepared for a
set of multiple known sources in their frequency band. This allows us to
estimate our future sensitivity to CW polarizations, in this most optimistic
case. Under these conditions and for the Crab pulsar in particular, we expect
signals of any polarization to become detectable for characteristic strain
amplitudes $h \gtrsim 3\times10^{-27}$ (Figs.\ \ref{fig:crab_scat} and
\ref{fig:st2d_signal_noise}); this threshold will vary among sources, due to
differences in position (sky location and orientation) and detector PSD at the
expected GW frequency (cf.\ e.g.\ \fig{lnbvsf}). Furthermore, the value of this
threshold will decrease linearly with the square-root of the observation time
\cite{Dupuis2005}.

A signal louder than the detection threshold will allow us to determine whether
its polarization content is consistent with GR or not, and the strength of this
statement will depend almost exclusively on the power of the non-GR component
(Figs.\ \ref{fig:crab_nGR_GR} and \ref{fig:crab_nGR_GR_i-ST}). In other words,
from a model-selection standpoint, the non-GR hypothesis will only be
unequivocally favored if the total power in non-GR modes is greater than the
threshold value, regardless of the strength of the GR modes. However, for
signals that do not satisfy this, we may always place upper limits on
nontensorial amplitudes and thus constrain deviations from GR; for instance,
\fig{hulvsf} presents the most optimistic projections for 95\%-credible upper
limits for scalar and vector amplitudes of CW signals from all pulsars in our
set ($\hul{s}\sim 4\times10^{-27}$ and $\hul{v}\sim 3\times10^{-27}$, in the
best case). As far as we are aware, these are the first generic estimates of
sensitivity to scalar and vector polarizations ever published \footnote{Note
that sensitivity estimates presented in \cite{Isi2015} were restricted to the
specific vector-only model of \cite{Mead2015}}.

From our projected upper limits, we have found that, at design sensitivity, the
LIGO-Virgo network will be generally less sensitive to continuous scalar
signals than to the individual vector or tensor modes by factors of 0.45--0.7,
depending on the location of the source (\fig{ulcomp}); this diminished
sensitivity to scalar modes stems from the quadrupolar nature of the detector
antenna patterns (\fig{aps} and Appendix \ref{app:uls}). Also, our injection
studies indicate that the upper limits on the amplitudes of nontensorial modes
will be roughly unaffected by the presence or absence of a tensor signal in the
data (\fig{crab_hbul_i-ST-VT}).

Although the results presented here made use of simulated Gaussian noise, the
procedure is identical for actual detector data. Furthermore, the
assumption of Gaussianity has been shown to hold relatively well for real CW
data \cite{Isi2015}, so the actual sensitivity limits should not be far from
those presented here. If the data are strongly non-Gaussian, however, one must
be careful in using $\ln\oddssn$ for detection purposes and may instead wish
to adopt one of the strategies suggested in Sec.\ \ref{sec:nongaussian}.

Another important limitation of our results is that here we only consider CW
signals emitted at $\fgw = 2\frot$, while it is to be expected that other
mechanisms (within GR or not) allow emission at other harmonics, $\fgw=\frot$
in particular. Yet, the only change required to account for this is to modify
the template in \eq{cw} to include terms at different harmonics; the ability to
do this already exists within our current infrastructure. We also assume that 
other aspects of the waves, like their speed, remain in agreement with the GR
prediction, an assumption that will be relaxed in a future study.

\begin{acknowledgments}
The authors would like to thank Ian Jones and Walter Del Pozzo for carefully
reading this manuscript and providing insightful suggestions; we also thank
Tjonnie Li and Carver Mead, as well as many colleagues in the LIGO Scientific
Collaboration Continuous Waves group, for many useful comments.
LIGO was constructed by the California Institute of Technology and
Massachusetts Institute of Technology with funding from the National Science
Foundation and operates under cooperative agreement PHY-0757058.
We are grateful for computational resources provided by Cardiff University, and
funded by an STFC grant supporting UK Involvement in the Operation of Advanced
LIGO.
MP is funded by the STFC under grant number ST/N005422/1.
Plots produced using \texttt{Matplotlib} \cite{Hunter2007}.
This paper carries LIGO Document Number LIGO-P1600305.
\end{acknowledgments}

\appendix
\section{Tensor models} \label{ap:tensor_modes}

\begin{figure}
\centering
\includegraphics[width=\columnwidth]{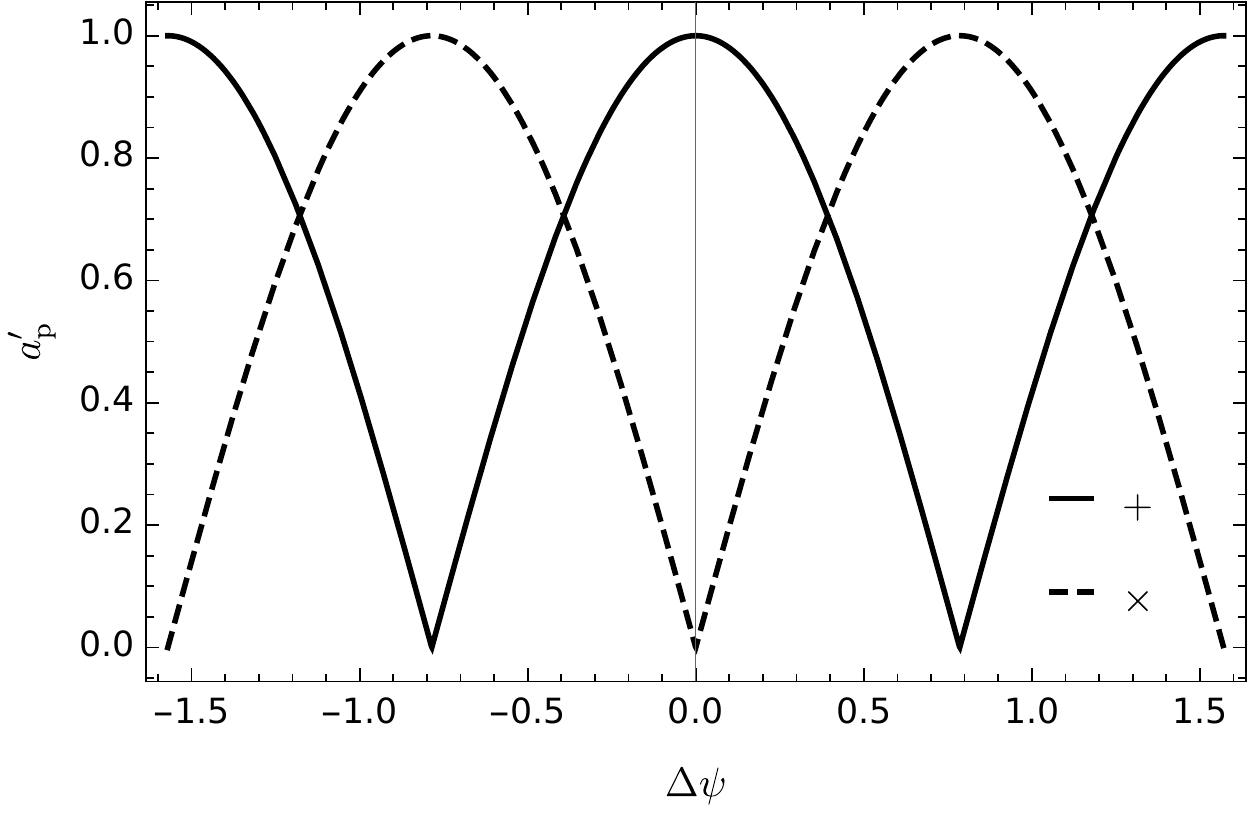}
\caption{{\em Effect of changing polarization angle}. Norm of the complex plus
($a'_+$, solid line) and cross ($a'_\times$, dashed line) weights after
rotating the source by $\Delta\psi$ in the plane of the sky, i.e.\ letting
$\psi\rightarrow\psi'=\psi+\Delta\psi$; this transformation is expressed in
Eqs.\ (\ref{eq:ap_psi}) and (\ref{eq:ac_psi}). In this case, we start from
$a_+=1$, $a_\times=0$ and $\psi=0$.}
\label{fig:rot_psi}
\end{figure}

A conceptual distinction can be drawn between the triaxial GR model and a
free-tensor model that includes $+$ and $\times$ but does not restrict their
relative amplitudes (denoted ``t''). The former has four free parameters
(overall amplitude, $h_0$; overall phase, $\phi_0$; inclination, $\iota$;
polarization, $\psi$) and corresponds to a signal template of the form [same as
\eq{lambda_gr}]:
\beq \label{eq:app_gr_template}
\Lambda_{\rm GR}(t) =\frac{1}{2} h_0 e^{i \phi_0}\left[\frac{1}{2}(1 +
\cos^2\iota)F_+(t; \psi) - i \cos\iota F_\times(t;\psi)\right].
\eeq
This is a reparametrization of the free-tensor model, which also has four
parameters (plus amplitude, $a_+$; cross amplitude, $a_\times$; plus phase,
$\phi_+$; cross phase, $\phi_\times$) and whose template is [same as
\eq{lambda_t}]:
\beq \label{eq:app_tensor_template}
\Lambda_{\rm t}(t) =\frac{1}{2} \left[ a_+e^{i\phi_+} F_+(t; \psi=0) + a_\times e^{i\phi_\times} F_\times(t;\psi=0) \right].
\eeq
If $\psi$ and $\iota$ are known, it is clear that the two models are different,
since $\hyp{GR}$ has two free parameters ($h_0$, $\phi_0$) and $\hyp{t}$ has
four ($a_+$, $a_\times$, $\phi_+$, $\phi_\times$). If the orientation is {\em
not} fixed, however, the two models span the same signal space. This is because
there is a degeneracy between $\psi$ and $a_+$, $a_\times$ due to the way the
antenna patterns transform under changes in $\psi$:
\beq \label{eq:Fp_psi}
F_+(t;\psi') = F_+(t;\psi) \cos 2\Delta\psi + F_\times(t;\psi) \sin
2\Delta\psi,
\eeq
\beq \label{eq:Fc_psi}
F_\times(t;\psi') = F_\times(t;\psi) \cos 2\Delta\psi - F_+(t;\psi)
\sin 2\Delta\psi,
\eeq
with $\psi'=\psi+\Delta\psi$. Eqs.\ (\ref{eq:Fp_psi}) and (\ref{eq:Fc_psi}) can
be derived from Eqs.\ (\ref{eq:Fp}) and (\ref{eq:Fc}) respectively, as in
\cite{Blaut2012} (or see, e.g., Sec.\ 9.2.2 in \cite{Thorne1987}).
Consequently, changing $\psi \rightarrow \psi'$ in \eq{app_tensor_template} is
equivalent to leaving $\psi$ fixed [at, say, $\psi=0$ as in
\eq{app_tensor_template}] while replacing the plus and cross complex amplitudes
by:
\beq \label{eq:ap_psi}
a_+' e^{i\phi_+'} = a_+ e^{i\phi_+} \cos 2\Delta\psi - a_\times
e^{i\phi_\times} \sin 2\Delta\psi,
\eeq
\beq \label{eq:ac_psi}
a_\times' e^{i\phi_\times'} = a_\times e^{i\phi_\times} \cos 2\Delta\psi + a_+
e^{i\phi_+} \sin 2\Delta\psi.
\eeq
This is illustrated in \fig{rot_psi}.

\begin{figure*}
\centering
\subfloat[GR, fixed orientation]{\label{fig:gr_post_300_fixed}\includegraphics[width=0.49\textwidth]{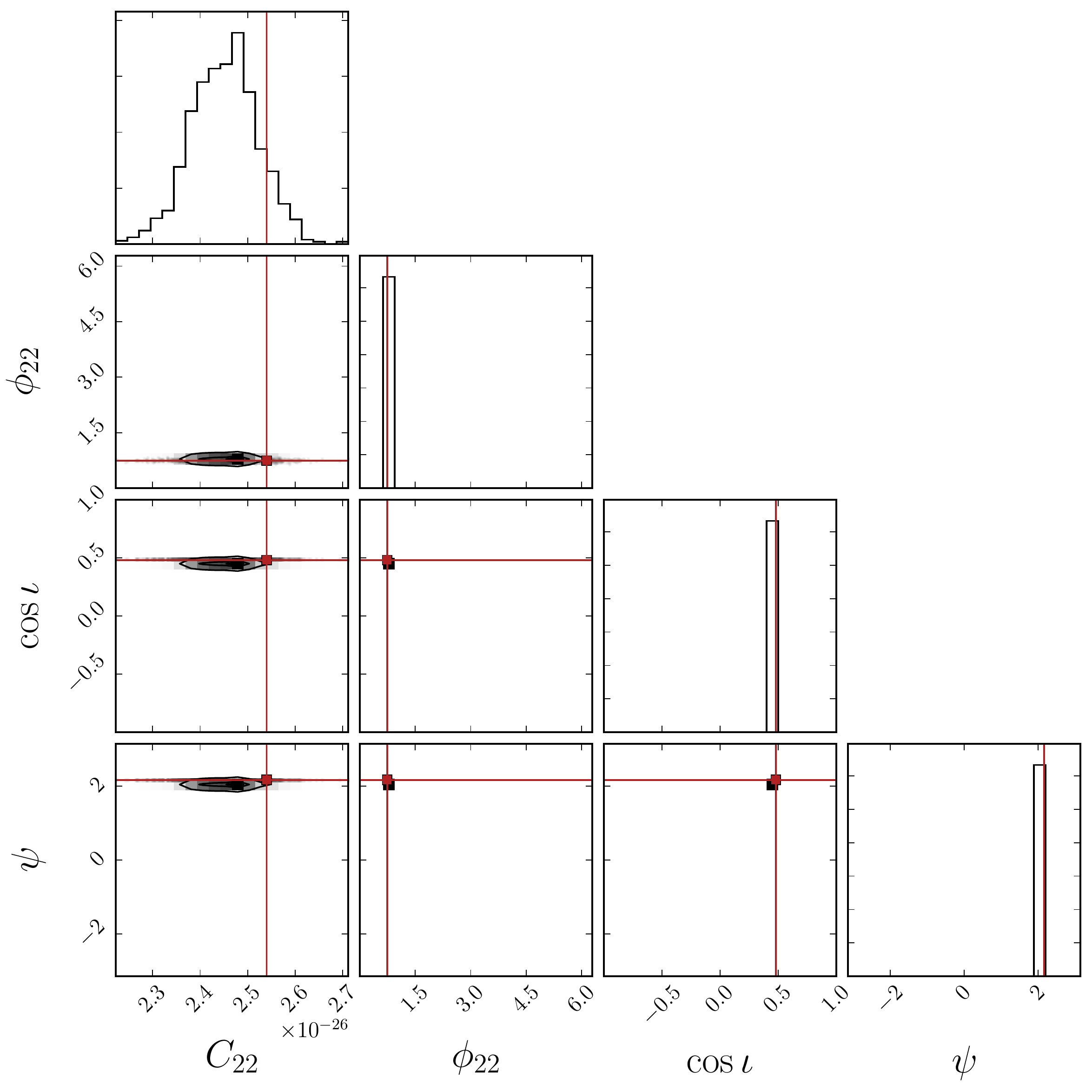}}\hfill
\subfloat[Free tensor, fixed orientation]{\label{fig:t_post_300_fixed}\includegraphics[width=0.49\textwidth]{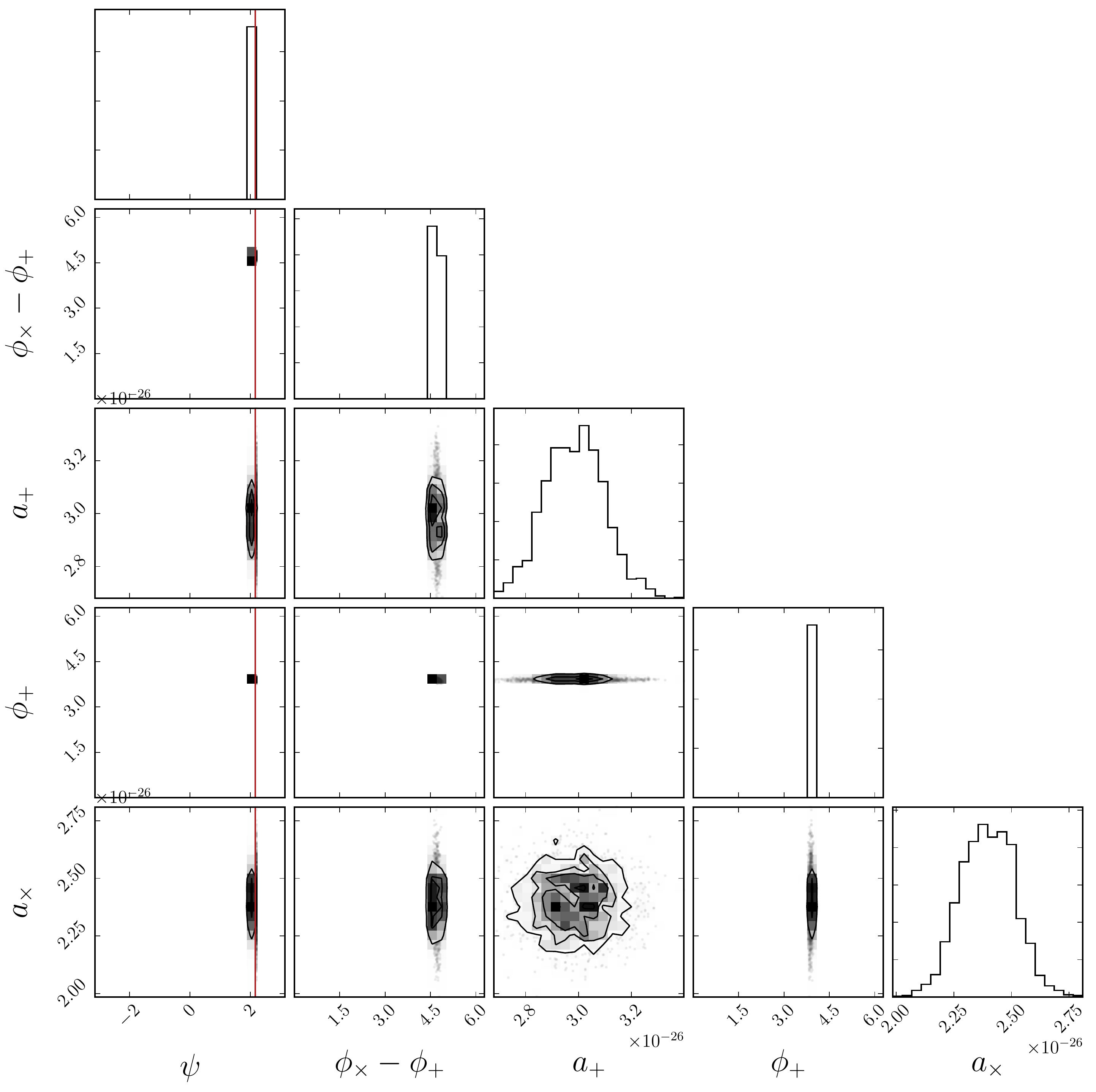}}\\
\subfloat[GR, unfixed orientation]{\label{fig:gr_post_300_unfixed}\includegraphics[width=0.49\textwidth]{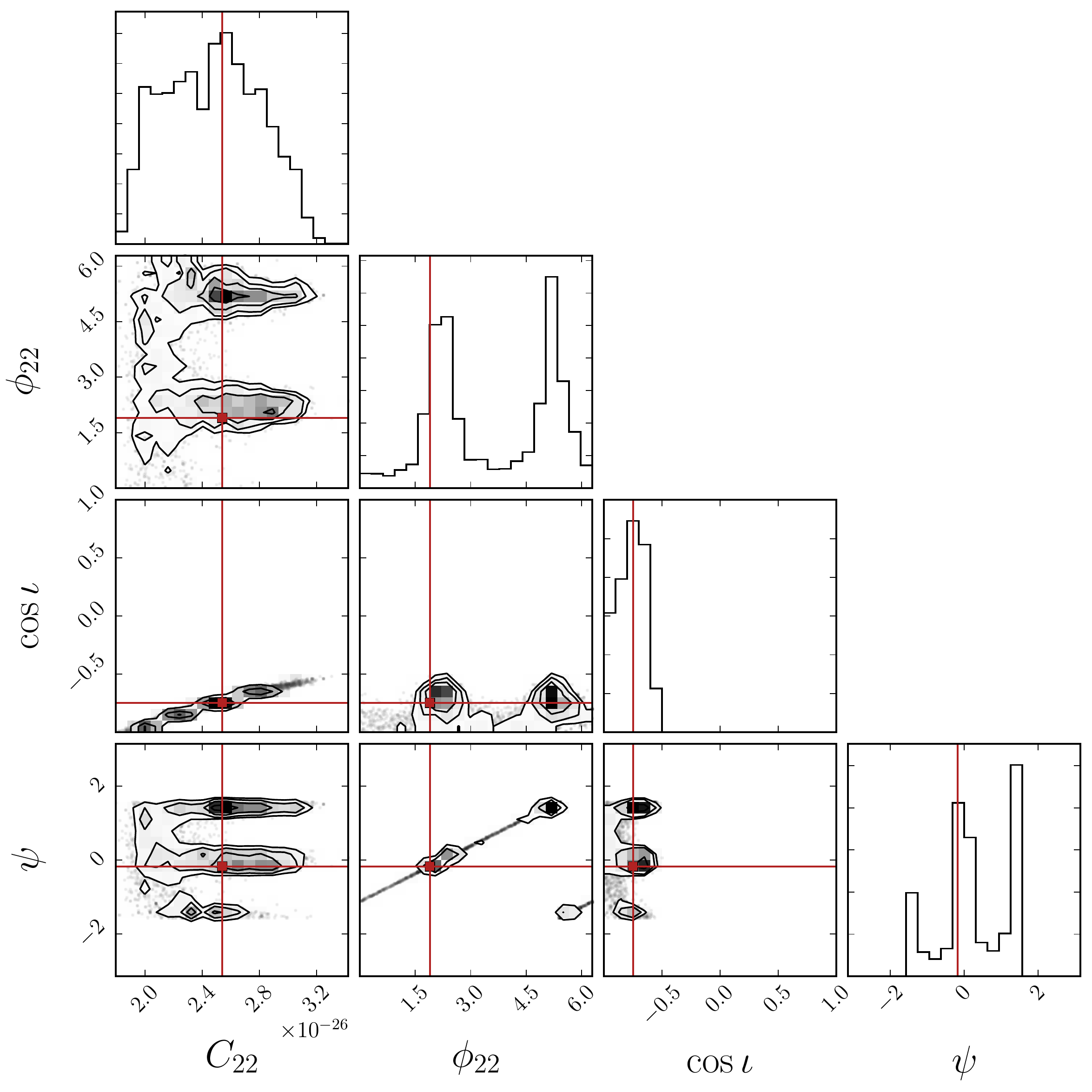}}\hfill
\subfloat[Free tensor, unfixed
orientation]{\label{fig:t_post_300_unfixed}\includegraphics[width=0.49\textwidth]{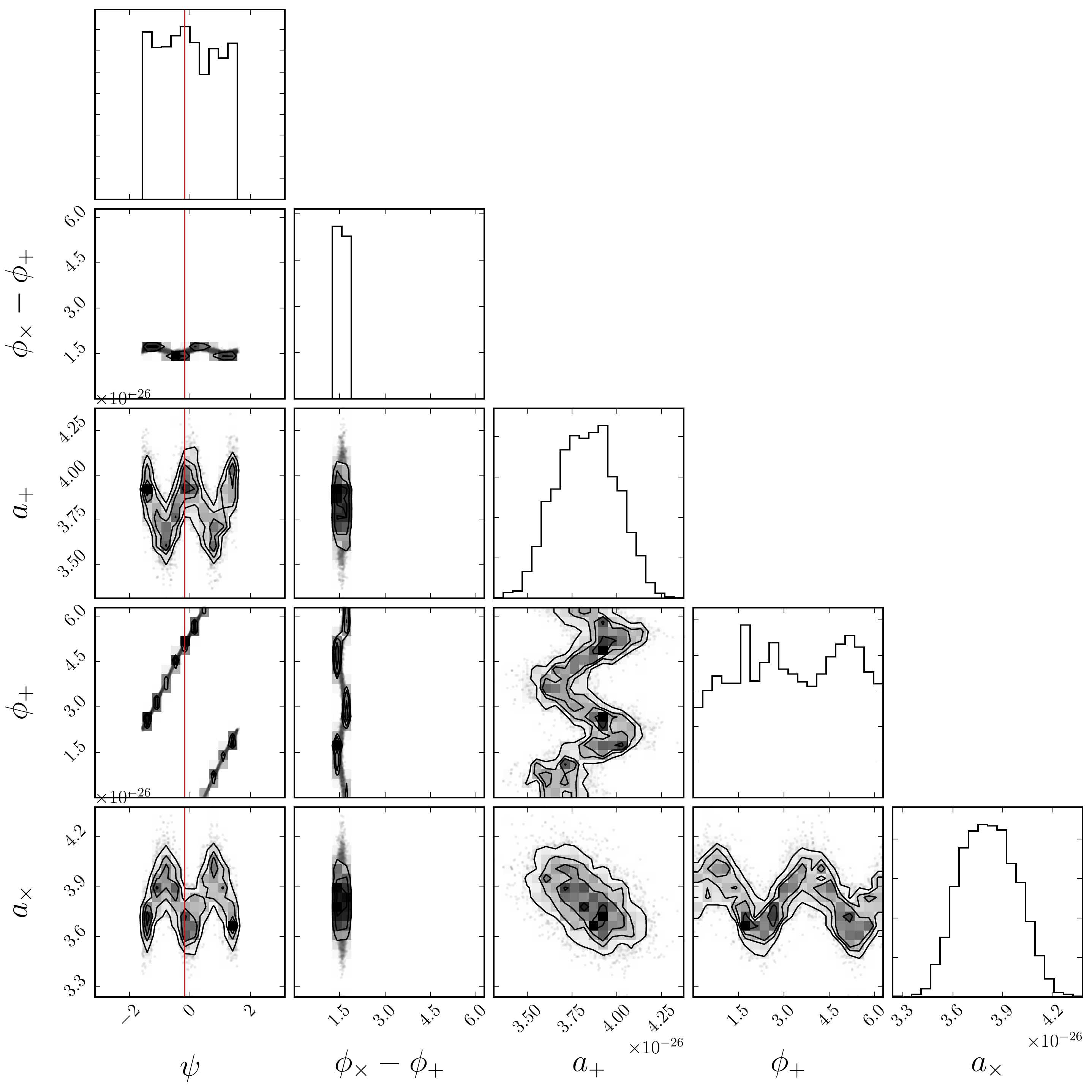}}
\caption{{\em Tensor posteriors in presence of signal}. Posterior PDFs for
parameters of $\hyp{GR}$ (left) and $\hyp{t}$ (right) with fixed (top) and
unfixed (bottom) source orientation ($\psi$, $\iota$). Each panel consists of a
{\em corner plot} displaying the two-dimensional posteriors for each pair of
parameters as indicated by the $x$ and $y$ labels, with the diagonals showing a
histogram of the one-dimensional PDF for each parameter [i.e.\ the 1D PDF
obtained after marginalization of the multidimensional posterior PDF all other
quantities, as in \eq{post}]. The data analyzed contain signals with parameters
indicated by the red lines; note that $C_{22}=h_0/2$ is the quantity that was
actually used to parametrize GR triaxial amplitudes in the code
\cite{Pitkin2017}. In both (a) and (b), $\cos\iota$ and $\psi$ are fully known,
and their resolution in these plots is limited by binning only. These plots
were produced using the \texttt{corner.py} package \cite{Foreman-Mackey2016}.} 
\label{fig:tensor_post_300}
\end{figure*}

\begin{figure*}
\centering
\includegraphics[width=0.49\textwidth]{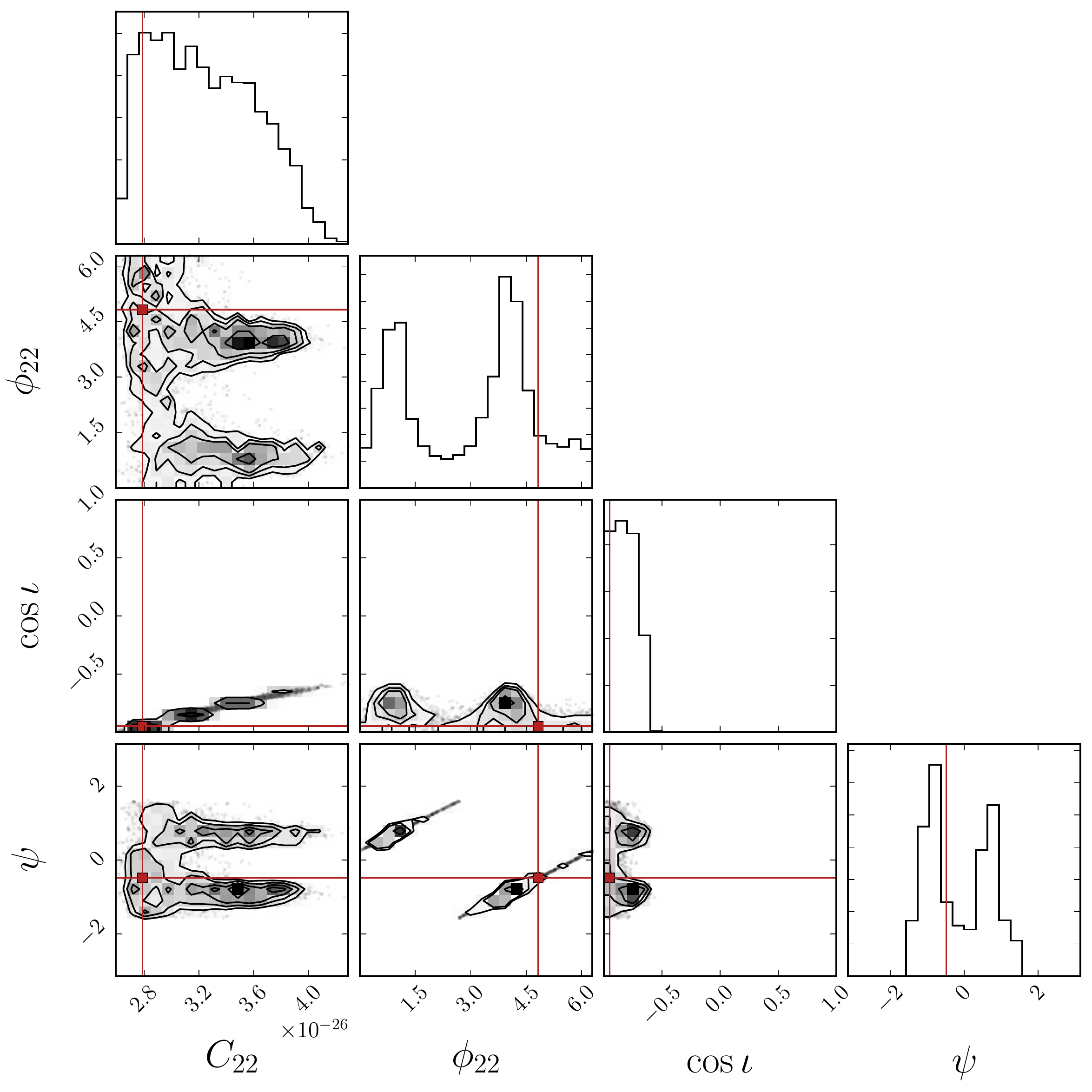}\hfill
\includegraphics[width=0.49\textwidth]{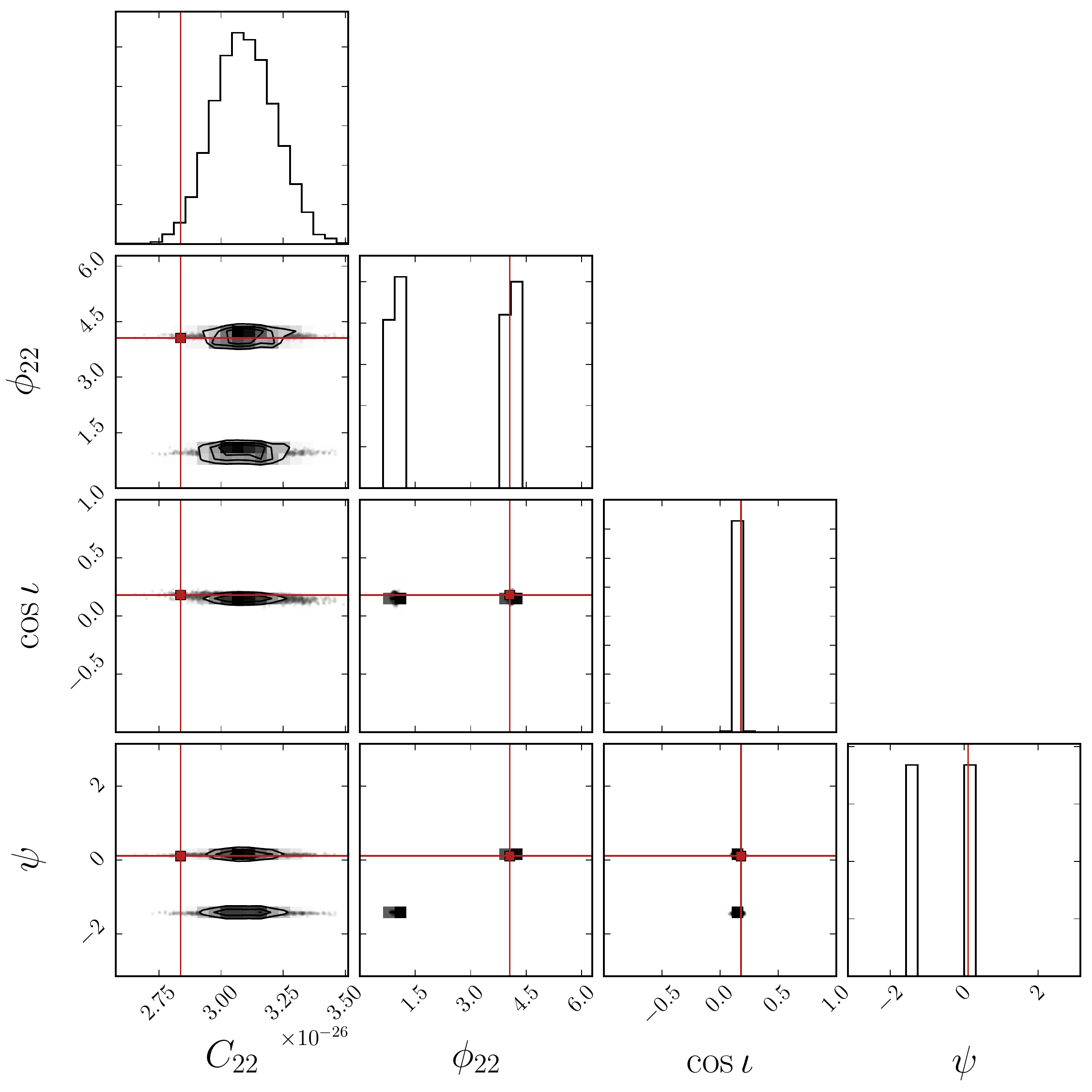}
\caption{{\em Effect of inclination}. Posterior PDFs for parameters of
$\hyp{GR}$ with unfixed source orientation ($\psi$, $\iota$). Each panel
consists of a {\em corner plot} displaying the two-dimensional posteriors for
each pair of parameters as indicated by the $x$ and $y$ labels, with the
diagonals showing a histogram of the one-dimensional PDF for each parameter.
The data sets analyzed contain signals with parameters indicated by the red
lines; note that $C_{22}=h_0/2$ is the quantity that was actually used to
parametrize GR triaxial amplitudes in the code \cite{Pitkin2017}. On the left,
the injected signal corresponds to a face-off source ($\cos\iota\approx-1$),
making it difficult to constrain the polarization angle $\psi$; on the right,
the injection has similar amplitude but corresponds to an edge-on source
($\cos\iota\approx 0$), making it easy to constrain $\psi$ [modulo $\pi/2$ due
to the $2\Delta\psi$ dependence of Eqs.\ (\ref{eq:Fp_psi}) and
(\ref{eq:Fc_psi})]. These plots were produced using the \texttt{corner.py}
package \cite{Foreman-Mackey2016}.}
\label{fig:gr_post_cosiota}
\end{figure*}

\begin{figure*}
\centering
\subfloat[Fixed orientation]{\includegraphics[width=0.5\textwidth]{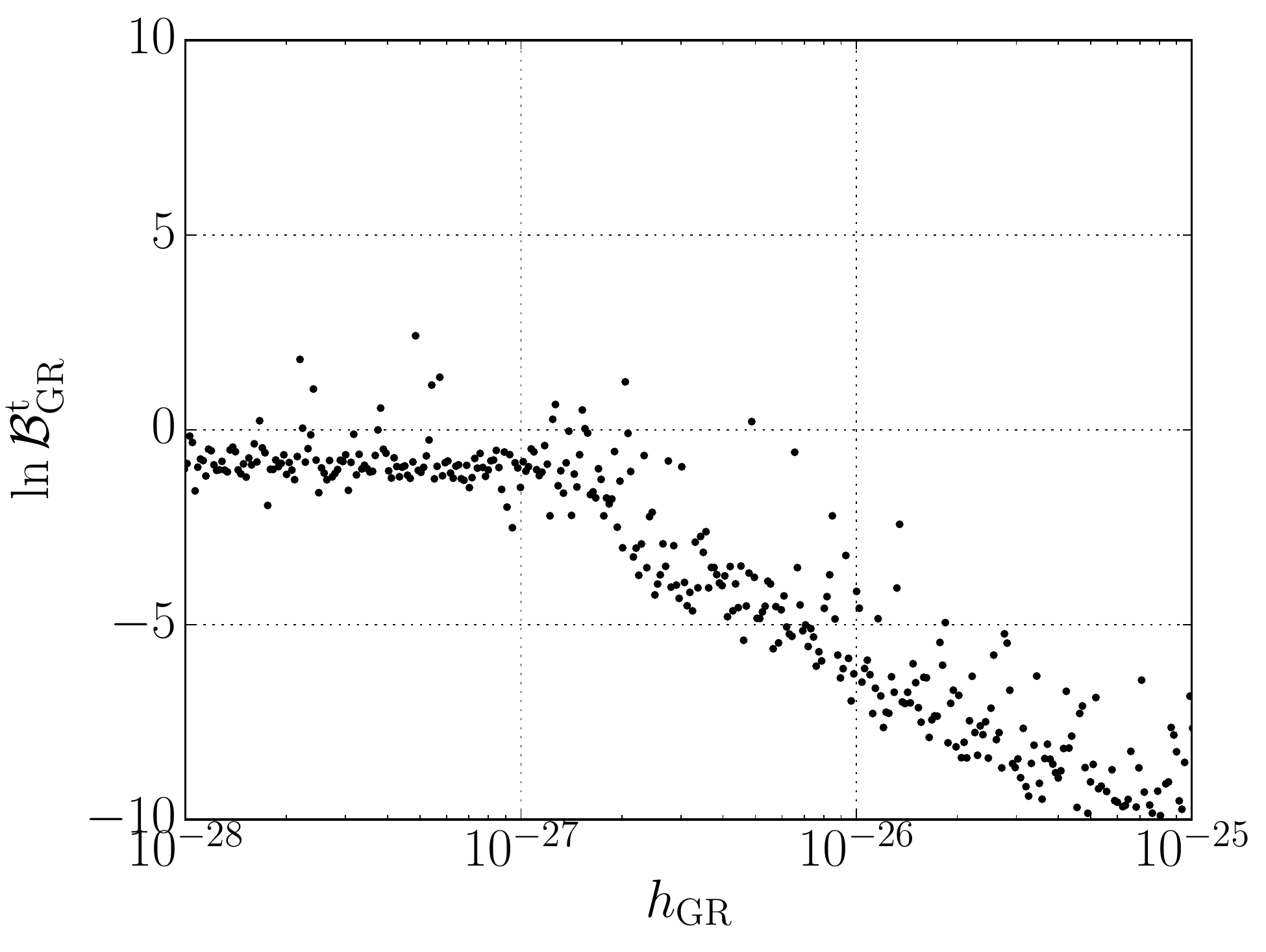}}%
\hfill
\subfloat[Unfixed orientation]{\includegraphics[width=0.5\textwidth]{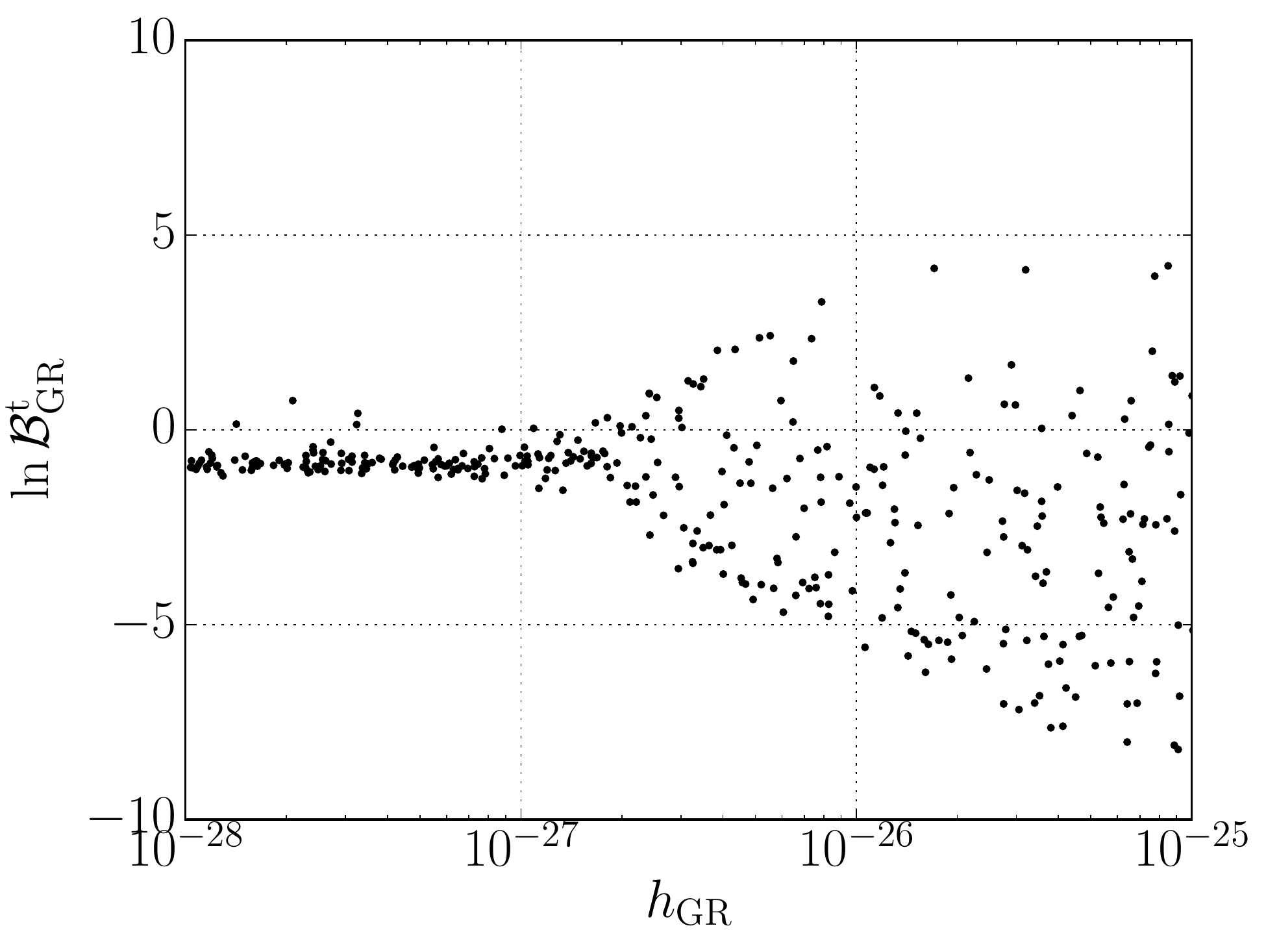}}
\caption{{\em Free-tensor vs GR}. Natural logarithm of the Bayes factor
comparing $\hyp{t}$ to $\hyp{GR}$, as a function of GR injection amplitude for
fixed (left) and unfixed (right) source orientation. On the left, the analysis
correctly gives preference to $\hyp{GR}$ for signals above the detection
threshold; on the right, however, the analysis is unable to satisfactorily
distinguish between $\hyp{t}$ and $\hyp{GR}$, due to the orientation
degeneracies discussed in appendix \ref{ap:tensor_modes}.}
\label{fig:tensor_gr}
\end{figure*}

These rotational properties are easily understood by recalling that GW
polarizations can be defined in {\em any} frame, although a given signal might
look more or less simple given the choice of frame. Eqs.\ (\ref{eq:Fp_psi}) and
(\ref{eq:Fc_psi}) provide the transformation between frames that are coaligned
except for a rotation of $\Delta\psi$ around their $z$-axes. Because waveform
predictions, e.g.\ \eq{app_gr_template}, are made in {\em specific} frames, it
is important to orient the wave frame appropriately when working with a given
theory and emission mechanism. However, if the signal parametrization is
independent of any theory, e.g.\ \eq{app_tensor_template}, one is free to pick
any frame (i.e.\ any $\psi$).

The relationship between the different tensor model parameters is reflected in
the posterior probability plots of \fig{tensor_post_300}. For fixed
orientation, both the triaxial (a) and free-tensor (b) analyses accurately
determine the amplitude and phase of the injected signal. In panel (b), $a_+$
and $a_\times$ are constrained to lie within a region consistent with $h^2_{\rm
inj} = a_+^2 + a_\times^2$ and $a_+/a_\times = (1+\cos^2\iota)/(2\cos\iota)$,
for an effective injection amplitude given by $h_{\rm inj}^2 = h_0^2
(1+\cos^2\iota)^2/4+h_0^2\cos^2\iota$, as in \eq{ht}. When the orientation is
allowed to vary, we observe the expected correlations between the recovered
triaxial amplitude and the orientation parameters in panel (c); in this case,
$\psi$ and $\cos\iota$ will also become correlated, as better shown in
\fig{gr_post_cosiota}.

The degeneracy between $\psi$ and $a_+$, $a_\times$ is particularly evident in
\fig{t_post_300_unfixed}, where the one-dimensional PDF for $\psi$ shows that
this parameter cannot be constrained, even for a loud signal. Furthermore,
joint posteriors between $\psi$ and $a_+$ \& $a_\times$ confirm that this is
due to the degeneracy from Eqs.\ (\ref{eq:ap_psi}) and (\ref{eq:ac_psi}), as
seen by comparing these two-dimensional PDFs to \fig{rot_psi}. Physically, this
is a consequence of the fact that we are free to orient the polarization frame
as we wish. 

Because their signal templates are degenerate when $\psi$ and $\iota$ are
allowed to vary, the distinction between $\hyp{GR}$ and $\hyp{t}$ is not really
meaningful for unfixed orientation. This can be seen from the values of
$\ln\bayes{t}{GR}$ in the cases of known and unknown orientations, as in
\fig{tensor_gr}. On the left panel, $\hyp{GR}$ is defined with specific values
of $\psi$ and $\cos\iota$ that match the injections; on the right, the
$\hyp{GR}$ priors allow $\psi$ and $\cos\iota$ to range over their full ranges,
and the injections are performed with random values of both. When the
orientation is fixed, $\hyp{GR}$ will always be preferred to $\hyp{t}$ for
resolvable signals because of its lower Occam's penalty; however, that is not
true for unfixed orientation. Note that, in the strictest sense, the two
hypotheses are not logically equivalent unless their parameter priors are
related by the Jacobian of the coordinate transformation between the two
parametrizations, Eqs.\ (\ref{eq:app_gr_template},
\ref{eq:app_tensor_template}); only in that case all regions of signal space
are treated equally by $\hyp{GR}$ and $\hyp{t}$. This explains the variation in
values of $\ln\bayes{t}{GR}$ on the right of \fig{tensor_gr}.

If one knew the source orientation and one believed that the {\em only} viable
mechanism for producing CWs at the assumed frequency in GR is the triaxial
model embodied by \eq{app_gr_template}, then one could include the free-tensor
hypothesis and all of its derivatives (i.e.\ t, st, vt, stv) in the non-GR set
$\tilde{M}$, on top of $\{\rm s, v, sv, GR+s, GR+v, GR+sv\}$. Doing so would
mean treating a tensor-only signal that does not conform to
\eq{app_gr_template} as evidence of a GR violation, rather than of a different
emission mechanism within GR. Given the many simplifications intrinsic to the
triaxial model, however, having that much confidence in its validity seems
unwarranted; hence we choose to not take that approach.

\section{Amplitude priors} \label{app:priors}

Previous CW Bayesian searches targeted to known pulsars have always applied a
flat prior on the signal amplitude parameter \cite{o1cw}. This is because flat
priors, if wide enough, cause the posterior to be only determined by the
likelihood (up to normalization), yielding more conservative upper limits on
the signal strength. Furthermore, unlike with priors uniform in the logarithm
of the quantity, upper limits derived with flat priors will generally not
depend on the limits set by the prior (again, assuming the range allowed
extends from zero amplitude to some large value that does not truncate the
likelihood).

\begin{figure}
\centering
\includegraphics[width=\columnwidth]{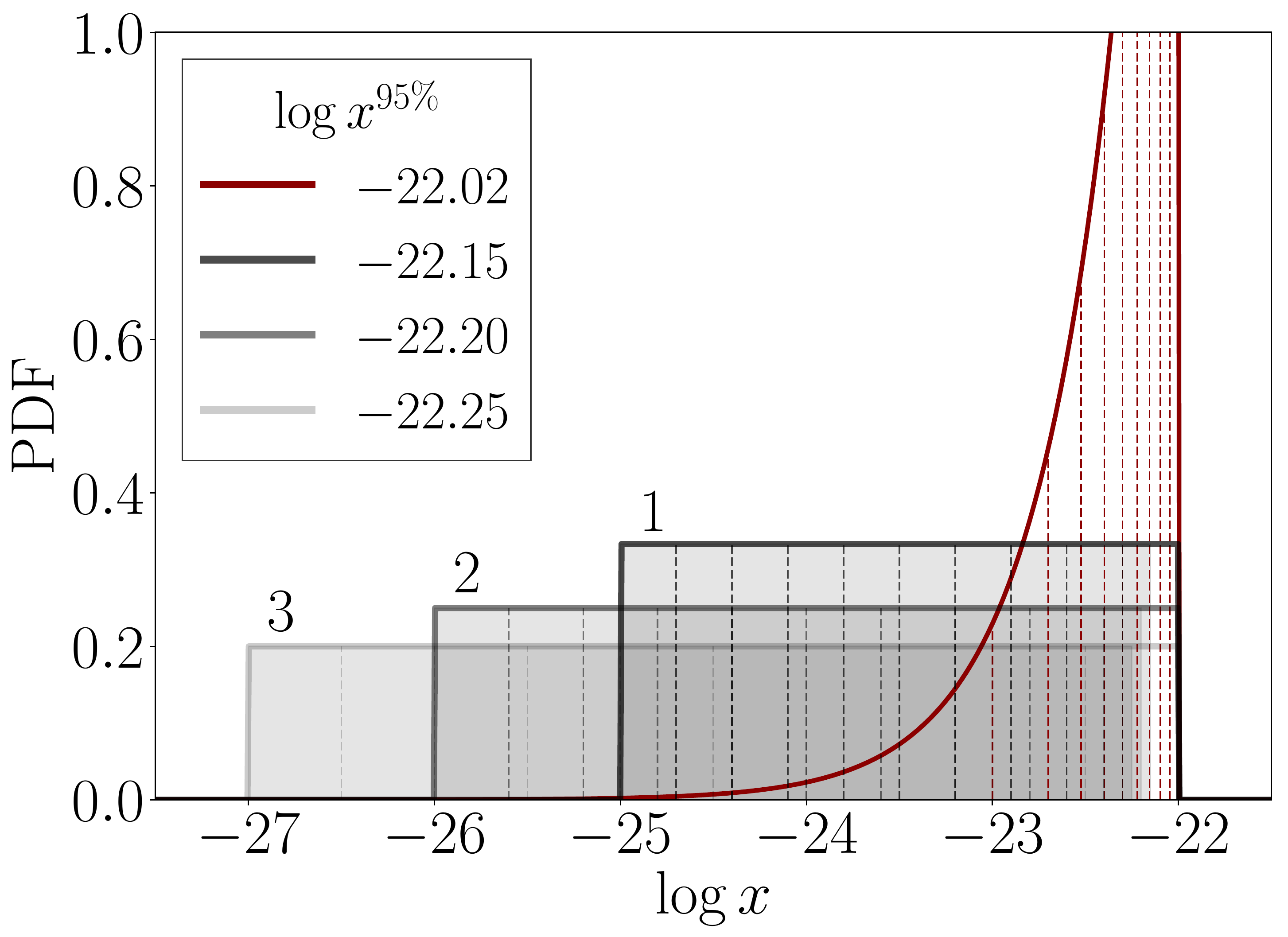}
\caption{{\em Log-uniform prior and upper limits}. For a 1D random variable
$x$, we show the probability densities corresponding to a uniform likelihood
with upper cutoff $\log\xmax=-22$ (red) and log-uniform priors with different
lower cutoffs ($\log\xmin=-25$ for box 1, $\log\xmin=-26$ for box 2 and
$\log\xmin=-27$ for box 3). Vertical dashed lines mark areas of equal
probability mass for each distribution. The combined effect of the likelihood
and each of the prior distributions is to produce 95\%-credible upper limits on
$x$ with values shown in the legend. The value obtained using only the
likelihood corresponds to that obtained with a uniform prior with a broad
enough range. As expected from \eq{xul_change}, the upper limit is not very
sensitive to the lower bound set by the prior.}
\label{fig:loguniform_ul_ex}
\end{figure}

Upper limits obtained using log-uniform priors (uniform in the logarithm of
the quantity) will, generally, be dependent on the range of the prior, although
not strongly. For example, consider a one-dimensional problem on some positive
parameter $x$. For simplicity, further assume we have a flat likelihood between
$x=0$ and an upper cutoff at $x=\xmax$; then, $\xmax$ will necessarily also be
an upper bound for the posterior. Because the likelihood is uniform, below the
cutoff the posterior will be determined, up to normalization, by the prior
only, i.e. for $x<\xmax$,
\beq
p(x\mid\data,\hyp{}) \propto p(x\mid\hyp{}).
\eeq
Now consider a log-uniform prior $p(x\mid\hyp{}) \propto {\rm
d}(\log{x})\propto 1/x$, with a lower bound $\xmin$, such that $0<\xmin<\xmax$.
Because such prior is uniform in the $\log x$, this implies that the
95\%-credible upper limit on $x$ will be given by:
\begin{align}
\log\xul &= \log\xmin + 0.95(\log{\xmax}-\log{\xmin}) \nonumber \\
&= \log \left(\xmax^{0.95}/\xmin^{0.95-1} \right).
\end{align}
Since $\xmax$ is set by the likelihood (by construction), if the prior is
changed by rescaling $\xmin$ by a factor $\alpha$,
\beq
\xmin \rightarrow \xmin'= \alpha \xmin,
\eeq
then, for a given set of data, the upper limit becomes $\xul_{\alpha}$,
satisfying:
\beq \label{eq:xul_change}
\xul_{\alpha} /\xul = \alpha^{0.05}.
\eeq
Thus, the dependence of the upper limit on the range defined by the log-uniform
prior is quite weak, as illustrated in \fig{loguniform_ul_ex}. This explains
why upper limits obtained with a log-uniform prior differ only by a factor
of a few from those obtained with a flat one, as seen in \fig{crab_hul_i-none}.

\begin{figure}
\centering
\includegraphics[width=\columnwidth]{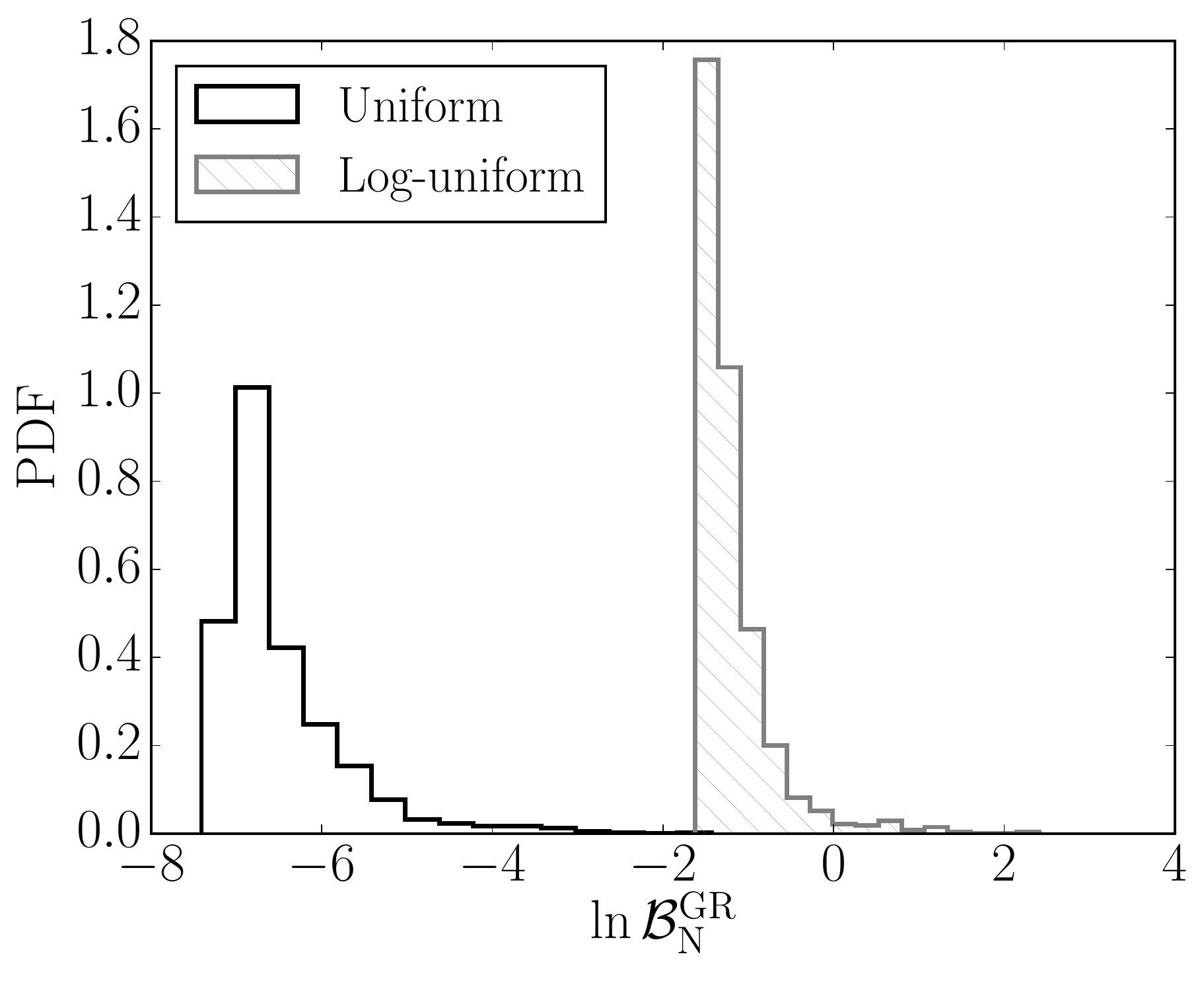}
\caption{{\em Log-uniform vs flat amplitude priors}. The logarithm of the GR vs
noise Bayes factor is computed for \red{1000} instantiations of Crab pulsar
noise. For the GR amplitude $h_0$, we apply priors uniform in the quantity
(black) and uniform in the logarithm of the quantity (hatched gray). The flat
prior causes one to more strongly favor the noise model, due to a larger
implicit Occam's penalty.}
\label{fig:lnb_flatvslog}
\end{figure}

However, the flat priors do not properly represent our ignorance of the scale
of the signal amplitude. This problem manifests itself in negative Bayes
factors that too quickly favor the noise hypothesis if no loud signal is
clearly present, rather than reflecting our expectation that a signal might be
hiding under the noise. This can be seen in \fig{lnb_flatvslog}, where we show
the distributions of $\ln\bayes{GR}{N}$, obtained for several noise-only data
instantiations for the Crab pulsar, corresponding to flat and log-uniform
priors in the GR amplitude parameter, $h_0$; a uniform prior results in lower
values of $\ln\bayes{GR}{N}$ that strongly favor $\hyp{N}$. This behavior is
not specific to the GR model.

For most of our analysis, we choose to apply priors uniform in the logarithm of
all amplitude quantities. However, for the sake of consistency with previous 
searches and in order to make our limits more conservative, we also present
upper limits produced using flat amplitude priors, as shown in 
\fig{crab_hul_i-none}.

\section{Numerical error} \label{app:errors}

The fractional numerical error in the computation of the natural logarithm of
the evidence by nested sampling is usually estimated by: 
\beq \label{eq:error}
\delta \left[ \ln \evidence\right] \sim \sqrt{H/N_{\rm live}}, 
\eeq
where $N_{\rm live}$ is the number of of live points and $H$ is the information
gained in the analysis:
\beq
H \equiv \int_\Theta \posterior ~ \ln \frac{\posterior}{\prior} {\rm
d}\vec{\theta},
\eeq
a quantity that is easy to estimate from the output of the nested sampling code
\cite{Skilling2006, Keeton2011}.

\begin{figure}
\centering
\includegraphics[width=\columnwidth]{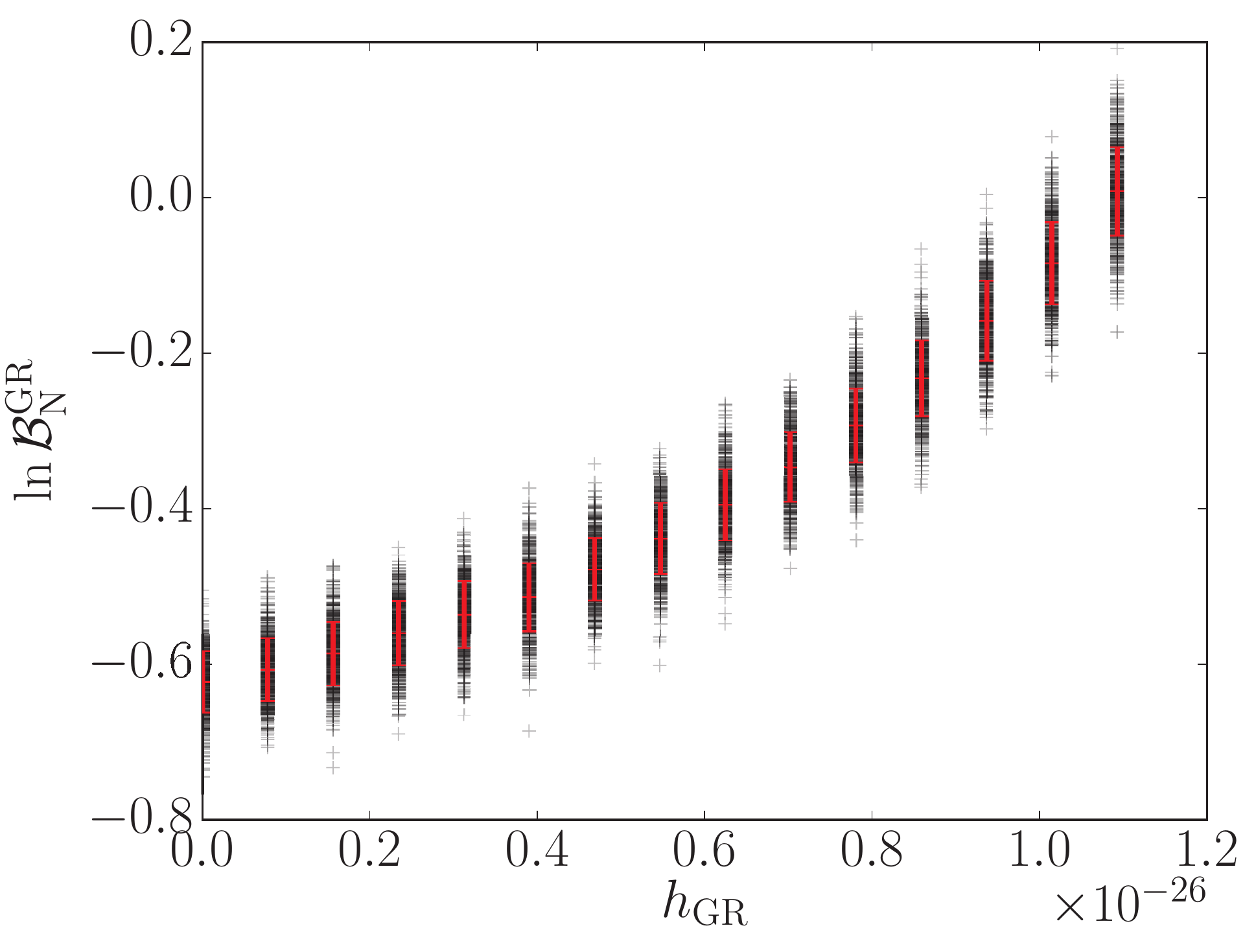}
\caption{{\em Numerical error in Bayes factor computation}. The logarithm of
the GR vs noise Bayes factor is computed \red{500} times for different values
of injected GR signal amplitude. The noise realization is not varied between
computations with the same injection strength, only the seed for the random
number generator used by the nested sampling algorithm. The red bars mark one
standard deviation around the mean.}
\label{fig:deltaz}
\end{figure}

\begin{figure}
\centering
\includegraphics[width=\columnwidth]{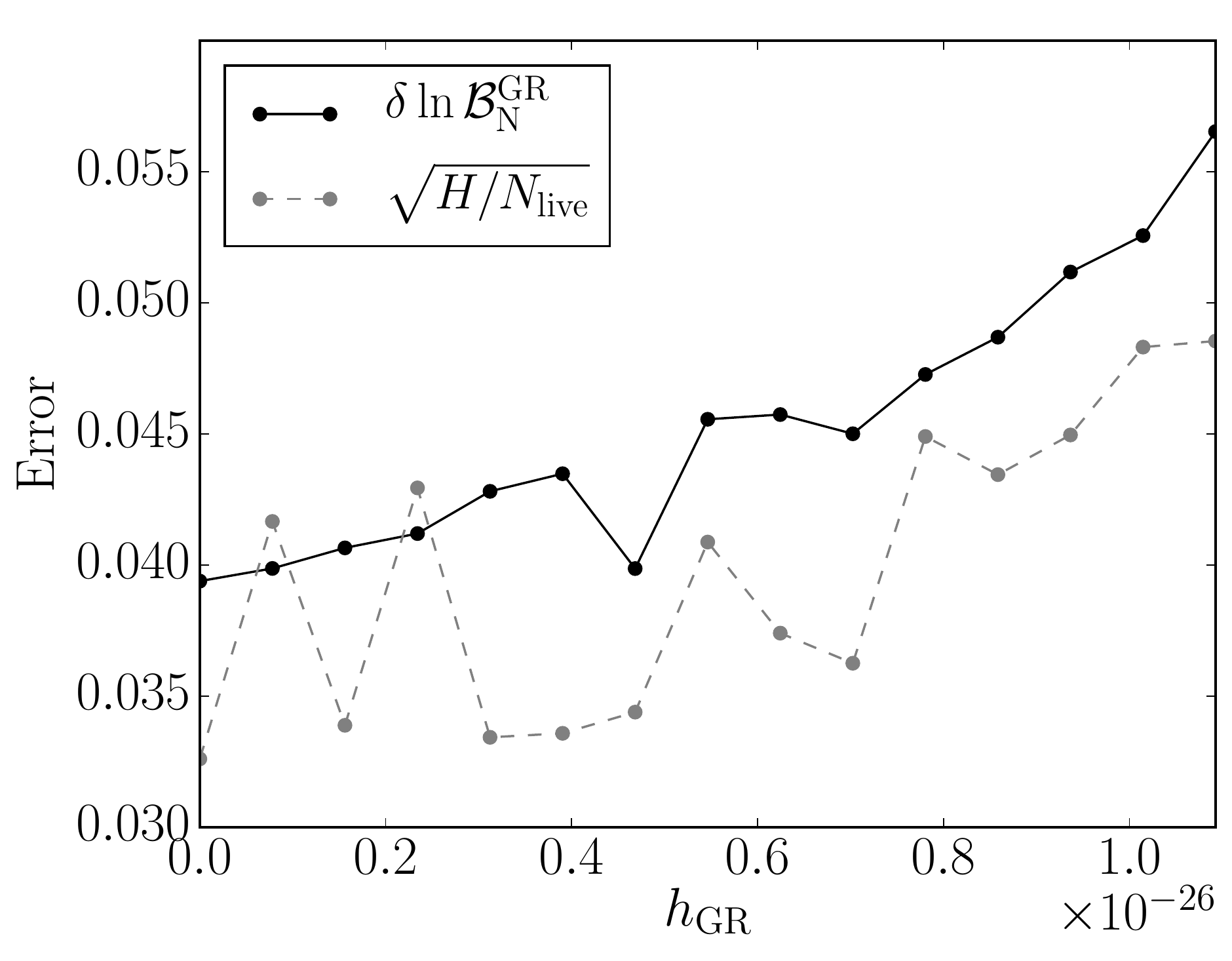}
\caption{{\em Observed error and prediction}. Error in the computation of the
logarithm of the GR vs noise Bayes factor as a function of injected GR signal
amplitude. The solid, black curve corresponds to measured standard deviations
from the computation of $\ln\bayes{GR}{N}$ \red{500} times per injection
strength (red bars in \fig{deltaz}). The dashed, gray curve shows the
theoretical prediction for the error in the logarithm of the evidence,
\eq{error}.}
\label{fig:deltaz_info}
\end{figure}

An example of the actual statistical error as function of SNR is presented in
Figs.\ \ref{fig:deltaz} \& \ref{fig:deltaz_info}, where the injected GR signal
amplitude serves as proxy for $\rho$ (for fixed PSD). From these plots it
becomes apparent that, although the actual error might exceed the estimator of
\eq{error}, its absolute magnitude is quite small and should not affect our
results. In any case, \eq{error} indicates that any level of accuracy may be
achieved by increasing the number of live points (at the cost of increased
computational burden). For more details on the numerical error of the nested
sampling algorithm in \texttt{LALInference}, we refer the reader to Sec.\ IVB
of \cite{Veitch2010}.

\section{Upper-limit ratios} \label{app:uls}

When comparing upper limits for the different modes, as in \fig{ulcomp}, two
scalings become apparent: first, the $+$, $\times$, x, and y upper limits
are, on average, more stringent than those for the scalar polarization
by a factor of ${\sim}1.8$; second, the upper limits on $\hT$ ($\hT$) are a
factor of ${\sim}1.3$ larger than those on the individual $+$ and $\times$ (x
and y) amplitudes.

The scaling between the scalar upper limit and those for the other individual
strain amplitudes can be accounted for by the decreased sensitivity of
quadrupolar GW detectors to scalar waves. For a single instrument (that is, not
a network), this can be appreciated visually from \fig{aps}, by noting that for
most sky locations the magnitude of the scalar response is considerably less
than for the other modes.

To properly evaluate the effect of the detector geometry on the analysis,
however, it is necessary to look at the relative SNRs of unit-amplitude scalar,
vector and tensor GWs from a given source, as they are received by the detector
network under consideration (H1, L1, V1) after some fixed observation time.
Assuming all detectors have comparably noise levels, the network SNR can be
proxied by the root-mean-square (RMS) amplitude of the effective network
antenna patterns, defined by
\beq
F^{\rm rms}_{p, \,{\rm net}} = \sqrt{ \frac{1}{T} \int_0^T \sum_d F_{p,d}^2(t)},
\eeq
for each polarization $p$, some long observation time $T$, and where the sum is
over detectors $d$. [Here we have fixed the source and detector parameters so
that the $F_p$'s of Eqs.\ (\ref{eq:Fp}--\ref{eq:Fb}) are now just simple
functions of time.] We may then compute this for all five polarizations and for
multiple sources to obtain a sky-average of the ratio of the scalar RMS antenna
pattern to those of the other polarizations. We find this ratio to be roughly
${\sim}0.55$ for all polarizations, in agreement with \fig{ulcomp}, since we
should expect
\beq
\left\langle \frac{F^{\rm rms}_{\rm s, \,net}}{F^{\rm rms}_{p, \, {\rm net}}}
\right\rangle \sim \left\langle \frac{h^{95\%}_p}{\hul{s}}  \right\rangle,
\eeq
where the average $\langle\cdot\rangle$ is taken over multiple sources
distributed across the sky.

The relation between the $\hT$ ($\hV$) upper limits and those for their
component amplitudes, $+$ and $\times$ (x and y), can be easily understood by
noting that, if using flat priors and in the absence of signal, the
marginalized posteriors for each of the component amplitudes ($h_+$,
$h_\times$, $h_{\rm x}$, $h_{\rm y}$) will roughly be described by a one-sided
normal distribution. Consequently, it can be shown that posterior for the
square-root of the sum of the squares of two of these quantities will be given
by a {\em chi distribution} with two degrees of freedom. Considering the
definitions of Eqs.\ (\ref{eq:ht}, \ref{eq:hv}). It is straightforward to show
(numerically or analytically) that this explains the observed factor of
${\sim}1.3$ difference between $\hul{t}$ ($\hul{v}$) and $\hul{+}$ or
$\hul{\times}$ ($\hul{x}$ or $\hul{y}$).

\bibliography{gw,statistics}

\end{document}